\documentclass[11pt]{article}
\usepackage[margin=1in]{geometry}
\usepackage{amsmath,amssymb,amsthm}
\usepackage{hyperref}
\usepackage{setspace}
\usepackage{graphicx}
\usepackage{subcaption}
\usepackage{float}
\usepackage{xcolor}
\usepackage{natbib}
\usepackage{pdflscape}
\usepackage{epigraph}
\usepackage{tabularx}
\usepackage{booktabs}
\usepackage{ragged2e}
\usepackage{fvextra}

\usepackage{soul}
\usepackage{setspace}

\sethlcolor{yellow} % default highlight (yellow)

\newcommand{\hlcyan}[1]{%
  \begingroup
  \sethlcolor{cyan}
  \hl{#1}
  \endgroup
}

\definecolor{ContentColor}{HTML}{2F855A} % green
\definecolor{StyleColor}{HTML}{3182CE}   % blue
\definecolor{TiltColor}{HTML}{DD6B20}    % orange

\newtheorem{assumption}{Assumption}

\newtheorem{remark}{Remark}

\title{Content vs.\ Form: What Drives the Writing Score Gap Across Socioeconomic Backgrounds? A Generated Panel Approach}

\author{
Nadav Kunievsky\footnote{Knowledge Lab, University of Chicago}
\and
Pedro Pertusi\footnote{Insper, Institute of Education and Research}\thanks{We thank Maria Goldshtein for extremely helpful comments on an earlier iteration of this paper and Dawoon Jeong for helpful discussion.}
}
\date{\today}

\begin{document}
\maketitle

% \begin{enumerate}
%     \item robustness (- centered, 6 values, +1 SAT (limit to 2-5) V  
%     \item Appendix Decomposition V
%     \item Notes for figure - stopped at figure 8 V 
%     \item Tables (Main Table with standard errors, robustness table with standard errors) V 
%     \item Our Prompt examples
%     \item Writing Prompts
% \end{enumerate}

\begin{abstract}
Students from different socioeconomic backgrounds exhibit persistent gaps in test scores, gaps that can translate into unequal educational and labor-market outcomes later in life. In many assessments, performance reflects not only what students know, but also how effectively they can communicate that knowledge. This distinction is especially salient in writing assessments, where scores jointly reward the substance of students’ ideas and the way those ideas are expressed. As a result, observed score gaps may conflate differences in underlying content with differences in expressive skill. A central question, therefore, is how much of the socioeconomic-status (SES) gap in scores is driven by differences in what students say versus how they say it. We study this question using a large corpus of persuasive essays written by U.S. middle- and high-school students. We introduce a new measurement strategy that separates content from style by leveraging large language models to generate multiple stylistic variants of each essay. These rewrites preserve the underlying arguments while systematically altering surface expression, creating a “generated panel” that introduces controlled within-essay variation in style. This approach allows us to decompose SES gaps in writing scores into contributions from content and style. We find an SES gap of 0.67 points on a 1–6 scale. Approximately 69\% of the gap is attributable to differences in essay content quality, Style differences account for 26\% of the gap, and differences in evaluation standards across SES groups account for the remaining 5\%. These patterns seems stable across demographic subgroups and writing tasks. More broadly, our approach shows how large language models can be used to generate controlled variation in observational data, enabling researchers to isolate and quantify the contributions of otherwise entangled factors.

\end{abstract}
\newpage
\epigraph{Style is the dress of thoughts; and a well-dressed thought, like a well-dressed man, appears to great advantage.}
{Philip Stanhope, Earl of Chesterfield, \textit{Letters to His Son}, Letter CXXVIII, January 21. 1751}
\section{Introduction}

For many of us, generating an idea and communicating it are distinct processes. Even when an idea feels clear internally, expressing it in writing can be difficult, particularly in settings where the writing is judged. We may have a good idea and a coherent answer, but we might struggle to express it in a manner that effectively communicates our reasoning and addresses all relevant points.

The ability to form ideas and clearly articulate them in writing are skills that develop over time, shaped by educational background, access to resources, and other socioeconomic circumstances. Differences in these factors can create systematic disparities in how students from different social groups tackle questions and express their thoughts on paper. Given that substantial grade gaps exist across students from different socioeconomic backgrounds \citep[e.g.,][]{hanushek2019achievement,morgan2024explaining,reardon2011widening}, this naturally raises a question: do these gaps primarily reflect differences in the content of what students write, or do they stem from differences in how these ideas are expressed?

From the perspective of educational fairness, a central question emerges: to what extent do observed score differences reflect differences in the substance of students' ideas—their reasoning, argumentation, and use of evidence—versus differences in their ability to express those ideas in forms that evaluators reward? If score gaps are driven primarily by differences in expression that stem from unequal access to particular linguistic resources or familiarity with academic conventions, then interventions focused on writing style and mechanics might substantially reduce disparities. Conversely, if gaps reflect deeper differences in analytical skills and content knowledge, then addressing them will require more fundamental changes in curriculum and instruction. Distinguishing between these possibilities is essential for designing equitable and effective educational policy.

To answer these questions, we need to separate differences in "what is said" from differences in "how it is said"—to decompose the observed score gap into components reflecting content and style. But doing so poses substantial measurement challenges. First, content and style are not directly observable as separate quantities.irst, content and style are not directly observable as separate quantities. Ideas are expressed through words, which makes it hard to measure the role of ideas separately from the words that express them. Second, even if we had separate measures of content and style quality, the two are deeply entangled in real writing. Students who develop stronger analytical skills typically also develop more sophisticated expression. Those with access to better educational resources acquire both richer vocabularies and more complex reasoning strategies. Conversely, students who face linguistic barriers or whose home discourse patterns differ from academic conventions may struggle to express sophisticated ideas in forms that evaluators recognize as "clear" or "well-organized." This entanglement means that content and style do not vary independently, making it difficult to assess their separate contributions to observed score differences.

Traditional empirical approaches struggle with these challenges. One might attempt to control for observable stylistic features—sentence length, vocabulary sophistication, grammatical correctness—and attribute the remaining variation to content. But such controls rest on strong assumptions about which features of writing reflect "how" versus "what," and risk either absorbing content-related variation into the style measures (if the controls are too comprehensive) or leaving substantial style-related variation in the residual (if the controls are too limited). Alternatively, one might ask human coders to rate content and style separately. But human judgments are inherently subjective, difficult to standardize across many essays, and expensive to scale. Moreover, raters often cannot cleanly separate substance from presentation: when ideas are expressed poorly, raters may perceive them as weaker or less developed than they actually are, confounding the two dimensions.

We propose a novel measurement strategy that leverages recent advances in large language models. Rather than attempting to separate content and style through statistical controls or subjective coding, we generate multiple stylistic variants of each essay using an LLM. Specifically, for each original essay, we produce several rewrites designed to alter surface expression—word choice, sentence structure, grammatical polish—while preserving the underlying arguments, evidence, and logical structure. These rewrites create a "generated panel" data in which the same substantive content appears in different stylistic realizations.

This panel structure allows us to decompose the observed score gap into three components: differences in content (what students argue), differences in style (how they express those arguments), and differences in scoring functions (how essays are evaluated across groups). By examining how scores vary across rewrites of the same essay, we can isolate the component of the score that responds to stylistic variation while holding content fixed. Conversely, by averaging over rewrites to eliminate within-essay stylistic noise, we can recover a measure of each essay's content that is comparable across students, even when their original writing styles differ systematically.

To operationalize this approach, we first estimate group-specific scoring functions from the human-assigned scores in our data. We train separate predictive models for high- and low-SES students using rich features that capture both semantic content (through text embeddings) and stylistic properties (through large set of linguistic variables measuring syntactic complexity, lexical diversity, and cohesion). These learned scoring functions represent how essays are evaluated under current practices for each group, allowing us to predict scores for any essay—including the LLM-generated rewrites, which were not seen by human graders. Combined with the rewrite panel, these predicted scores enable us to separately identify the contributions of content, style, and scoring rules to the observed gap. As the rewrites preserve content and induce a common shift in the style component—this design permits a clean separation of these three components.

We apply this method to a large dataset of persuasive essays written by U.S. middle- and high-school students, linked to measures of socioeconomic status. On average, high-SES students score 0.67 points higher than low-SES students on a 1–6 scale. Our decomposition reveals that approximately 69\% of this gap is associated with differences in content, 26\% with differences in style, and the remaining 5\% with differences in scoring functions across groups.

This pattern is stable across demographic subgroups and most writing prompts. Among both white and non-white students, and among both males and females, content differences account for roughly two-thirds to three-quarters of the score gap. The main systematic variation appears across grade levels: in higher grades, the share of the gap attributed to style increases, rising from around 25\% in grade 6 to over 40\% in grade 11. This shift may reflect increasing differentiation in students’ stylistic choices in later grades, either because students specialize in different styles of writing or because some students are more successful at adapting to the highly rewarded ‘school’ style.

% It is important to emphasize that our decomposition is primarily descriptive. It partitions the observed gap under the current joint distribution of essays, students, and scoring practices. The components correspond to natural "levers"—the ideas students express, the styles they use, and the standards by which they are judged—but interpreting these components as the effects of interventions requires additional assumptions about how the system would respond to changes. We return to this point in our empirical framework.

% Within students, we find that content and style are positively correlated: essays with stronger arguments tend to be more polished. However, this correlation is weaker among low-SES students ($\rho = 0.19$) than among high-SES students ($\rho = 0.38$), suggesting that the mapping from ideas to effective expression is less consistent for students who may face linguistic distance from academic norms or uneven access to writing instruction.

These findings have several implications for how we understand SES disparities in writing assessment. First, they suggest that observed score gaps are driven primarily by differences in the substance of student writing—the quality of arguments, the coherence of reasoning, the relevance and use of evidence—rather than by superficial features of presentation. To the extent that these substantive differences reflect real inequalities in critical thinking, analytical training, and access to rich curricular content, addressing the gap will require more than teaching students to write grammatically correct sentences or format their essays according to conventions. It will require attention to the deeper educational experiences that shape students' capacity to construct and organize compelling arguments.

At the same time, the non-trivial contribution of style—roughly one-quarter of the gap—underscores that mastery of academic discourse conventions remains an important and consequential dimension of writing performance. Students who control the lexical, syntactic, and organizational features valued in school writing receive higher scores, even when their underlying ideas are comparable. To the extent that these conventions are themselves socially patterned—acquired more readily by students whose home language use aligns with school norms, or whose access to instruction includes explicit attention to register and style—disparities in expression can amplify differences in opportunity and reinforce existing hierarchies.

Our paper also speaks to an emerging policy question: what would happen to score gaps if all students had access to LLM-based writing tools that could "polish" their essays before submission? Our results suggest that this would not close the gap: rewriting low-SES essays with an LLM prompt raises their average predicted scores, but a substantial disparity remains because the content component dominates. Most of the score difference reflects what students argue and how they structure and develop their reasoning. This matters for equity debates about AI writing assistance. If stylistic conventions unfairly penalize students’ ideas, polishing tools could reduce one source of disadvantage; but if they mainly shift style while leaving content differences unchanged, their impact on overall gaps will be limited, and unequal access could introduce new advantages. More fundamentally, treating technological standardization of expression as an equity fix risks obscuring the underlying inequalities in content development and reinforcing a narrow, standardized conception of what counts as “good” writing.

Beyond the specific findings on writing assessment, this paper makes two broader contributions. Methodologically, it demonstrates how large language models can function as research instruments, not just as end-user tools for text generation, but as devices for creating controlled variation in observational data. The generated panel design we introduce is applicable to any setting in which substance and presentation are confounded and researchers need to assess their separate contributions. Potential applications include studying how the presentation of job market papers affects hiring decisions, how framing shapes persuasive communication, or how linguistic style influences the reception of creative and professional work.

\paragraph{Related Literature} It's well-documented that essay-scoring differences across social groups are significant and persistent \citep{nationsreport2012}. A large education and applied linguistics literature studies why these gaps arise, emphasizing that performance in school writing reflects not only ideas and reasoning but also mastery of \emph{academic language} and discourse conventions—features that are unevenly distributed across socioeconomic backgrounds and schooling contexts. This perspective motivates our focus on separating “what is said” from “how it is expressed,” and connects to work that links linguistic features such as syntactic complexity, cohesion, lexical sophistication, and organization to human-assigned writing scores \citep{wang2023multi}. At the same time, the automated essay scoring (AES) literature documents that scoring models can place substantial weight on surface and stylistic properties, raising concerns about construct validity and about whether models reward form in ways that can amplify pre-existing inequalities (e.g., if academic register and polish are themselves socially patterned). Closely related, research on fairness in automated scoring evaluates whether scoring algorithms treat groups differently and shows that conclusions can depend on the fairness criterion used \citep{litman2021fairness}.

Methodologically, our paper builds on two recent strands. First, it draws on the growing “text-as-data” tradition in economics and political science, which treats text as a high-dimensional object that can be represented, summarized, and manipulated to study substantive questions. Our contribution fits this agenda by using text representations to construct interpretable “content” and “style” components and by using controlled textual edits to form counterfactuals. Second, it leverages recent progress in large language models (LLMs) as tools for generating \emph{rewrite panels}: for a fixed underlying essay, we create multiple stylistic variants intended to preserve meaning while shifting expression along targeted dimensions. This design provides a new way to probe which aspects of writing are rewarded by scorers, and to do so at scale while holding the underlying message approximately fixed. In this sense, LLM-based rewriting functions like a structured perturbation device that complements purely observational correlations between linguistic features and scores.

Our approach to measuring disparities also aligns with recent trends in economics and computer science that define disparities \emph{conditional on} a variable of interest \citep{dwork2012fairness, kusner2017counterfactual}. For instance, \cite{arnold2022measuring} examines disparities in judicial treatment by comparing bail decisions among individuals with similar predicted risk, and \cite{bohren2022systemic} studies group gaps in employment probabilities among workers of equal productivity. Such approaches highlight the normative aspect of measuring disparities across social groups by specifying which attributes should be held fixed. We adapt this logic to writing by treating the essay’s underlying content as the “attribute to be held fixed,” and then quantifying how much of the remaining score gap is attributable to non content factors. At the same time, our goal is not only to document differential impacts but also to identify the mechanisms that generate them. This objective connects to the decomposition tradition in economics, which explains observed gaps by constructing counterfactual distributions—fixing one component and modifying the relevant conditional distribution to isolate its contribution \citep{fortin2011decomposition}.

Finally, our design resonates with the explainable AI literature, which seeks to understand algorithmic predictions by modifying inputs and examining resulting changes in outputs. In text analysis, prior work has proposed targeted edits (e.g., flipping sentiment-bearing adjectives) to diagnose model behavior \citep{feder2021causalm, vig2020investigating}. Our use of rewriting is conceptually similar but more conservative in its claims: we treat the edits as a way to construct counterfactual score distributions and to decompose gaps, rather than as a causal estimate absent additional identifying assumptions (e.g., random assignment of essays to graders). Related input-modification approaches have also been used outside text. \cite{ludwig2023machine}, for instance, studies why defendant photos predict bail decisions by generating minimally perturbed images that induce large changes in predicted release probabilities, and then interpreting the differences. Analogously, our rewrite-based counterfactuals illuminate which dimensions of writing are most strongly rewarded by scoring systems, and how those rewards map into socioeconomic score gaps.

\paragraph{Roadmap.} The remainder of the paper proceeds as follows. Section 2 develops the empirical framework and identification strategy. Section 3 describes the data and variable construction. Section 4 presents descriptive statistics and assesses the performance of our scoring models and rewrite procedure. Section 5 reports the decomposition results and robusntess. Section 6 concludes.

\section{Empirical Framework}

We observe essays indexed by $i=1,\dots,N$. Let $G_i\in\{H,L\}$ denote group membership of essay $i$'s author (e.g SES). For each \emph{scoring concept} or \emph{ranker} $m\in\{H,L\}$, a deterministic scoring function $S^{(m)}$ maps a rendered text $x_{i}$ to a numeric score:
\[
s^{(m)}_{i} \;=\; S^{(m)}(x_{i}) \in \mathbb{R},
\]
where $x_{i}$ collects the full set of observed essay and author characteristics, including the text embedding and a rich set of style variables. In the framework below we treat $S^{(m)}$ as given. In the empirical implementation, $S^{(m)}$ is not observed directly but is estimated from the observed human scores. The predicted values from these models serve as estimates of $S^{(m)}(x_{i})$.
% ; with a slight abuse of notation we denote these estimated scoring functions by the same symbol $S^{(m)}$ in what follows.
% We observe essays indexed by $i=1,\dots,N$. Let $G_i\in\{H,L\}$ denote group membership of the essay $i$'s author (SES). For each \emph{scoring concept} or \emph{ranker} $m\in\{H,L\}$, a learned deterministic scoring function $S^{(m)}$ maps a rendered text $x_{ik}$ to a numeric score:
% \[
% s^{(m)}_{i} \;=\; S^{(m)}(x_{ik}) \in \mathbb{R}.
% \]
% where $x_i$ are the full set of the observed essay and author  charchtristics, including the text embedding and a very rich set of variables that captures the different style of each essay.

Write $C_i$ for the (latent) \emph{content} of essay $i$. Let $R_{i}$ denote the phrasing/style realization of essay $i$. We make the following assumption on the scoring function.

\begin{assumption}[Separable content and style]\label{ass:sep}
    For any scorer $m$ the essay scoring is separable in content and style:
    \begin{equation}
\label{eq:baseline}
s^{(m)}_{i}
=
\underbrace{\theta_m(C_i)}_{\text{Content index under $m$}}
+ \underbrace{\rho_m(R_{i})}_{\text{Style component under $m$}}.
\end{equation}
\end{assumption}

This assumption on the ranker implies that scoring varies additively with the content and with the style, where $\theta_m$ and $\rho_m$ are mappings from content and style, respectively, to the score under $m$. Notice that we allow for different rankers to weight content and style differently. For example, rankers for one group of students may be more responsive to style and less responsive to the actual content of the answer, while another may be more sensitive to the content. This ranking function can also allow for difference in the overall ranking by simply giving lower scores to the content and style, capturing cases of discrimination or differences in who is assigned to rank low SES students vs. High SES students. 

\subsection{Measuring What Drives The Gap in Scores }
We are interested to measure how much of the gap in scores between two groups is driven by difference in content and how much the gap is driven by difference in writing style. To do so we use Kitagawa--Oaxaca--Blinder decomposition approach (\cite{Kitagawa1955, Oaxaca1973, Blinder1973}).      
Suppose the policy grades H originals with $S^{(H)}$ and L originals with $S^{(L)}$. The observed gap is
\[
\Delta_{\mathrm{obs}}
:= \mathbb{E}_H[s^{(H)}_{i}] - \mathbb{E}_L[s^{(L)}_{i}].
\]
We suggest decomposing the gap as follows. Let the group means 
\[
A^{(m)}_G:=\mathbb{E}_G[\theta_m(C_i)], \quad U^{(m)}_G:=\mathbb{E}_G[\rho_m(R_{i})],
\]
The following identity holds:
\begin{equation}
\label{eq:H-identity}
\begin{aligned}
\Delta_{\mathrm{obs}}
&= \mathbb{E}_H\!\big[s^{(H)}_{i}\big] - \mathbb{E}_L\!\big[s^{(H)}_{i}\big] \;+\; \mathbb{E}_L\!\big[s^{(H)}_{i}\big] - \mathbb{E}_L\!\big[s^{(L)}_{i}\big] \\
&=
\underbrace{(A^{(H)}_H - A^{(H)}_L)}_{\text{content under $S^{(H)}$}}
+
\underbrace{(U^{(H)}_H - U^{(H)}_L)}_{\text{style (original) under $S^{(H)}$}}
+
\underbrace{\mathbb{E}_L\!\big[s_{i}^{(H)} - s_{i}^{(L)}\big]}_{\text{Ranker Tilting}} \, .
\end{aligned}
\end{equation}

The first bracket captures how much of the score gap is driven by differences in content quality across the two groups, when essays are ranked by the high-SES scorer. These differences reflect what students choose to argue about and how they structure their reasoning. For example, some students may present a well-developed causal argument with clear evidence, while others provide only loosely connected claims. Whether an argument is convincing, rational, and relevant—all of these contribute to this component.

The second bracket measures how much of the gap arises from differences in writing style, again evaluated through the lens of the high-SES scorer. This component reflects control of language, correctness of word choice, coherence, and other stylistic decisions students make in expressing their ideas. For instance, two students may present equally strong arguments, but one might write in a more polished and grammatically precise style. The gap here reflects how the scoring rule—estimated from high-SES writing—responds to stylistic features more common among low-SES writers.

Finally, the third component captures how much of the score difference is attributable to the rankers themselves, holding the underlying texts fixed. Even when reading the exact same set of essays, rankers functions may differ systematically in how they assign scores. For example, a low-SES student might receive a lower score simply because one scorer is more severe or holds implicit biases, or because high-SES and low-SES students happen to be assigned different rankers with different grading tendencies.

In our discussion, we mostly focus on their relative sizes—comparing the share of the writing-score gap explained by each component, rather than on the absolute sizes. 

% \begin{remark}
%     Why not simple decomposition by seperating to different varaibles - mainly as content and style go hand in hand.
% \end{remark}
\subsection{Identification}\label{sec:identification}

Equation \eqref{eq:H-identity} is an accounting identity, but its components are not directly observed. The objects
\[
A_G^{(m)}=\mathbb E_G[\theta_m(C_i)]
\qquad\text{and}\qquad
U_G^{(m)}=\mathbb E_G[\rho_m(R_{i})]
\]
involve the latent content index $\theta_m(C_i)$ and the latent style component $\rho_m(R_{i})$. Even under separability \eqref{eq:baseline}, these two primitives are not separately pinned down from a single observed score, because the decomposition is only defined up to an additive constant (shifting $\theta_m$ by a constant and shifting $\rho_m$ by the opposite constant leaves $s_{i}^{(m)}$ unchanged). Identification therefore requires an anchoring restriction that uses the within-essay panel of rewrites to create a common reference style environment.

% Let $k=0$ denote the original essay and let $k\in\mathcal K:=\{1,\dots,K\}$ index a fixed set of $K$ rewrites produced by an LLM. For each $k\in\mathcal K$ the rewrite  is intended to preserve the semantic content of essay $i$ while altering its stylistic realization. Denote by $C_{ik}$ the content of rewrite $k$ of essay $i$, by $R_{ik}$ the realized style of rewrite $k$ of essay $i$, and by $x_{ik}$ the rendered text of rewrite $k$ of essay $i$, and by $s_{ik}^{(m)} = S^{(m)}(x_{ik})$ the score of rewrite $k$ of essay $i$ by scorer $m$, where $k=0$ indicate the original. We make the following assumptions on the rewrites.
Let $k\in\{0,1,\dots,K\}$ index versions of essay $i$, where $k=0$ denotes the original and $k\in\mathcal K:=\{1,\dots,K\}$ indexes a fixed set of $K$ LLM-generated rewrites. Each rewrite is intended to preserve the semantic content of essay $i$ while altering its stylistic realization. Let $C_{ik}$ denote the content of version k of essay $i$, $R_{ik}$ its realized style, and $x_{ik}$ its rendered text. Let $s_{ik}^{(m)} = S^{(m)}(x_{ik})$ denote the score assigned to version $k$ of essay $i$ by scorer $m$. We make the following assumptions on the rewrites.
\begin{assumption}[Rewrite content fidelity]
\label{ass:id-fidelity}
For all $k\in\mathcal K$, the rewrite preserves content:
\[
C_{ik}=C_{i0} := C_i.
\]
\end{assumption}

Assumption \ref{ass:id-fidelity} formalizes the design goal that rewrites change the realization $R_{ik}$ while leaving latent content $C_i$ fixed, so that within-essay variation across $k\in\mathcal K$ can be attributed to style rather than content.

Our second assumption characterizes how the LLM rewrites enter the style component.

\begin{assumption}[Rewrite style homogeneity]
\label{ass:id-style}
For each ranker $m$ and rewrite $k\in\mathcal K$, the effect of rewrite $k$ on the score is an additive constant $\lambda_{k,m}$:
\[
\rho_m(R_{ik}) = \lambda_{k,m} + u_{ikm},
\]
where the residual satisfies
\[
\mathbb E[u_{ikm}\mid C_i,G_i] = \mathbb E[u_{ikm}\mid C_i]=0.
\]
\end{assumption}

Assumption \ref{ass:id-style} says that, conditional on the essay's content $C_i$, the effect of asking the LLM for rewrite type $k$ on the style component under ranker $m$ is a constant shift $\lambda_{k,m}$, up to idiosyncratic noise $u_{ikm}$ with mean zero. The second part rules out systematic differences in the rewrite noise across groups, conditional on content. This implies that for two essays with identical content—one from each group—the conditional mean of $u_{ikm}$ is the same for both essays, and equals zero.

Assumption 3 has a testable implication. It implies that rewrite type $k$ acts like a group-invariant vertical shift in scores (up to mean-zero noise conditional on content). In Section 4.3 we evaluate this implication directly using the rewrite panel: for each pair of rewrite types \(k,k^{\prime}\) we compute within-group mean score differences and then a difference-in-differences across SES, which should be approximately zero under additivity. The resulting “Difference-in-difference matrix” should be close to zero relative to the magnitude of rewrite-induced score changes. 

Given Assumptions \ref{ass:id-fidelity}–\ref{ass:id-style}, the content gap can be identified as follows. Define the rewrite-averaged score
\[
\bar{s}^{(m)}_i := \frac{1}{K} \sum_{k\in\mathcal K} s^{(m)}_{ik}.
\]
Using \eqref{eq:baseline} and Assumptions \ref{ass:id-fidelity}–\ref{ass:id-style} we obtain
\begin{align*}
\mathbb E_G[\bar{s}^{(m)}_i]
&= \mathbb E_G\Big[\theta_m(C_i) + \frac{1}{K}\sum_{k\in\mathcal K}\rho_m(R_{ik})\Big] \\
&= \mathbb E_G[\theta_m(C_i)] + \underbrace{\frac{1}{K}\sum_{k\in\mathcal K}\lambda_{k,m}}_{=: \,\bar\lambda_m},
\end{align*}
since $\mathbb E[u_{ikm}\mid C_i,G_i]=0$ by Assumption \ref{ass:id-style}. The constant $\bar\lambda_m$ does not depend on group membership. Therefore the difference in rewrite-averaged scores between groups exactly recovers the content gap under ranker $m$:
\[
A_H^{(m)} - A_L^{(m)}
= \mathbb E_H[\theta_m(C_i)] - \mathbb E_L[\theta_m(C_i)]
= \mathbb E_H[\bar{s}^{(m)}_i] - \mathbb E_L[\bar{s}^{(m)}_i].
\]
The rewrites thus serve to equalize the distribution of writing style across essays, allowing us to isolate differences in the way content is scored.

To identify the style component, define the essay-level deviation between the original score and the rewrite-averaged score under ranker $m$:
\[
d^{(m)}_i := s^{(m)}_{i0} - \bar{s}^{(m)}_i.
\]
Using \eqref{eq:baseline} and Assumptions \ref{ass:id-fidelity}–\ref{ass:id-style} again,
\begin{align*}
d^{(m)}_i
&= \big[\theta_m(C_i) + \rho_m(R_{i0})\big] - \Big[\theta_m(C_i) + \frac{1}{K}\sum_{k\in\mathcal K}\rho_m(R_{ik})\Big] \\
&= \rho_m(R_{i0}) - \frac{1}{K}\sum_{k\in\mathcal K}\rho_m(R_{ik}) \\
&= \rho_m(R_{i0}) - \bar\lambda_m - \frac{1}{K}\sum_{k\in\mathcal K}u_{ikm}.
\end{align*}
Taking expectations and using $\mathbb E[u_{ikm}\mid C_i,G_i]=0$ yields
\[
\mathbb E_G[d^{(m)}_i] = \mathbb E_G[\rho_m(R_{i0})] - \bar\lambda_m.
\]
The constant $-\bar\lambda_m$ cancels when we take differences across groups, so that
\[
U_H^{(m)} - U_L^{(m)}
= \mathbb E_H[\rho_m(R_{i0})] - \mathbb E_L[\rho_m(R_{i0})]
= \mathbb E_H[d^{(m)}_i] - \mathbb E_L[d^{(m)}_i].
\]

In words, subtracting the rewrite-averaged score from the original score nets out differences in content and the average shift induced by the rewrite procedure. The remaining group difference reflects how the scoring rule responds to the styles actually used by the two groups relative to the common benchmark style implemented by the LLM rewrites.

Finally, the ranker-tilting term in \eqref{eq:H-identity} is directly observed as
\[
\mathbb E_L\big[s_{i0}^{(H)} - s_{i0}^{(L)}\big],
\]
which compares how the two rankers score the same set of original low-SES essays.
\paragraph{Discussion.} The key objects in this identification argument are Assumptions~\ref{ass:id-fidelity} and~\ref{ass:id-style}. Taken together with separability \eqref{eq:baseline}, they say the following. First, rewrites change only the realization $R_{ik}$ but not the latent content $C_i$, so all within-essay variation across $k$ can be attributed to style. Second, conditional on $C_i$, the effect of requesting rewrite type $k$ under ranker $m$ can be summarized by a constant shift $\lambda_{k,m}$ in the style component, plus idiosyncratic noise with mean zero and a distribution that does not depend on group membership. Under these conditions, the rewrite-averaged score $\bar{s}^{(m)}_i$ is the sum of the content index $\theta_m(C_i)$ and a constant $\bar\lambda_m$ that is common across groups, so differences in $\bar{s}^{(m)}_i$ identify the content gap $A_H^{(m)}-A_L^{(m)}$. Similarly, the deviation $d^{(m)}_i = s^{(m)}_{i0}-\bar{s}^{(m)}_i$ nets out both content and the average rewrite shift, so differences in $d^{(m)}_i$ identify the style gap $U_H^{(m)}-U_L^{(m)}$. In this sense, the role of the rewrites is purely anchoring: they define a reference style environment under which content and style can be separated in a way that is comparable across groups.

    These assumptions and the identification argument impose discipline on how the rewrite operator should be constructed. First, to make content fidelity plausible, prompts to the LLM should explicitly emphasize preserving meaning, arguments, and logical structure, and only allow modifications to surface features. Second, Assumption~\ref{ass:id-style} is fragile to interactions between the rewrite mechanism and the essay itself. If the LLM’s rewrites respond strongly and systematically to the content of the essay, the induced differences in style will be correlated with $\theta_m(C_i)$ and will be absorbed into the content component. To mitigate this, prompts should aim either to produce rewrites in (approximately) a fixed writing style with limited sensitivity to content, while still generating enough variation to pin down the content component, or to generate rewrites that move essays symmetrically along a stylistic axis that is correlated with scores. In the latter case some rewrites tend to increase the score and some tend to decrease it for any given essay, so that
    \[
    \frac{1}{K}\sum_{k\in\mathcal K} \rho_m(R_{ik}) \approx 0
    \qquad\text{for all $i$,}
    \]
    and small residual deviations from zero will not materially affect the decomposition. In both designs the goal is that the rewrite operator induces a common reference distribution over styles, rather than introducing systematic group- or content-specific shifts that would contaminate the identified content and style components.

    In the appendix we show that without assumptions \ref{ass:sep} and \ref{ass:id-style} one can still construct a meaningful decomposition if one is willing to fix a particular neutralizing rewrite operator $T$. There we introduce an operator $T$ that maps each original essay to a neutralized version $T(\text{original text})$, and define, for a given scoring function $S$,
    \[
    \mu_G^{S,\mathrm{orig}} := \mathbb{E}\big[S(\text{original text})\mid G\big],
    \qquad
    \mu_G^{S,\mathrm{neu}} := \mathbb{E}\big[S(T(\text{original text}))\mid G\big].
    \]
    The observed gap can be decomposed as
    % \[
    % \Delta_{\mathrm{obs}} = \mu_H^{S^{(H)},\mathrm{orig}} - \mu_L^{S^{(L)},\mathrm{orig}}
    % \]
    \[
    \Delta_{\mathrm{obs}}
    =
    \underbrace{\big(\mu_H^{S^{(H)},\mathrm{neu}} - \mu_L^{S^{(H)},\mathrm{neu}}\big)}_{\text{Content}^{(H)}}
    +
    \underbrace{\Big[(\mu_H^{S^{(H)},\mathrm{orig}} - \mu_H^{S^{(H)},\mathrm{neu}}) - (\mu_L^{S^{(H)},\mathrm{orig}} - \mu_L^{S^{(H)},\mathrm{neu}})\Big]}_{\text{Style}^{(H)}}
    +
    \underbrace{\big(\mu_L^{S^{(H)},\mathrm{orig}} - \mu_L^{S^{(L)},\mathrm{orig}}\big)}_{\text{Scoring-function tilt (vs.\ $S^{(H)}$)}}.
    \]
    Here the ``content'' term measures the gap that would remain if all essays were rewritten into the same neutral style before scoring with $S^{(H)}$, the ``style'' term captures the differential premium of the original high- and low-SES writing styles relative to that neutral baseline under $S^{(H)}$, and the tilt term compares how $S^{(H)}$ and $S^{(L)}$ score the same low-SES originals.

    This baseline decomposition is valid for any scoring functions $S^{(H)}$ and $S^{(L)}$ and any rewrite operator $T$, without invoking the separability assumptions \eqref{eq:baseline} and \ref{ass:id-style}. It requires fewer structural assumptions, at the expense of committing to a particular baseline distribution over styles. It is therefore particularly useful for policy questions such as: ``If all students were allowed (or required) to pass their essays through an LLM-based neutralizer before grading, how much of the score gap would remain?'' When the separable model \eqref{eq:baseline} and Assumptions \ref{ass:id-fidelity}--\ref{ass:id-style} hold, and when the neutral rewrite $T$ can be interpreted as imposing the same reference style environment as the rewrite panel, the neutral content term $\mu_H^{S^{(H)},\mathrm{neu}} - \mu_L^{S^{(H)},\mathrm{neu}}$ coincides with the structurally identified content gap $A_H^{(H)} - A_L^{(H)}$ derived above. In this sense, the structural decomposition and the neutral rewrite-based decomposition are two views of the same underlying idea: using rewrites to define a common style benchmark and separating differences in content from differences in style and scoring functions.

\begin{remark}[When does the decomposition have a causal interpretation?]\label{remark:causal}
Up to this point, our decomposition is descriptive: it partitions the observed score gap into components indexed by content, style, and scoring functions under the observed joint distribution of essays. Each component suggests a natural ``lever'' (changing content, changing style, changing the scoring rule), but reading the terms as causal effects requires strong conditional independence assumptions about how the remaining components would behave under such interventions.

To fix ideas, consider a hypothetical policy that ``equates content'' by shifting the content distribution of low-SES students from $F_L^C$ to $F_H^C$. Let
\[
m_L^{\rho}(c) := \mathbb E[\rho(R_{i0}) \mid C_i=c, G_i=L]
\]
denote the mean of our style functional at content level $c$ for low-SES students. Under the status quo,
\[
\mathbb E_L[\rho(R_{i0})] \;=\; \int m_L^{\rho}(c)\, dF_L^C(c).
\]
Under a content-only policy that changes the marginal distribution of $C_i$ but leaves the conditional law of style given content unchanged for low-SES students, the mean style component would instead be
\[
\int m_L^{\rho}(c)\, dF_H^C(c).
\]
By contrast, our \emph{content} component is constructed holding the \emph{mean} style term $\mathbb E_L[\rho(R_{i0})]$ fixed when varying content. Therefore, the content component coincides with the causal effect of a feasible content-only policy only under an additional restriction ensuring that the style term is (approximately) insensitive to the induced change in the content mix---for example, that $m_L^\rho(c)$ is (approximately) constant in $c$ over the relevant support (or, more generally, that $\int m_L^\rho(c)\, dF_H^C(c)\approx \int m_L^\rho(c)\, dF_L^C(c)$). In realistic settings, content and style are typically intertwined, so equalizing content is more naturally interpreted as a descriptive reweighting of observed essays than as the effect of a feasible content intervention.\footnote{A further requirement is overlap: the policy is only well-defined on regions where $F_H^C$ assigns mass to content values attainable by low-SES students.}

An analogous issue arises for policies that “equate style”: our style component replaces $R_{i0}$ with an LLM-induced reference style while holding $C_i$ and the scoring functions fixed, and it is causal only if this intervention changes style alone—neither shifting expressed content nor changing how scorers respond once style is modified; since writing tools can affect what is articulated and graders may adapt to standardization, we read the style term as a mechanical association in the current environment. The same caveat applies to the scoring-function component: the “scoring-function tilt” compares how $S^{(H)}$ and $S^{(L)}$ grade the same texts and is causal only if scoring rules are stable objects and unifying them would not induce equilibrium responses (e.g., grader selection or institutional feedbacks).

% For these reasons, we interpret the decomposition primarily as a descriptive accounting of how the observed gap is currently associated with differences in content, style, and scoring. Under strong autonomy assumptions---interventions that affect only the targeted component and leave the remaining conditional distributions invariant---the same formulas can be given a causal interpretation with respect to well-defined policies. Our empirical analysis is agnostic about these stronger assumptions, and we therefore refrain from causal claims beyond this descriptive interpretation.
\end{remark}

\subsection{Estimation}

In the empirical analysis we implement the decomposition in two layers. We first estimate the scoring functions $S^{(H)}$ and $S^{(L)}$ from the human data, and then recover empirical counterparts of the content and style components using a fixed-effects decomposition of the resulting predicted scores. For each scoring concept $m \in \{H, L\}$, we train a gradient-boosted tree model (XGBoost\footnote{The XGBoost model uses 396 trees with a maximum depth of 4, a learning rate of 0.05, subsampling rates of 0.8, and L1/L2 regularization set to 0.055 and 0.026, respectively. These parameters were selected by performing a simple randomized hyperparameter search to tune the XGBRegressor, focusing on key parameters that control model capacity and regularization. The search varied the number of trees, tree depth, and L1/L2 regularization strengths, with regularization terms sampled from log-uniform distributions to cover multiple orders of magnitude. Model performance was evaluated using RMSE as the optimization metric. Hyperparameters were selected via cross-validation over the full dataset, using shuffled $K$-fold cross-validation with three folds.}) to predict the human-assigned score using the feature vector $x_{i0}$ of the original essay. This feature vector includes the text embedding, the full set of style variables, student characteristics, and an indicator for the essay prompt. Exact definition of the variable is in section \ref{sec:data}. This delivers a fitted scoring rule $\hat S^{(m)}(\cdot)$ for each $m$. With a slight abuse of notation, we write $S^{(m)}$ for these estimated functions in what follows and treat them as fixed when we construct the decomposition. For each essay $i$ and each version $k$ (the original and all rewrites), we compute the predicted score
\[
\hat s^{(m)}_{ik} := S^{(m)}(x_{ik}),
\]
which serves as the empirical analogue of $s^{(m)}_{ik}$ in the framework. To avoid overfitting, predicted scores are obtained via sample splitting, similar to cross-validating: the scoring model is trained on 80\% of the data and used to predict scores on the remaining 20\%. This procedure is repeated so that every observation is scored using a model trained on data that excludes it.

Given these predicted scores, we recover estimates of the latent content and style components via a fixed-effects regression that mirrors the additive structure in \eqref{eq:baseline}. For each $m$ we estimate
\[
\hat s^{(m)}_{ik} = \alpha^{(m)}_i + \sum_{k\in \{1,...,K\}} \gamma^{(m)}_k \mathbf{1}\{k\} + \varepsilon^{(m)}_{ik},
\]
where $\alpha^{(m)}_i$ is an essay fixed effect and the dummies $\mathbf{1}\{k\}$ indicate the rewrite type, with the original $k=0$ absorbed into the fixed effect. Under Assumptions~\eqref{eq:baseline}, \ref{ass:id-fidelity}, and \ref{ass:id-style}, the fixed effect $\alpha^{(m)}_i$ provides an estimate of the content index $\theta_m(C_i)$ up to an additive constant, while the fitted style component for version $k$ of essay $i$ is
\[
\hat\rho_m(R_{ik}) := \hat s^{(m)}_{ik} - \hat\alpha^{(m)}_i = \hat\gamma^{(m)}_k + \hat\varepsilon^{(m)}_{ik}.
\]
We then construct sample analogues of the group-specific content and style indices by averaging these estimated components over essays. For each $G\in\{H,L\}$ and each $m$ we define
\[
\widehat{A}^{(m)}_G := \frac{1}{N_G}\sum_{i:G_i=G} \hat\alpha^{(m)}_i,
\qquad
\widehat{U}^{(m)}_G := \frac{1}{N_G}\sum_{i:G_i=G} \hat\rho_m(R_{i0}),
\]
where $\hat\rho_m(R_{i0})$ is the estimated style component for the original version of essay $i$. By construction, $\widehat{A}^{(m)}_G$ is the empirical analogue of $\mathbb E_G[\theta_m(C_i)]$ and $\widehat{U}^{(m)}_G$ is the analogue of $\mathbb E_G[\rho_m(R_{i0})]$, up to constants that cancel when we take differences across groups. The estimated content and style components of the gap under $S^{(H)}$ are therefore given by
\[
\widehat{\text{Content}}^{(H)} = \widehat{A}^{(H)}_H - \widehat{A}^{(H)}_L,
\qquad
\widehat{\text{Style}}^{(H)} = \widehat{U}^{(H)}_H - \widehat{U}^{(H)}_L,
\]
while the ranker-tilting term is estimated directly from the scoring functions as
\[
\widehat{\text{Tilt}} = \frac{1}{N_L}\sum_{i:G_i=L}\big(\hat s^{(H)}_{i0} - \hat s^{(L)}_{i0}\big),
\]
which compares how the two learned scoring rules grade the same set of original low-SES essays.

Inference on these components is obtained by bootstrapping the entire procedure at the essay level. In each bootstrap replication we resample essays with replacement, 
% re-estimate the scoring functions $S^{(H)}$ and $S^{(L)}$ on the resampled data, recompute predicted scores for originals and rewrites, 
re-estimate the fixed-effects model, and recompute the empirical decomposition. The dispersion of the resulting bootstrap distribution for each component provides standard errors and confidence intervals that account jointly for uncertainty in the first-stage learning of the scoring functions and in the second-stage decomposition based on the fixed effects.

% \begin{enumerate}
%     \item Estimation  V 
%     \item Estimating the ranking function  V 
%     \item What motivates the usage of two types of rewrites V 
%     \item Talking about the decomposition in the appendix and what it gives compare to this one V 
%     \item When the decomposition is "causal"  V 
% \end{enumerate}

\section{Data}\label{sec:data}
In this section, we describe our main data sources and how we construct the data. We begin with our primary dataset of student essays, PERSUADE 2.0. We then explain how we construct the content variables. We then describe how we construct the style variables, which capture stylistic choices in writing. Finally, we outline how we produce the essay rewrites.

\subsection{The PERSUADE 2.0 dataset}

PERSUADE 2.0 (Persuasive Essays for Rating, Selecting, and Understanding Argumentative and Discourse Elements) is a large-scale, open-source corpus designed to advance research into argumentative writing quality, discourse elements, and their effectiveness \citep{crossley2024large}. The corpus was created to address the need for systematic analysis of student argumentation and to support the development of unbiased computational algorithms for assessing writing quality. PERSUADE 2.0 builds upon the PERSUADE 1.0 corpus by adding holistic essay quality scores and effectiveness ratings for individual discourse elements. The dataset contains essays written in response to 15 different prompts across two distinct writing tasks: independent writing (8 prompts) and source-based writing (7 prompts, each with a single source text).

The dataset comprises 25,996 argumentative essays written by students in grades 6-12 across the United States. The corpus reflects the diversity of the U.S. student population, with writers identifying as White (45\%), Hispanic (25\%), Black (19\%), and other racial/ethnic groups. Essays are linked to detailed demographic information including gender and race/ethnicity. Notably, 20,759 essays (approximately 80\% of the corpus) include data on student eligibility for federal assistance programs, which we use as an indicator of socioeconomic status (SES). All essays in the dataset have been assigned holistic quality scores by trained expert raters.

The holistic scores were assigned using a standardized 1-6 point SAT essay scoring rubric, where 6 indicates "clear and consistent mastery of writing." Expert raters were assigned to rate the essays underwent prompt-specific training and employed a double-blind rating process, with 100\% adjudication by a third rater when necessary. The rubric for source-based essays was slightly modified to include evaluation of how effectively students incorporated evidence from source texts. Inter-rater reliability before adjudication demonstrated strong agreement, indicating consistent and reliable quality judgments across the corpus. \citep{crossley2024large}

\subsubsection{``Content'' Variables}
To represent the semantic content of the PERSUADE essays, we embed each entire essay using contextual representations from a pre-trained BERT encoder \citep{devlin2019bert}. We use bert-base-uncased, which is pre-trained on large-scale general-domain English text and is therefore well-suited to the naturalistic language in these essays. Because BERT accepts sequences of up to 512 WordPiece tokens, essays exceeding this length are truncated to the first 512 tokens. For each essay, we take the final-layer embedding of the special \([CLS]\) token, which is commonly used as a pooled summary representation of the input sequence. This procedure maps each essay to a fixed-dimensional vector in a shared high-dimensional space. Prior work shows that contextual BERT representations encode higher-level abstractions—such as topic, intent, and relational meaning—beyond surface lexical overlap \citep{devlin-etal-2019-bert,reimers-gurevych-2019-sentence,ethayarajh-2019-contextual,tenney-etal-2019-bert,jawahar-etal-2019-bert,clark-etal-2019-bert,rogers-etal-2020-primer,karpukhin-etal-2020-dense}. We therefore interpret distances between \([CLS]\) vectors as reflecting conceptual relatedness between essays rather than mere word overlap.

\subsubsection{``Style'' Variables}
To quantify stylistic dimensions of the essays, we employ computational linguistic tools from \href{https://www.linguisticanalysistools.org/}{the Suite of Automatic Linguistic Analysis Tools} \citep{potter2025assessing}. These tools enable systematic measurement across three key dimensions of writing style. In total, our stylistic representation comprises 355 linguistically variables, spanning syntactic, lexical, and discourse-level dimensions.

\paragraph{Syntactic complexity and sophistication} We assess the structural properties of student writing using TAASSC (Tool for the Automatic Analysis of Syntactic Sophistication and Complexity, version 1.3.8; \cite{kyle2016measuring}). This tool captures two complementary aspects of syntax: traditional complexity metrics such as clause subordination and phrase elaboration, alongside usage-based sophistication measures that reflect developmental difficulty. The sophistication metrics draw on frequency and contingency patterns from large-scale corpora, notably the Corpus of Contemporary American English, to identify constructions that are statistically less common and therefore more challenging to acquire \citep{kyle2016measuring, potter2025assessing}. Under this framework, rarer syntactic patterns signal greater linguistic maturity. From TAASSC, we extract 149 measures, including indices such as mean clause length, frequency-weighted measures of multi-clause constructions, verbal dependency counts, and proportions of finite and non-finite structures.

\paragraph{Lexical variation} The diversity of vocabulary within each essay is measured using TAALED (Tool for the Automatic Analysis of Lexical Diversity; \citep{kyle2021assessing, zenker2021investigating}). This tool generates indices that capture the degree of lexical variety employed by writers. Prior validation work demonstrates strong correspondence between TAALED metrics and expert assessments of vocabulary richness, establishing these measures as reliable proxies for stylistic development \citep{kyle2021assessing, potter2025assessing}. TAALED contributes 38 lexical diversity and sophistication variables, including moving-average type–token ratios, measures of how common or rare words are in large reference corpora, and word-length–based indices that reflect vocabulary development.

\paragraph{Textual cohesion} We evaluate the connectivity of ideas using TAACO (Tool for the Automatic Analysis of Cohesion, version 2.1.3; \cite{crossley2016taaco, crossley2019taaco2}). TAACO quantifies how explicitly writers signal relationships among concepts at multiple scales: adjacent word and sentence connections, paragraph-level coherence, and document-wide integration. The indices encompass various cohesive devices including repetition of lexical items, semantic similarity, and explicit connectives that guide reader comprehension. TAACO provides 168 cohesion-related variables, which include local and global cohesion metrics such as lexical overlap between adjacent sentences, semantic similarity across paragraphs, and the use of explicit connectives. 

Collectively, these three tools, TAASSC, TAALED, and TAACO, capture stylistic features that characterize how writers communicate rather than, providing dimensions of expression, organization, and connectivity.

\subsection{Rewrites}

As discussed in Section \ref{sec:identification}, separating content from style requires generating stylistic variation while holding content fixed. For this purpose, we construct two sets of rewrites.

\begin{table}[!h]
    \small
    \setlength{\tabcolsep}{6pt}
    \renewcommand{\arraystretch}{1.25}    \begin{tabularx}{\textwidth}{@{}p{2.4cm}X@{}}
\toprule
\textbf{Version} & \textbf{Text} \\
\midrule
Original&
Some schools require students to complete summer projects to assure they continue learning during their break. In other words, teachers believe students should continue learning after the school year ends. Some might argue that summer projects should be teacher-designed so the student will have the ability to learn over the summer. Summer projects should be student-designed because this will allow the student to have more free time, it will allow less stress for the student, and they should decide for themselves stuff relating to the project. \\
\midrule
Rewrite SAT-1 &
Some schools \hl{said students gotta do} summer projects so learning doesn't stop when it's summer. Teachers think students \hl{gotta keep learning} after school is done. \hlcyan{Some people think summer projects should be made by teachers so students can keep learning.} Summer projects should be made by students \hl{cause this way it give students more free time}, \hl{make the students have less stress}, and \hl{cause the students should pick things about the project by themself}. \\
\midrule
Rewrite SAT-3 &
Some schools \hl{say students should do} summer projects so they can keep learning during the break. This means teachers think kids \hl{should keep learning} even when school is over. \hlcyan{Some people think summer projects should be made by teachers so students can learn in the summer.} But summer projects should be made by students \hl{because it will give them more free time}, \hl{it will make them less stressed}, and they \hl{should pick the details of the project themselves}. \\
\midrule
Rewrite SAT-6 &
Some schools \hl{mandate} the completion of summer projects to ensure that students remain engaged in learning during their break. In essence, educators hold the belief that \hl{learning should continue} beyond the academic year. \hlcyan{While some might contend that summer projects should be designed by teachers in order to provide a structured opportunity for summer learning, summer projects ought to be student-designed.} This approach \hl{enables students to enjoy greater free time}, \hl{reduces their levels of stress}, and \hl{empowers them to make decisions} about their own work. \\
\midrule
Neutral & Some schools \hl{mandate that} students complete summer projects to \hl{ensure} they continue learning during their \hl{vacation}. \hlcyan{In essence}, teachers believe students should \hl{engage in learning beyond the academic year}. Some \hl{argue} that summer projects should be \hl{crafted by teachers} so that students can \hl{actively} learn throughout the summer. \hlcyan{However}, summer projects should be student-designed because this \hl{would give} students more free time, \hl{reduce their stress levels}, and allow them to \hl{make decisions about the project themselves}. \\
\midrule
Legend &\textbf{\hl{Yellow}}: lexical register changes (informality, word choice, simplification).  \\ & \textbf{\hlcyan{Cyan}}: discourse-level or argumentative restructuring \\
\bottomrule
\end{tabularx}

    \caption{Example of SAT Rewrites}
    \label{tab:TabExample}
    \begin{minipage}{\linewidth}
    \footnotesize
    \justifying
    \textit{Note:} The table illustrates differences across rewrite levels. The original text is a paragraph from one student’s essay in response to a question about summer projects. SAT-1 is the lowest level (we prompt the model to rewrite the text in a low-level writing style), while SAT-6 is the highest. “Neutral” shows the rewritten version when we prompt the model to rewrite the text without any specific stylistic instructions. Yellow highlights indicate changes in lexical register, and cyan highlights indicate argumentative restructuring.  
    \end{minipage}
\end{table}

\paragraph{SAT Score-Based Rewrites.} For each original essay, we generate six rewrites by applying prompts that target specific SAT score tiers and operationalize each tier using the SAT rubric dimensions that primarily reflect language and style (e.g., “the essay exhibits skillful use of language” versus “the essay displays fundamental errors in vocabulary”). The prompts instruct the model to adjust only these stylistic components, such as word choice, sentence fluency, grammar/mechanics, and cohesion, to match the target tier, while keeping the essay’s substantive content and meaning fixed. The resulting rewrites are intended to span a quality gradient along a single ``writing quality’’ axis, yielding six versions per essay aligned with distinct quality levels. Table \ref{tab:TabExample} illustrates how the same student-written paragraph is rewritten at different levels. As the SAT level increases, the rewrites become more academic and formal in tone.

\paragraph{Neutral Rewrites/GPT Baseline.} We also generate rewrites using a single neutral prompt that serves as a control without explicit stylistic instructions. Rather than varying the prompt, we generate six outputs per essay by sampling repeatedly from the model conditional distribution of rewrites, on the same input. These texts are therefore stylistically more uniform, reflecting the model's default (``neutral'') writing style. Table \ref{tab:TabExample} illustrates how the model rewrites the same student-written paragraph when prompted to rewrite without any specific stylistic instructions.

\vspace{1em} 

All rewrites are generated using GPT-4o (via Azure), with \texttt{temperature = 1} and \texttt{max\_tokens = 2048}. The exact prompts are reported in Appendix \ref{app:RewritePrompt}.To satisfy Assumption \ref{ass:id-fidelity}, rewrites must preserve the underlying content of the original essay. We therefore subject each rewritten essay to an additional verification step using GPT-4o, which compares the rewritten text to the original and evaluates whether the substantive content has been altered (see Appendix \ref{app:RewritePrompt} for the verification prompt). Rewrites that are flagged as altering content are discarded from the analysis. Across the six SAT based rewrite samples, the vast majority of generated texts are retained after verification. Specifically, the number of rejected rewrites (i.e., flagged as altering content) is 54 (0.27\%) for rewrtie SAT of level 1, 38 (0.19\%) for level 2, 37 (0.19\%) for level 3, 676 (3.38\%) for level 4, 107 (0.54\%) for level 5, and 791 (3.96\%) for level 6. For the no-description/neutral rewrite condition, 329 rewrites (1.59\%) are rejected. Overall, rejection rates remain low across all rewrites, indicating that essay content in the large maintained in the majority of cases.

Using an LLM for content verification is appropriate in this setting because the task of assessing semantic equivalence between two texts is precisely the type of judgment for which large language models have been shown to perform reliably, and because the verifier is used only as a binary filter rather than as a source of continuous measurement. This design limits the scope for systematic bias from the verification step and ensures that identification relies on content-preserving rewrites by construction.

\section{Descriptive Statistics}
In this section, we first describe the data and the score gaps between high- and low-SES students in our sample. We then present our estimated scoring function, and finally we discuss the rewrites.

\subsection{Descriptive Statistics and The Score Gap}

Table \ref{tab:descriptive_stats} reports descriptive statistics for our sample. The sample is roughly balanced by socioeconomic status, with 46.5\% of students classified as low SES. Women make up 51.2\% of the sample, with a slightly higher share among low-SES students. Racial composition differs sharply by SES: 24.7\% of low-SES students are White, compared with 63.0\% among high-SES students. Finally, although the PERSUADE data cover grades 6–12, our SES-linked sample is concentrated in grades 6, 8, 10, and 11, where coverage is richest.

\begin{figure}[!h]
    \centering
    \includegraphics[width=\linewidth]{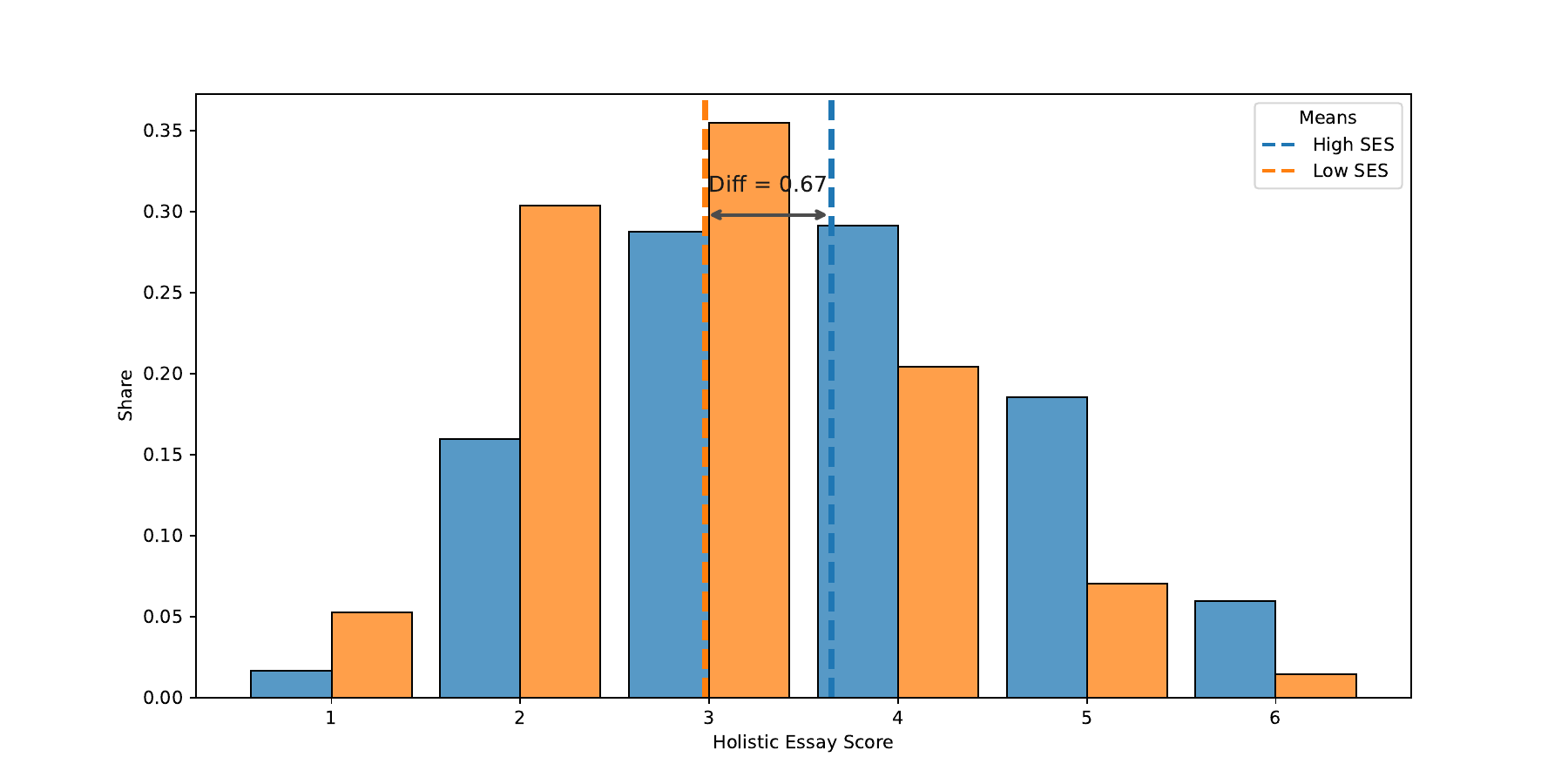}
    \caption{The Score Distribution For High and Low SES students}
    \label{fig:mainGap}

    \begin{minipage}{\linewidth}
    \footnotesize
    \justifying
    \textit{Note:} The figure shows overlaid histograms of the observed holistic essay score (x-axis, on the 1–6 scale) for high-SES and low-SES students, with the y-axis reporting the share of essays at each score. Dashed vertical lines mark the group means, and the plot annotates the mean difference between groups.
    \end{minipage}
\end{figure}

Next, we move the descire the score gaps. Figure \ref{fig:mainGap} shows the score gap we seek to explain by displaying the score distributions for high and low SES students. The dashed lines denote the mean score within each group. On average, high-SES students score about 0.67 points higher than low-SES students on the 1–6 scale.

Beyond the difference in means, the distributions exhibit distinct shapes. Scores for low-SES students are more concentrated at the lower end of the scale, particularly at scores of 2 and 3. In contrast, scores for high-SES students arec more dispersed, with the modal score at 4. High-SES students are also substantially more likely to receive top scores of 5 or 6, whereas such scores are relatively rare among low-SES students  (24.5\% vs. 8.5\%). Overall, the low-SES score distribution is shifted leftward relative to that of high-SES students.

\begin{table}[!h]
\centering
\resizebox{\textwidth}{!}{%
  \begin{tabular}{lllll}
\toprule
 & Full Sample & Low SES & High SES & Gap SES (High-Low) \\
\midrule
N & 20759 & 9643 & 11116 &  \\
Holistic Essay Score (Outcome) & \shortstack{3.337 \\ (0.008)} & \shortstack{2.979 \\ (0.011)} & \shortstack{3.648 \\ (0.011)} & \shortstack{0.669 \\ (0.016)} \\
Female & \shortstack{0.512 \\ (0.003)} & \shortstack{0.526 \\ (0.005)} & \shortstack{0.500 \\ (0.005)} & \shortstack{-0.026 \\ (0.007)} \\
White & \shortstack{0.452 \\ (0.003)} & \shortstack{0.247 \\ (0.004)} & \shortstack{0.630 \\ (0.005)} & \shortstack{0.384 \\ (0.006)} \\
6th Grade & \shortstack{0.066 \\ (0.002)} & \shortstack{0.073 \\ (0.003)} & \shortstack{0.060 \\ (0.002)} & \shortstack{-0.012 \\ (0.003)} \\
8th Grade & \shortstack{0.461 \\ (0.003)} & \shortstack{0.421 \\ (0.005)} & \shortstack{0.495 \\ (0.005)} & \shortstack{0.075 \\ (0.007)} \\
9th Grade & \shortstack{0.001 \\ (0.000)} & \shortstack{0.001 \\ (0.000)} & \shortstack{0.001 \\ (0.000)} & \shortstack{0.000 \\ (0.000)} \\
10th Grade & \shortstack{0.304 \\ (0.003)} & \shortstack{0.376 \\ (0.005)} & \shortstack{0.242 \\ (0.004)} & \shortstack{-0.134 \\ (0.006)} \\
12th Grade & \shortstack{0.149 \\ (0.002)} & \shortstack{0.101 \\ (0.003)} & \shortstack{0.190 \\ (0.004)} & \shortstack{0.089 \\ (0.005)} \\
Word Count (Volume) & \shortstack{412.621 \\ (1.348)} & \shortstack{368.372 \\ (1.756)} & \shortstack{451.006 \\ (1.932)} & \shortstack{82.635 \\ (2.610)} \\
Avg Word Length (Lexical Sophistication) & \shortstack{4.346 \\ (0.002)} & \shortstack{4.295 \\ (0.003)} & \shortstack{4.390 \\ (0.003)} & \shortstack{0.095 \\ (0.004)} \\
Mean Length of Clause (Syntactic Complexity) & \shortstack{12.128 \\ (0.021)} & \shortstack{11.911 \\ (0.032)} & \shortstack{12.316 \\ (0.027)} & \shortstack{0.405 \\ (0.042)} \\
Mean Length of T-Unit (Sentence Complexity) & \shortstack{28.954 \\ (0.085)} & \shortstack{28.721 \\ (0.127)} & \shortstack{29.157 \\ (0.113)} & \shortstack{0.436 \\ (0.170)} \\
Mean Verbal Dependencies & \shortstack{6.107 \\ (0.007)} & \shortstack{5.992 \\ (0.010)} & \shortstack{6.207 \\ (0.009)} & \shortstack{0.215 \\ (0.013)} \\
Proportion of Infinitives & \shortstack{0.151 \\ (0.000)} & \shortstack{0.148 \\ (0.001)} & \shortstack{0.154 \\ (0.001)} & \shortstack{0.006 \\ (0.001)} \\
Proportion of Non-Finite Clauses & \shortstack{0.413 \\ (0.001)} & \shortstack{0.400 \\ (0.001)} & \shortstack{0.424 \\ (0.001)} & \shortstack{0.023 \\ (0.001)} \\
Lexical density (tokens) & \shortstack{0.438 \\ (0.000)} & \shortstack{0.433 \\ (0.000)} & \shortstack{0.443 \\ (0.000)} & \shortstack{0.011 \\ (0.001)} \\
Lexical diversity (MTLD; all words) & \shortstack{55.275 \\ (0.089)} & \shortstack{54.212 \\ (0.134)} & \shortstack{56.198 \\ (0.118)} & \shortstack{1.986 \\ (0.178)} \\
Lexical diversity (MATTR; lemmas) & \shortstack{0.739 \\ (0.000)} & \shortstack{0.737 \\ (0.000)} & \shortstack{0.741 \\ (0.000)} & \shortstack{0.004 \\ (0.001)} \\
Content-word overlap (adjacent sentences) & \shortstack{0.161 \\ (0.000)} & \shortstack{0.160 \\ (0.001)} & \shortstack{0.163 \\ (0.001)} & \shortstack{0.003 \\ (0.001)} \\
Causal connectives (reason \& purpose) & \shortstack{0.012 \\ (0.000)} & \shortstack{0.013 \\ (0.000)} & \shortstack{0.012 \\ (0.000)} & \shortstack{-0.001 \\ (0.000)} \\
Nominalizations & \shortstack{11.316 \\ (0.082)} & \shortstack{8.912 \\ (0.095)} & \shortstack{13.401 \\ (0.125)} & \shortstack{4.489 \\ (0.157)} \\
Pronoun-to-noun ratio & \shortstack{0.305 \\ (0.001)} & \shortstack{0.312 \\ (0.002)} & \shortstack{0.299 \\ (0.001)} & \shortstack{-0.013 \\ (0.002)} \\
\bottomrule
\end{tabular}
}
\caption{Descriptive Statistics}
\label{tab:descriptive_stats}
\end{table}

The SES score gap appears across demographic subgroups and grades. The gap is larger among non-White students (0.746) than White students (0.569), and similar for males (0.657) and females (0.682). By grade, the gap is smallest in grade 6 (0.238) and remains around 0.48-0.58 in grades 8–11. 

% \textbf{Add differences in scores for SES by gender, race and Grade}

Figure \ref{fig:scoreGap_by_prompt} in the Appendix shows that this gap is relatively stable across writing prompts. The estimated differences range from 0.73 points for the “Mandatory Extracurricular Activities” prompt to 0.41 for “Community Service” and 0.23 for “A Cowboy Who Rode the Wave.” The consistency of the gap across assignments and topical domains suggests that the observed score differences are not driven by differences in specific area, but are likely to exists across wide array of domains.

\subsection{The Scorer Functions}
Figure \ref{fig:lowHighPredictor} assesses the performance of the ranking model separately for low- and high-SES students. The figure plots the expected holistic score conditional on the model’s predicted score (x-axis). Intuitively, if the ranker is unbiased and successfully captures the underlying scoring rule, the average realized score should coincide with the prediction, placing the conditional expectation along the 45-degree line. The figure shows that this relationship holds closely for both low and high SES scorers. Deviations from the 45-degree line are small and statistically non significant, indicating limited systematic bias and suggesting that the model captures the underlying ranking function well for both groups. %Figure XX \textbf{(Pedro - where the figure)} in the Appendix provides complementary evidence using a clipped confusion matrix, in which predicted scores are rounded to their core values. The resulting matrix exhibits a strong concentration of mass along the main diagonal, consistent with accurate ranking and limited misclassification.

Figure \ref{fig:r2AcrossSES} reports the $R^2$ values for the two group-specific predictors. Both predictive models explain a large share of the variation in holistic scores (0.735 for low-SES students and 0.764 for high-SES students), indicating strong predictive performance in both groups and suggesting that our set of variables captures most of the systematic variation in scores.

\begin{figure}[!htbp]
    \centering
    \begin{subfigure}{0.48\textwidth}
        \centering
        \includegraphics[width=\linewidth]{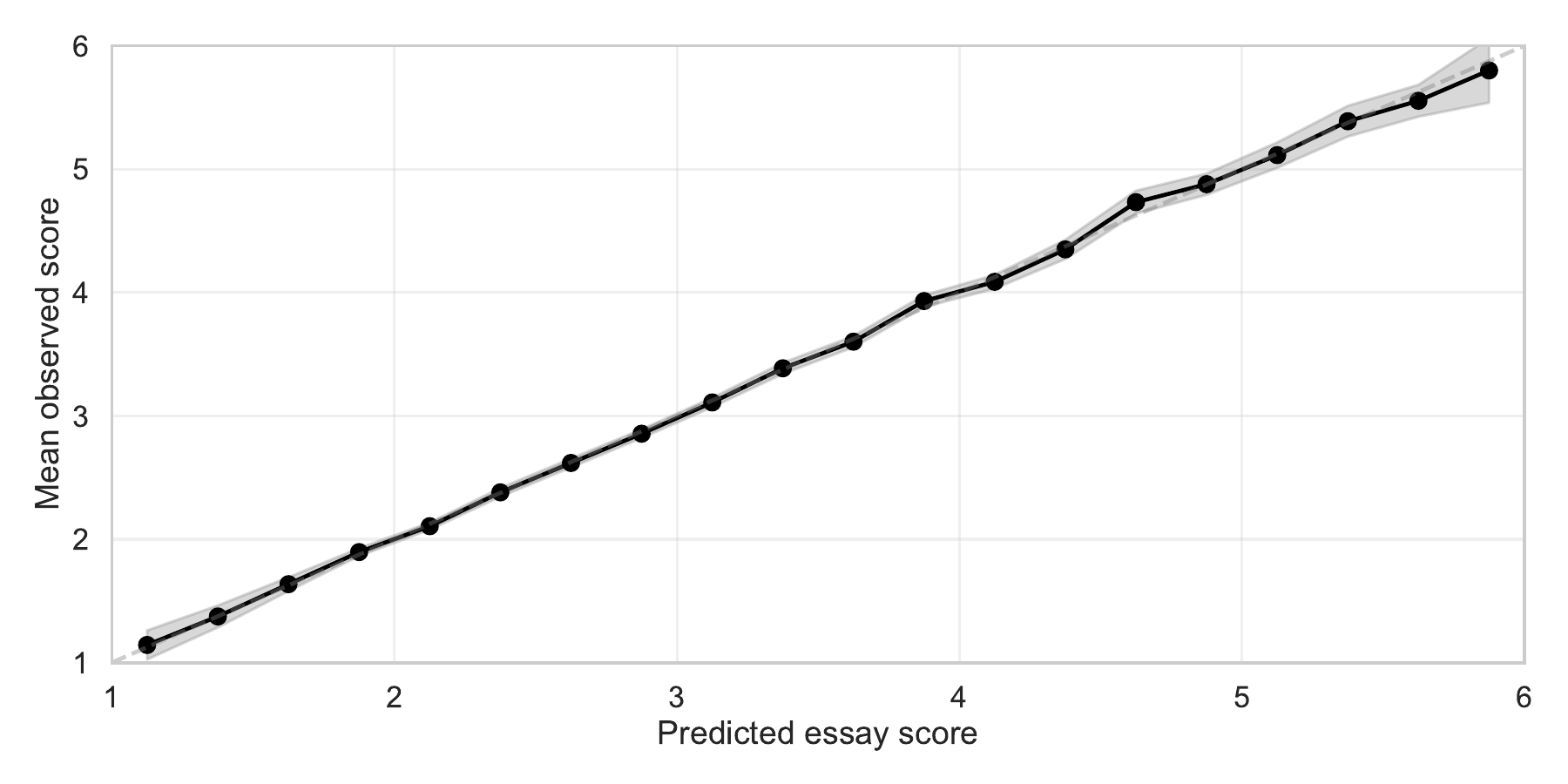}
        \caption{Low SES Score Function}
        \label{fig:one}
    \end{subfigure}
    \hfill
    \begin{subfigure}{0.48\textwidth}
        \centering
        \includegraphics[width=\linewidth]{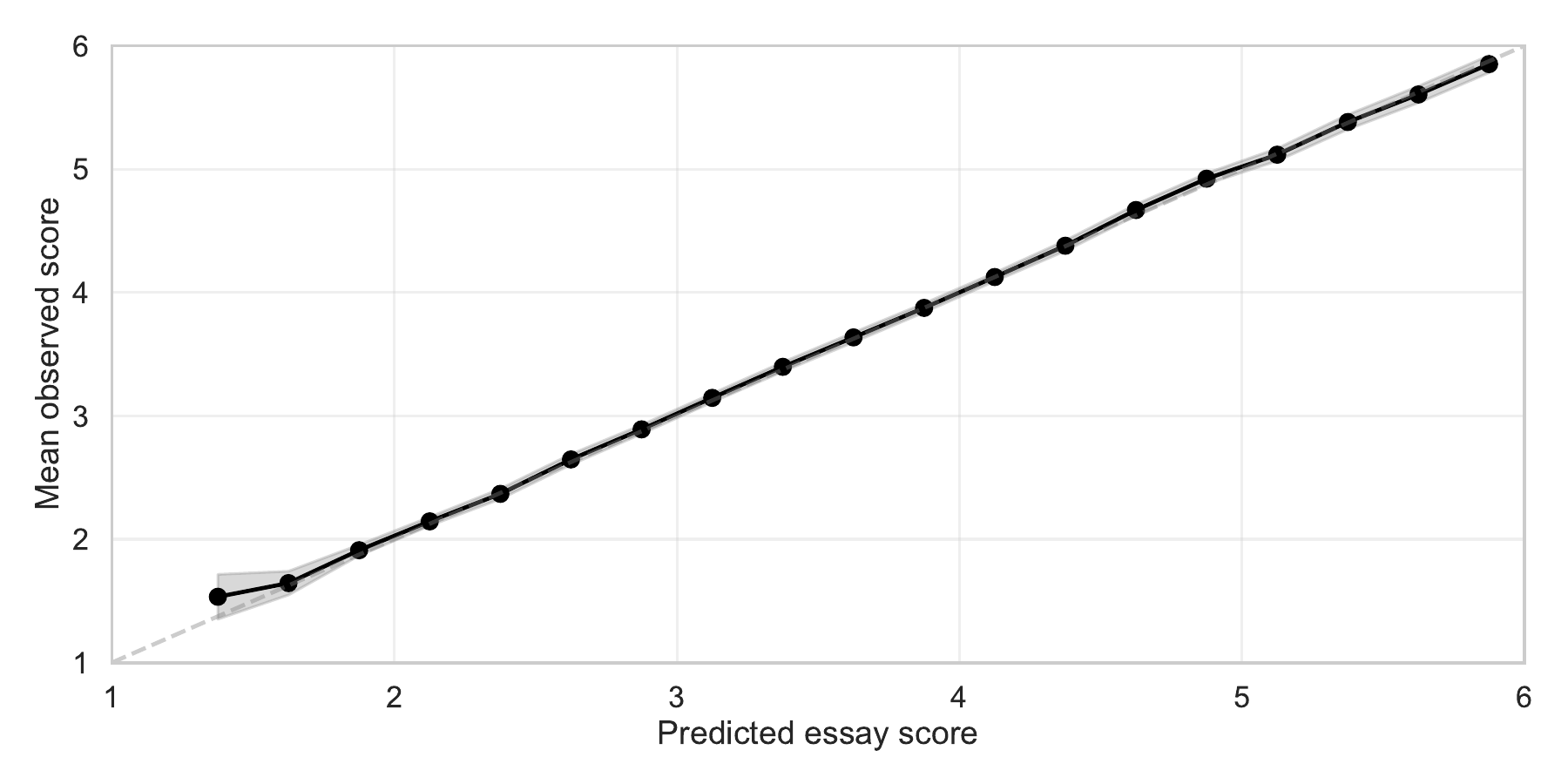}
        \caption{High SES Score Function}
        \label{fig:two}
    \end{subfigure}
    \caption{Low and High SES predictor Bias}\label{fig:highVsLowRankers}
    \begin{minipage}{\linewidth}
    \footnotesize
    \justifying
    \textit{Note:} The figure shows “score functions” for two models in separate panels: (a) a low-SES score function and (b) a high-SES score function. In each panel, the x-axis is the model’s predicted essay score and the y-axis is the mean observed holistic score; plotted points summarize the relationship by bins (with uncertainty bands/intervals), and a 45-degree line is included as a reference
    \end{minipage}
\end{figure}

Figure \ref{fig:highVsLowRankers} compares the rankings assigned by the high- and low-SES models for the same essays. The figure plots binned averages of the high-SES predicted scores against the corresponding low-SES predicted scores. This relationship directly reflects the “tilting” of the score distribution across groups. Large deviations from the 45-degree line would indicate substantial differences in how the two models rank essays, and thus a larger contribution of ranking differences to the overall score gap. In contrast, points lying close to the 45-degree line imply broadly similar rankings across groups. The figure shows that across most of the support, high-SES predicted scores lie slightly above the 45-degree line. This indicates that, on average, the high-SES model assigns marginally higher ranks to the same essays than the low-SES model. These differences are small in magnitude and increase modestly at higher score levels, suggesting limited but systematic tilting rather than large discrepancies in ranking.

\begin{figure}[!htbp]
    \centering
    \begin{subfigure}[t]{0.48\textwidth}
        \centering
        \includegraphics[width=\linewidth]{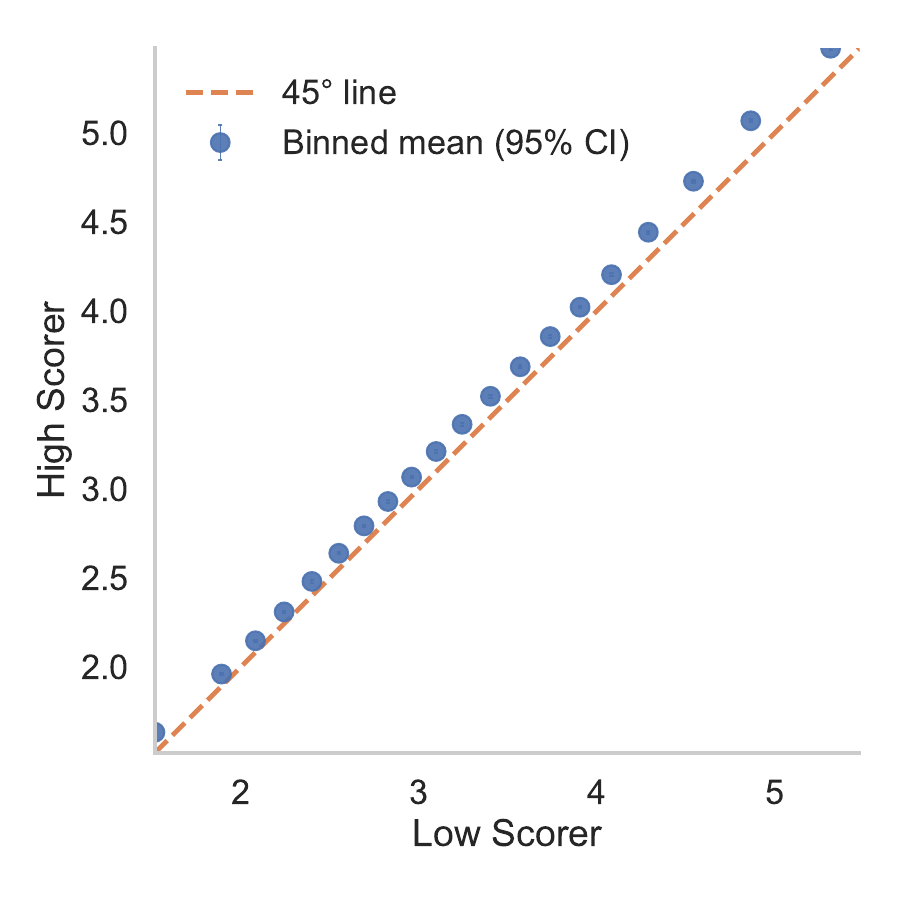}
        \caption{Low vs.\ High Predictors}
        \label{fig:one}
    \end{subfigure}
    \hfill
    \begin{subfigure}[t]{0.48\textwidth}
        \centering
        \includegraphics[width=\linewidth]{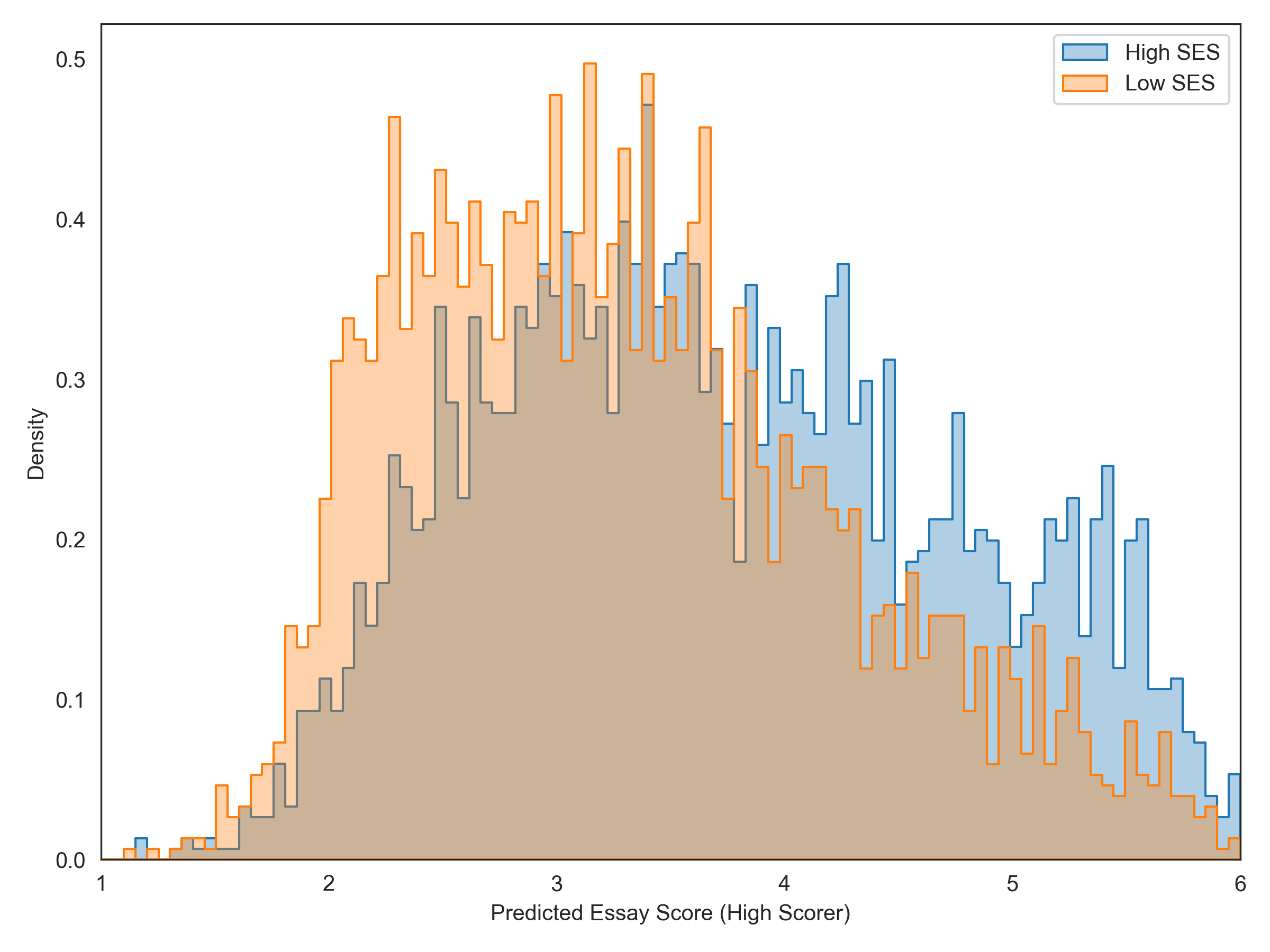}
        \caption{Predicted Scores Distribution (High Scorer)}
        \label{fig:distribution_of_prediction}
    \end{subfigure}
    \caption{Predicted Scores Distribution (High Scorer)}
    \label{fig:lowHighPredictionDistribution}
    % \justified
    \begin{minipage}{\linewidth}
    \footnotesize
    \justifying
    \textit{Note:} The figure has two panels summarizing predictions from the “High Scorer.” Panel (a) plots binned mean predicted scores from the High Scorer (y-axis) against predicted scores from the Low Scorer (x-axis), with a 45-degree reference line and 95\% confidence intervals around binned means. Panel (b) shows overlaid histogram distributions of the High Scorer’s predicted essay scores for high-SES and low-SES students.
    \end{minipage}
\end{figure}

Finally, Figure \ref{fig:distribution_of_prediction}\footnote{The corresponding distribution under the low-scorer model is shown in Figure \ref{fig:predictedLowScorerDist} in the Appendix.} plots the distribution of predicted essay scores for high- and low-SES students using the same (high-SES) scoring function and the full set of explanatory variables. Even under this common scoring rule, the predicted distribution for low-SES students is systematically shifted to the left, placing substantially more mass on lower scores relative to high-SES students. This comparison isolates differences in inputs rather than differences in how inputs are rewarded. The persistence of a sizable distributional gap under a fixed scoring function therefore indicates that disparities in the scoring rule alone cannot account for the observed SES gap. Instead, the figure reinforces the conclusion that differences in underlying content and style features play a central role.

% \begin{figure}[!h]
%     \centering
%     \includegraphics[scale=0.5]{figuresFinal/descriptives/highVsLowSESPredictor.pdf}
%     \caption{Caption}
%     \label{fig:highVsLowRankers}
% \end{figure}

\subsubsection{Content and Style}
\begin{figure}[!h]
    \centering
    \includegraphics[width=0.8\linewidth]{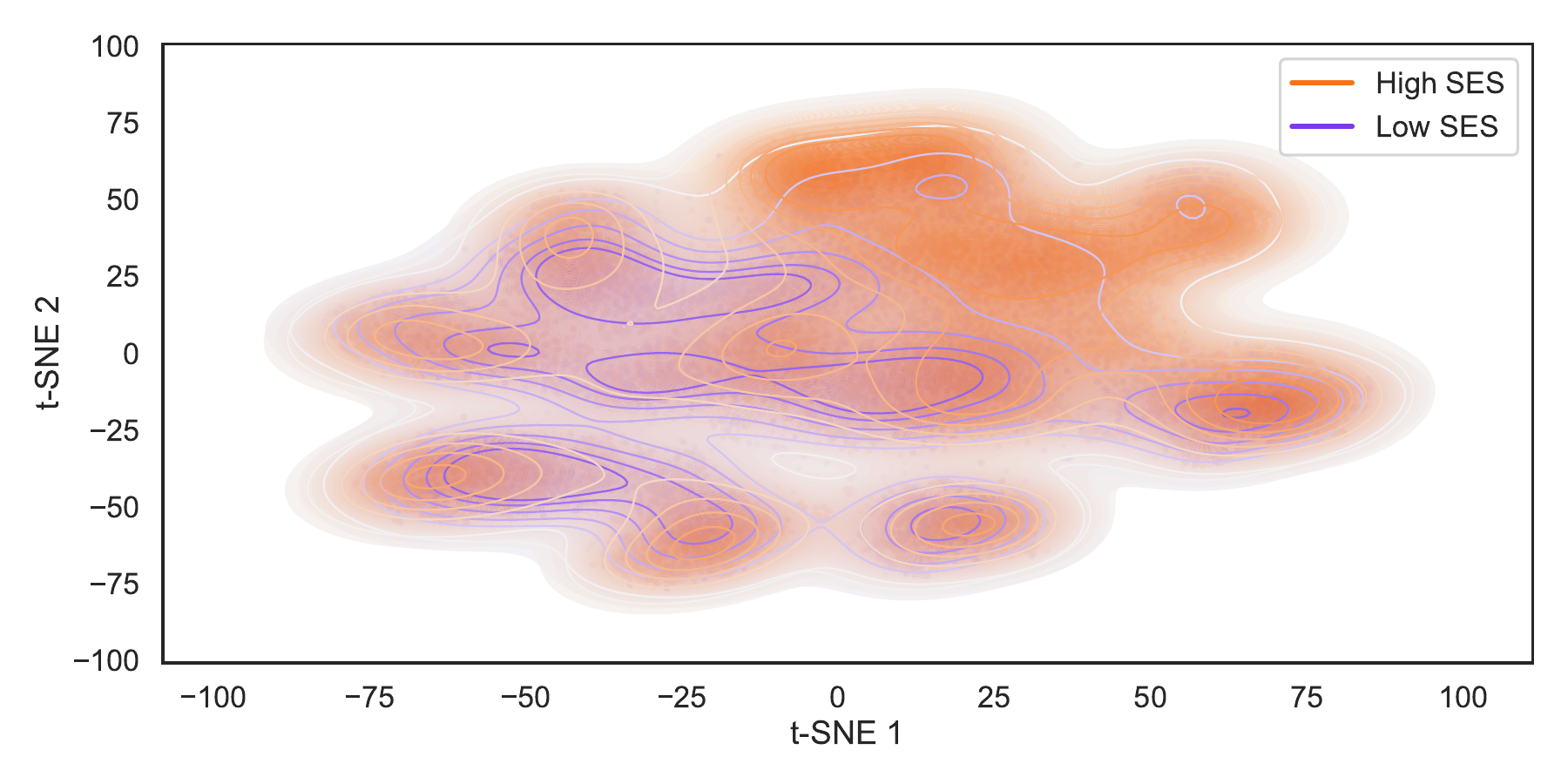}
    \caption{Density of t-SNE Projection of Text Embeddings by Socioeconomic Status}
    \label{fig:tsneDistAll}
    \begin{minipage}{\linewidth}
    \footnotesize
    \justifying
    \textit{Note:} The figure shows a two-dimensional t-SNE projection of text embeddings with axes labeled t-SNE 1 and t-SNE 2. It overlays density shading/contours for high-SES and low-SES essays (pooled across prompts), allowing visual comparison of where each group’s essays concentrate in the embedded space.
    \end{minipage}
\end{figure}

We now turn to examine differences in the  content and style variables. Figure \ref{fig:tsneDistAll} displays the density of a two-dimensional projection of the text embeddings, obtained using the t-SNE algorithm \citep{vanDerMaaten2008tsne}, pooling across all writing prompts. The figure shows that high- and low-SES students exhibit distinct regions of concentration in several prompts. To the extent that the embeddings capture semantic differences across topics, this pattern indicates that the writing of high- and low-SES students occupies different regions of the content space, suggesting systematic differences in the ideas they choose to express. We also perform an exercise in which we predict the SES of essay authors using the embeddings. We find strong separation, achieving an AUC of 0.705, again implying that content differs between high- and low-SES students.

Figure \ref{fig:tsneGridPrompt} in the Appendix explores these distributions across the 12 different writing prompts. We again find that the observed content differences are not driven by any single prompt, but instead appear consistently across prompts. This pattern suggests that high- and low-SES students systematically differ in how they respond to these questions, rather than reacting differently to a particular prompt.

Turning to style, table \ref{tab:descriptive_stats}, shows a sample of the style variables in our sample\footnote{We include an explanation of each style variable that appears in table \ref{tab:descriptive_stats}, in Appendix \ref{app:styleVariables}}. The table shows that high-SES students write essays that look systematically more “academic” and elaborated than low-SES students along multiple stylistic dimensions. A major difference between high and low SES is sheer volume: high-SES essays are about 83 words longer on average (451 vs. 368, roughly 22\% longer), which goes hand-in-hand with higher-scoring writing and may mechanically allow more development of ideas. Beyond length, high-SES essays use slightly more lexically sophisticated and information-dense language: average word length is higher (4.390 vs. 4.295), lexical density is modestly higher (0.443 vs. 0.433), and lexical diversity is higher as measured by MTLD\footnote{MTLD (Measure of Textual Lexical Diversity) is a length-robust measure of vocabulary variety that tracks how quickly word repetition accumulates as a text unfolds. Higher MTLD indicates more diverse vocabulary because longer segments are needed before the text’s type–token ratio falls below a fixed threshold.
} (56.2 vs. 54.2), with a smaller but consistent increase in lemma-based MATTR (0.741 vs. 0.737)\footnote{MATTR (Moving-Average Type–Token Ratio) measures lexical diversity by computing the type–token ratio within a fixed-length sliding window and averaging it across the text. Higher MATTR indicates more varied vocabulary, and using a fixed window makes it less sensitive to overall text length than the raw type–token ratio.}. High-SES essays also show somewhat greater syntactic complexity: clauses are longer (MLC 12.316 vs. 11.911), T-units are longer (29.157 vs. 28.721), verbs carry more syntactic attachments (mean verbal dependencies 6.207 vs. 5.992), and non-finite constructions appear more often (infinitives 0.154 vs. 0.148; non-finite clauses 0.424 vs. 0.400), consistent with more embedding and syntactic compression. Looking at nominalization, we find a pronounced gap between high- and low-SES essays. High-SES essays contain substantially more nominalizations (13.4 vs. 8.9; a difference of 4.49, roughly a 50\% increase), consistent with a more abstract, noun-heavy style characteristic of academic prose. In contrast, discourse cohesion and explicit causal signaling differ little: adjacent-sentence content-word overlap is nearly identical (0.163 vs. 0.160), and causal connectives are if anything slightly less frequent in high-SES writing (0.012 vs. 0.013). Finally, high-SES essays rely a bit less on pronouns relative to nouns (pronoun-to-noun ratio 0.299 vs. 0.312), indicating more explicit noun-based reference, which often reads as clearer and more formal.

\begin{figure}[!htbp]
    \centering
    \begin{subfigure}{0.48\textwidth}
        \centering
        \includegraphics[width=\linewidth]{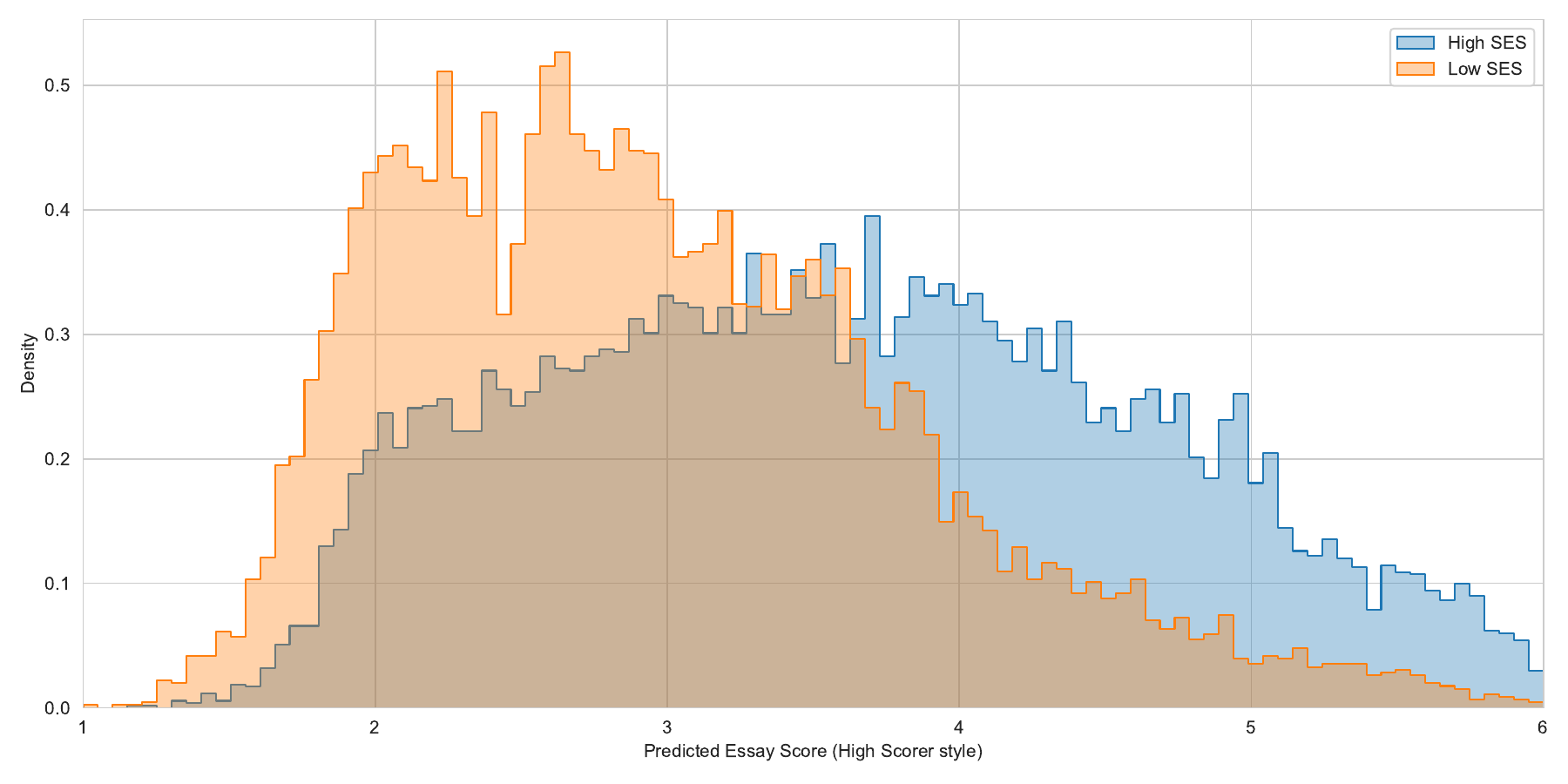}
        \caption{Prediction Distribution Using \emph{Only} Style Variables}
        \label{fig:stylePredictionDist}
    \end{subfigure}
    \hfill
    \begin{subfigure}{0.48\textwidth}
        \centering
        \includegraphics[width=\linewidth]{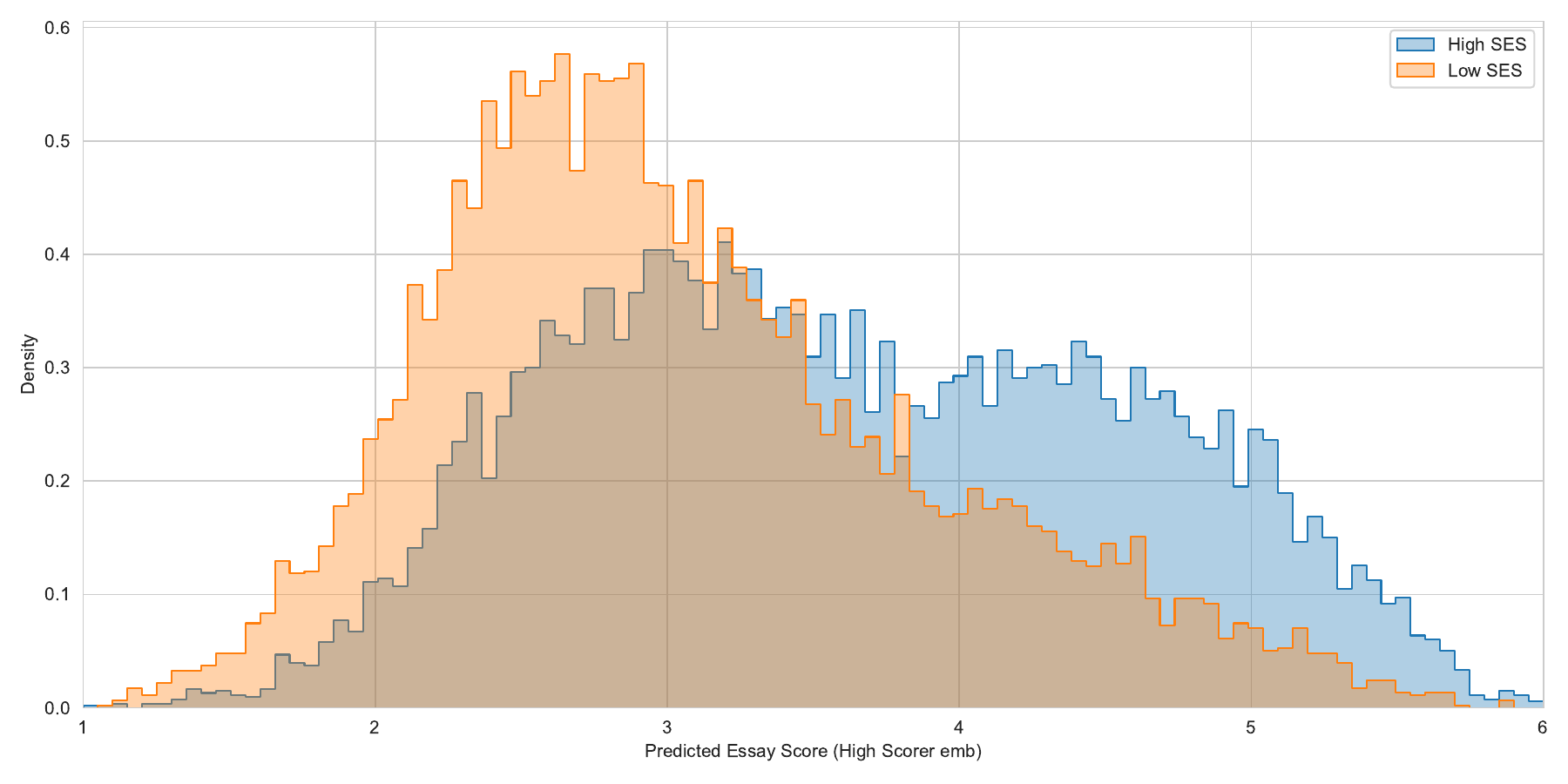}
        \caption{Prediction Distribution Using \emph{Only} Embedding Variables}
        \label{fig:embPredictionDist}
    \end{subfigure}
    \caption{Low and High SES predictor Bias}
    \label{fig:lowHighPredictor}
    \begin{minipage}{\linewidth}
    \footnotesize
    \justifying
    \textit{Note:} The figure shows distributions of predicted essay scores from two different prediction models, split by SES. Panel (a) overlays high-SES and low-SES histograms of predicted scores from a model using only style variables; panel (b) does the same for a model using only embedding variables.
    \end{minipage}
\end{figure}

These descriptive patterns, indicating higher levels of writing among high-SES students, also show up in a single composite style score. Figure \ref{fig:stylePredictionDist} plots the distribution of predicted-score probabilities from a model trained only on the style variables. High-SES essays are spread more evenly across the score range, whereas low-SES essays are disproportionately concentrated at low predicted holistic scores. This separation is qualitatively similar to what we see using the full feature set (all variables and embeddings), though the style-only model produces a more pronounced pile-up at the bottom among low-SES essays.

Comparing figure \ref{fig:stylePredictionDist} and \ref{fig:embPredictionDist} demonstrates our main challenge in isolating the role of  content and the style in shaping outcomes for students. Figure \ref{fig:embPredictionDist} shows the predicted value distribution based on the embedding only. As we can see the implied distribution are very similar to to the style distribution, implying that the two may carry similar infomration.  Figure \ref{fig:lowScorer_PredictionCorrelation}explore this more and shows the correlation between the scores assigned using predictors who were trained only on a subset of the variables (all variables, embedding, all only style varaibles, TAACO, TAASSC, TAALED). The figure shows very high correlation in predicted values, indicating that the different set of variables carry similar information on on the holistic score.

\begin{figure}[!h]
    \centering
    \includegraphics[width=\linewidth]{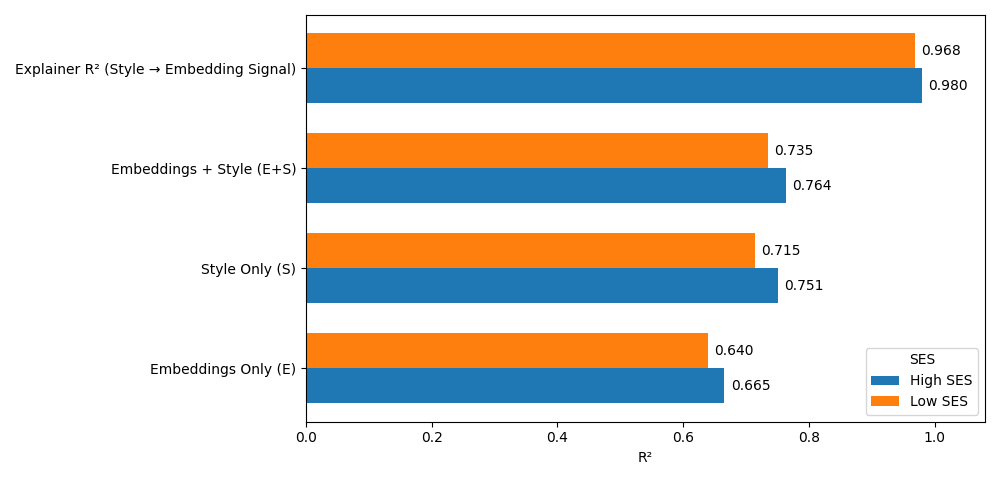}
    \caption{$R^2$ Across Explanatory Variables and Information Redundancy}
    \label{fig:r2AcrossSES}
    \begin{minipage}{\linewidth}
    \footnotesize
    \justifying
    \textit{Note:} The figure shows a horizontal bar chart of \(R^2\) values (x-axis) for multiple model specifications (y-axis categories: Embeddings only, Style only, Embeddings + Style, and an “Explainer \(R^2”\), that captures the fraction of variance in the holistic score explained by the content-only model that is also explained by the style-only model. Each specification has separate bars for high-SES and low-SES, and the numeric \(R^2\) values are printed at the ends of the bars.
    \end{minipage}
\end{figure}

Figure \ref{fig:r2AcrossSES} makes the redundancy between “content” (embeddings) and our style measures especially clear. Using embeddings alone already predicts a large share of score variation: $R^2=0.64$ for low-SES students and 0.66 for high-SES students. Style variables on their own perform slightly better ($R^2=0.71$ and $0.75$, respectively). But combining content and style adds only a modest incremental gain (to $0.73$ for low SES and $0.764$ for high SES) despite doubling the information set.

This small marginal improvement implies that content and style are strongly correlated in the data: much of what style “explains” is already encoded in the semantic/content representation, and vice versa. That pattern is intuitive. More nuanced ideas often require more precise lexical and syntactic control, so stronger content tends to co-occur with more “academic” style. Likewise, some ideas are inherently harder to express tersely—complex arguments naturally demand longer sentences and more elaborate structure—so stylistic sophistication can be partly a byproduct of the content being conveyed. More so, although we treat embeddings as capturing semantic meaning, contextual representations also encode surface and syntactic regularities that reflect stylistic choices \citep{jawahar2019does}, reinforcing that content and style are difficult to cleanly separate in practice.

To quantify redundancy between style features and content (embedding) features, we measure how much of the variance explained by content is also explained by style. Let $R_S^2$ denote the $R^2$ from regressing the holistic score on style variables only, $R_C^2$ the $R^2$ from regressing the score on content variables only, and $R_{SC}^2$ the $R^2$ from regressing the score on both sets of variables. Define the incremental contribution of content beyond style as $\Delta_C \equiv R_{SC}^2 - R_S^2$. Then $R_C^2 - \Delta_C$ is the portion of the content-only explained variance that is not unique to content: it is variance that content explains in isolation, but that becomes redundant once style is included because style already accounts for it.

We summarize this redundancy via
\[R^2_{\mathrm{expl}}
=\frac{R_C^2-\Delta_C}{R_C^2}
=\frac{R_S^2+R_C^2-R_{SC}^2}{R_C^2}.\]
The numerator, $R_C^2-\Delta_C = R_S^2+R_C^2-R_{SC}^2$, is the standard commonality (shared explained variance) of style and content with respect to holistic scores. Dividing by $R_C^2$ expresses this overlap as a fraction of the explanatory power of the content model. Thus, $R^2_{\mathrm{expl}}$ measures the share of the content model’s signal that is recoverable from style.

In Figure \ref{fig:r2AcrossSES}, we find $R^2_{\mathrm{expl}}\approx 0.98$, for both high and low SES models, implying that about 98\% of the variance in scores explained by content features is also explained by style features, leaving only about 2\% as uniquely incremental content signal beyond style. Overall, this indicates that style and content features contain highly overlapping information about the score.

\subsection{Rewrites}\label{sec:rewrites}

% We next turn to the analysis of the rewritten essays. Figure \ref{fig:rewriteAverageScore} shows the average predicted score for each rewritten essay style, for low or high SES students. The figure shows that our approach of moving along an axis works, as increasing the SAT is indeed increasing the grade. The figuer also shows the standard GPT approach improves the average score, putting it at around similar level to SAT level 4. The figure shows that in general, rewriting the text does not close or eliminate the gap, supporting our guess that some of the gap by differences associated with content, and not just writing style. Therefore a policy that only aims to equate expression levels won't close the gap entirely. 

% We also note that for SAT 1 and SAT 2, the scores actually drop below the original text for both groups. It shows the LLM isn't just "fixing up" the essays and it's successfully complying with the instruction to write "worse" text. The fact that the gap still exists at SAT 1 reinforces that the our scorer function is detecting a difference in the underlying ideas and content.

\begin{figure}[!h]
    \centering
    \includegraphics[width=\linewidth]{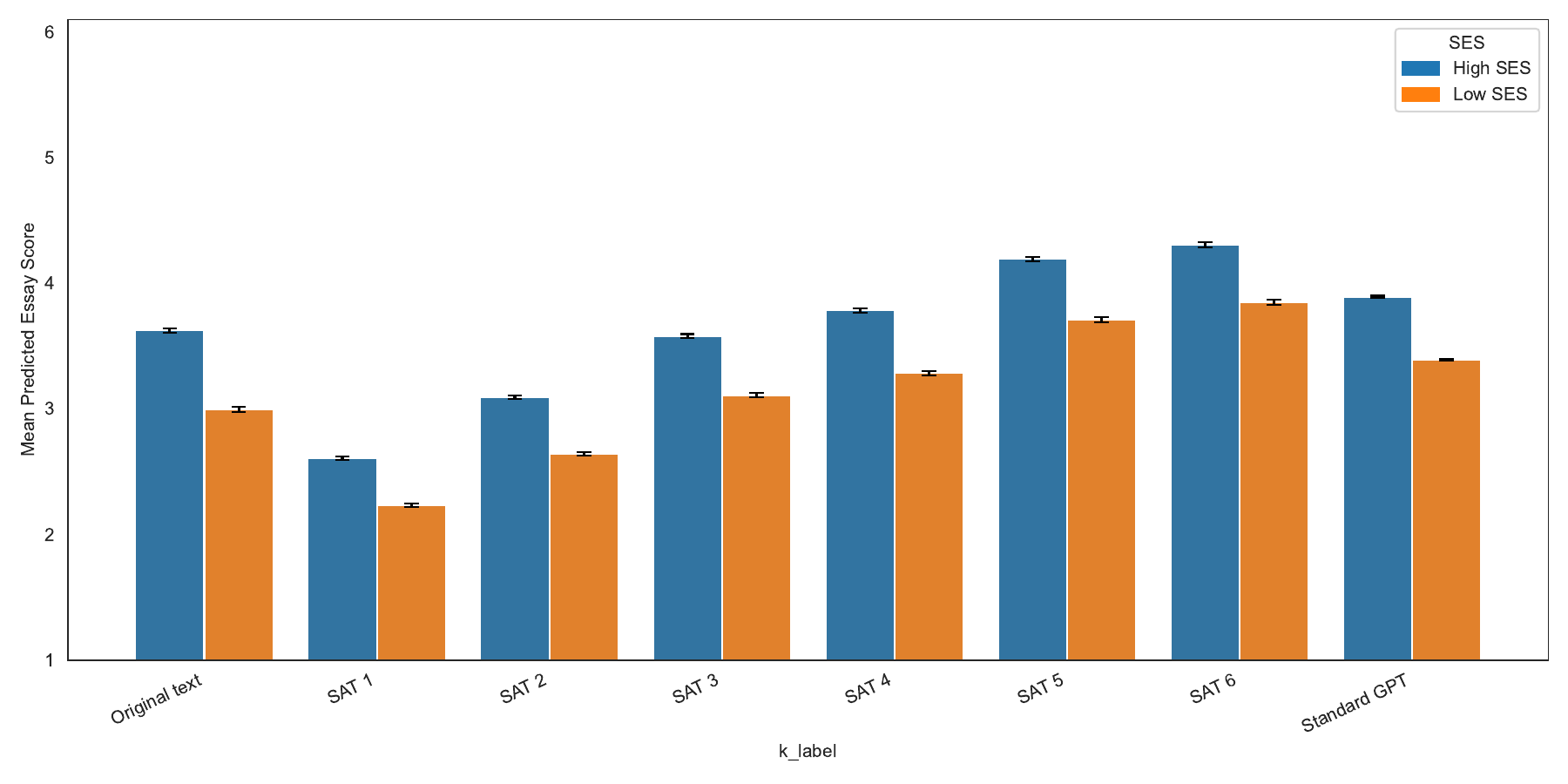}
    \caption{Mean Scores By Rewrite style}
    \label{fig:rewriteAverageScore}
    \begin{minipage}{\linewidth}
    \footnotesize
    \justifying
    \textit{Note:} The figure shows mean predicted essay scores (y-axis) for the original essays and multiple rewrite conditions (x-axis categories: Original text, SAT 1–SAT 6, and Standard GPT). For each rewrite condition, it displays side-by-side bars for high-SES and low-SES with error bars around the mean.
    \end{minipage}

\end{figure}

We next turn to the rewritten-essay analysis. Figure \ref{fig:rewriteAverageScore} plots the mean predicted holistic score for each rewrite style, separately for low- and high-SES students. The figure validates our ``move along a writing axis'' design: as the target SAT level increases from 1 to 6, average predicted scores rise monotonically for both groups, indicating that the rewrites successfully shift writing quality in the intended direction. At the lower end, SAT-1 and SAT-2 rewrites score below the original essays for both groups, confirming that the model is not mechanically ``improving'' text but can also degrade it when instructed, with our scorer responding accordingly.

The figure also helps calibrate the standard ``GPT-polish'' baseline. On average, the generic GPT rewrite raises predicted scores to roughly the same level as our SAT-4 rewrites, suggesting that this commonly used approach corresponds to a sizable upward move along the style axis.

Importantly, none of the rewrite regimes comes close to eliminating the SES gap. Even when both groups are pushed toward a common reference style, a substantial score difference remains. This persistence is consistent with the view that part of the observed gap reflects differences in content---ideas, argument structure, and substance---rather than expression alone.

These patterns speak directly to Assumption \ref{ass:id-style}. As we vary the target style from SAT level 1 to 6, scores for both groups move in parallel: the transformation shifts the level of predicted scores up or down, but leaves the High--Low gap approximately unchanged. This ``parallel shift'' is what we would expect if the rewrite operates primarily through a style channel that enters additively and similarly for both groups, while content differences remain as a stable wedge. Appendix Figure \ref{fig:kdensity_by_k} provides complementary distributional evidence. It plots kernel density estimates of the predicted-score distribution for each rewrite level. The densities shift upward or downward with the target SAT level but remain similar in shape, consistent with the rewrites acting approximately as a constant (additive) style shift.

\begin{figure}[!h]
    \centering
    \includegraphics[width=\linewidth]{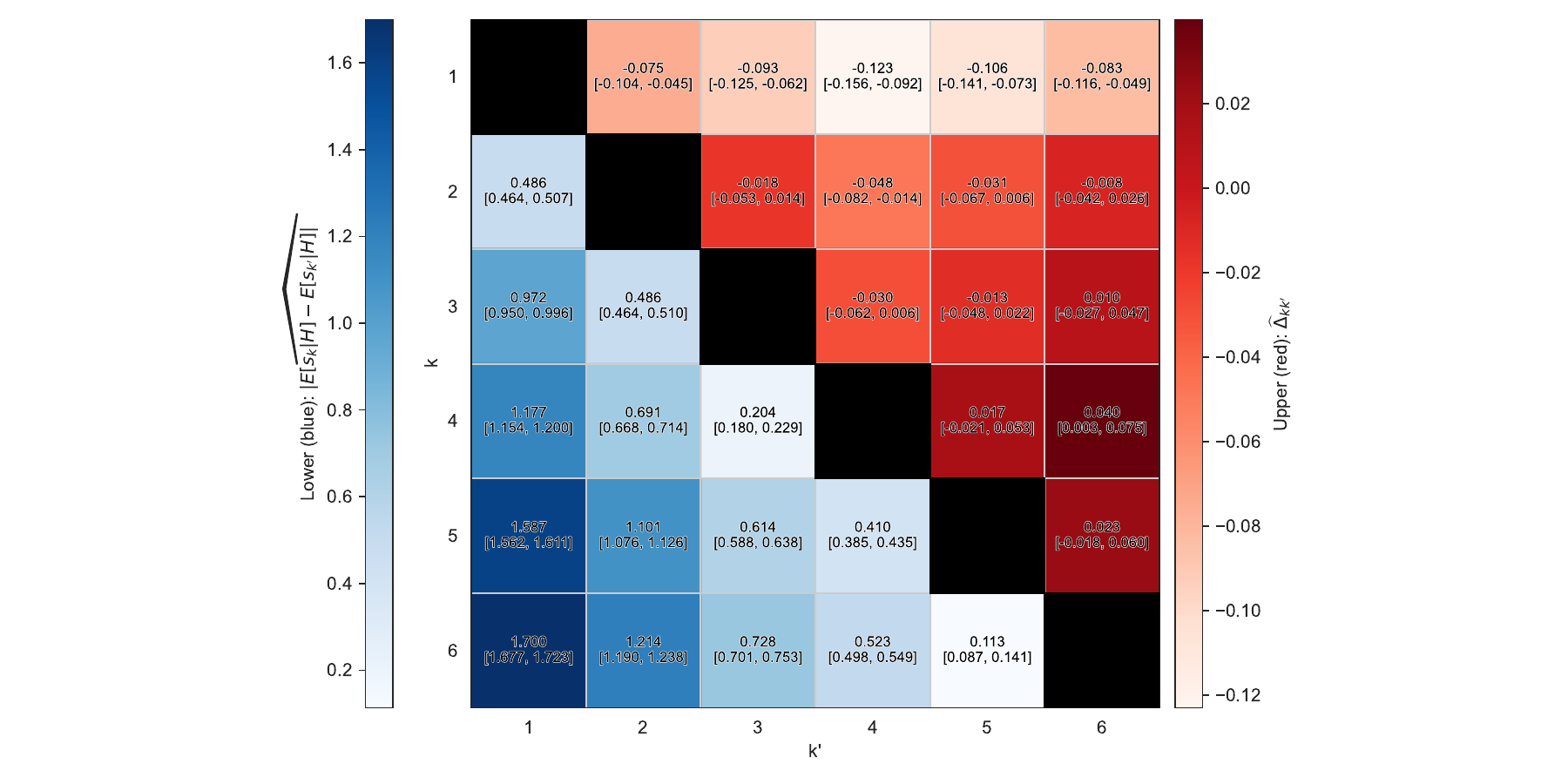}
    \caption{Mean Score for each Rewrite Type}
    \label{fig:diffInDiff}
        \begin{minipage}{\linewidth}
    \footnotesize
    \justifying
\textit{Note:} The figure presents a comparison matrix across SAT rewrite levels. Rows are indexed by $k$ (labeled 1--6 on the left; the target rewrite SAT score), and columns are indexed by $k'$ (labeled 1--6 along the bottom). The lower triangle (blue cells, where $k>k'$) reports the magnitude of the within--High-SES difference in mean scores between rewrite level $k$ and rewrite level $k'$, i.e.,$\left|\Delta^{k,k'}_{H}\right|
\quad \text{where} \quad
\Delta^{k,k'}_{H} = \mathbb{E}\!\left[s_{ik}-s_{ik'} \mid H\right]$. Blue color intensity corresponds to the cell value using the left (blue) colorbar (larger values $\Rightarrow$ darker blue). Each blue cell reports (i) the point estimate (first line) and (ii) a bracketed confidence interval (second line), computed via a bootstrap with 2{,}000 repetitions. The upper triangle (red cells, where $k<k'$) reports a difference-in-differences comparison across SES groups for the rewrite pair $(k,k')$, denoted $\Delta_{k,k'}$ (the ``upper (red)'' quantity). Specifically, it compares the within-group rewrite difference for High SES to the corresponding within-group rewrite difference for Low SES for the same pair $(k,k')$. Red color intensity follows the right (red) colorbar (more negative values $\Rightarrow$ darker red; values closer to zero $\Rightarrow$ lighter). As in the lower triangle, each red cell reports a point estimate with a bracketed confidence interval underneath, computed via a bootstrap with 2{,}000 repetitions.
    \end{minipage}

\end{figure}

% To examine this further, we preform a "diff-in-diff" style analysis, comparing the difference between between the rewrites, within group, therefore controlling for within group differences in content, and then comparing these differences across groups. Under assumption \ref{ass:id-style} the difference in difference should be zero. To see that notice that  
% \[
% \Delta^{kk'}_G = E[s_{ik} - s_{ik'}|G] = E[e_{ik} - e_{ik'}|G] = \lambda_{k} - \lambda_{k'},
% \]
% is independent from the group membership. 

% Figure \ref{fig:diffInDiff} shows the results. The upper triangle in red shows the results from the diff-in-diff, where the lower triangle in bleu shows the absolute value of $\Delta^{kk'}_H$. As we can see the all the values in red are reltivly small and close to zero, indiciating that the diff-in-diff support our assumption. more so, the differences are even smaller when compared with the absolute value of The rewrite premium, which are order of magnitude over the differences in red. 

To probe Assumption \ref{ass:id-style} more sharply, we run a difference-in-differences diagnostic that isolates the \emph{incremental effect of a rewrite level} net of within-group content. The first difference compares two rewrite levels within the same SES group, so any stable group-specific content differences cancel. The second difference then compares these within-group rewrite effects across High- and Low-SES students. If rewrites operate as a common, additive style shift, our Assumption \ref{ass:id-style}, this difference-in-differences should be zero.

Formally, for group $G\in\{H,L\}$ define the within-group rewrite contrast
\[
\Delta^{kk'}_G \;:=\; \mathbb{E}\!\left[s_{ik}-s_{ik'}\mid G\right].
\]
Under Assumption \ref{ass:id-style}, rewriting from $k'$ to $k$ changes the score only through the style component, so
\[
\Delta^{kk'}_G
=\mathbb{E}\!\left[\rho_m(R_{ik})-\rho_m(R_{ik'})\mid G\right]
=\lambda_{m,k}-\lambda_{m,k'},
\]
which does not depend on group membership. Consequently, the difference-in-differences contrast
\[
\widehat{\Delta}^{kk'}_H-\widehat{\Delta}^{kk'}_L
\]
should be close to zero for all $(k,k')$.

Figure \ref{fig:diffInDiff} reports these contrasts across all pairs of rewrite levels. The lower triangle (blue) shows the magnitude of the within--High-SES rewrite premium $|\Delta^{kk'}_H|$, which is large: moving from lower to higher SAT targets shifts predicted scores by several tenths to well over a point. The upper triangle (red) shows the corresponding difference-in-differences, $\widehat{\Delta}^{kk'}_H-\widehat{\Delta}^{kk'}_L$. Overall, these red entries cluster tightly around zero across the matrix, with only a modest tendency for comparisons involving the lowest rewrite level ($k=1$) to be slightly larger in magnitude. Moreover, they are an order of magnitude smaller than the underlying rewrite premia in blue. Taken together, the figure provides strong, direct evidence for Assumption \ref{ass:id-style}: rewrites substantially move score levels, but they do so in a nearly parallel way across groups, leaving the High--Low gap essentially unchanged.

%%%%%%%%%%%%%%%%%%%%%%%%%%%%%%%%%
\section{Decomposition Results}
\begin{figure}[!h]
    \centering
    \includegraphics[width=\linewidth]{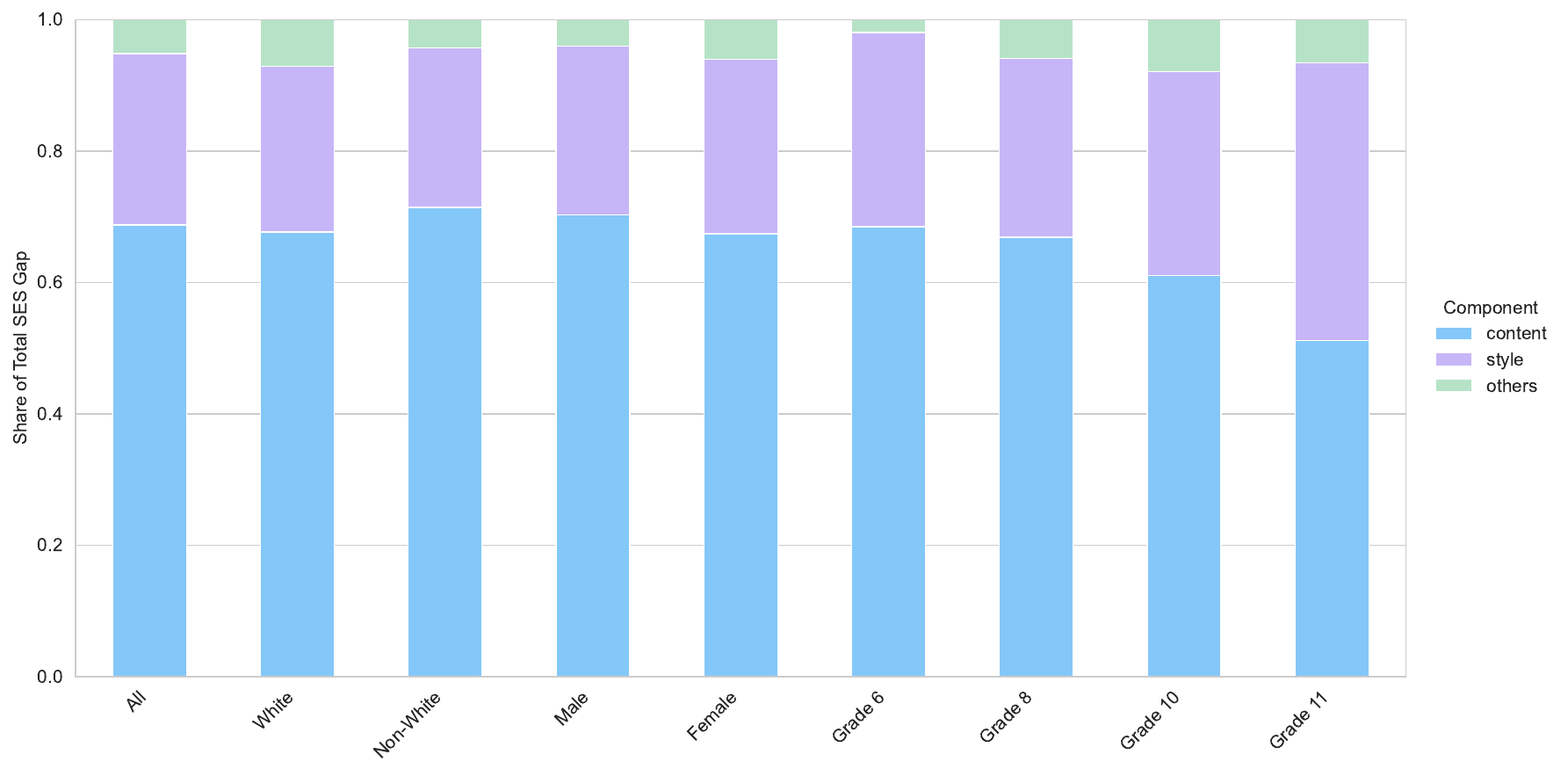}
    \caption{Main Result - Content and Style Decomposition}
    \label{fig:decomMain}
    \begin{minipage}{\linewidth}
    \footnotesize
    \justifying
    \textit{Note:} The figure shows a set of stacked bars that break the overall high–low SES difference in writing scores into three components—content, style, and others—and reports these components as shares of the total SES gap within each subgroup. Each bar corresponds to a subgroup (All; White and Non-White; Male and Female; and grade groups such as Grade 6, 8, 10, and 11), and the y-axis runs from 0 to 1 so that the full height of a bar represents 100\% of that subgroup’s SES gap. Within each bar, the colored segments indicate the fraction attributed to the content component, the fraction attributed to the style component, and the remaining fraction grouped as “others” (the residual portion of the decomposition).  The plot is displaying how the gap is allocated across components rather than the gap’s size in score points. For exact values and standard errors see Table  \ref{tab:decompositionTableMain}, in the Appendix.        
    \end{minipage}

\end{figure}

In this section we discuss our decomposition results using the SAT rewrites. Figure \ref{fig:decomMain} summarizes our central decomposition of the high–low SES writing score gap into three parts: (i) content differences, (ii) style differences, and (iii) the scoring-function differences (“tilt”) component, capturing other unexplained contributors to the gap.\footnote{Full results, with bootstrap standard errors can be found in table \ref{tab:decompositionTableMain} in the Appendix.}. This structure is designed to separate “what is being said” from “how it is said,” and allow us to learn about how each of these components contribute to the realized writing score gap. 

The first takeaway is that most of the score gap is attributed to content, with style playing a sizable but secondary role, and the tilt component typically smaller. In the pooled sample (“All”), the content share accounts for 68.8\% of the total SES gap, the style share accounts for 26.1\% of the total gap, leaving 5.2\% for the scorer function tilt. These results suggests that the SES gap is driven primarily by substantive differences in what students say (i.e. arguments, reasoning, and the relevance and organization of evidence), rather than by surface how they say it. At the same time, the style share is nontrivial: even holding fixed the underlying content benchmark induced by the rewrite panel, differences in writing style, phrasing, grammatical control, and coherence explain an meaningful portion of little more than quarter of the gap. Interpreted through a policy lens, this implies that interventions aimed purely at improving expression (or mechanically standardizing it) could potentially reduce, but not eliminate, the SES gap.

\begin{figure}[!htbp]
  \centering
  \includegraphics[width=\linewidth]{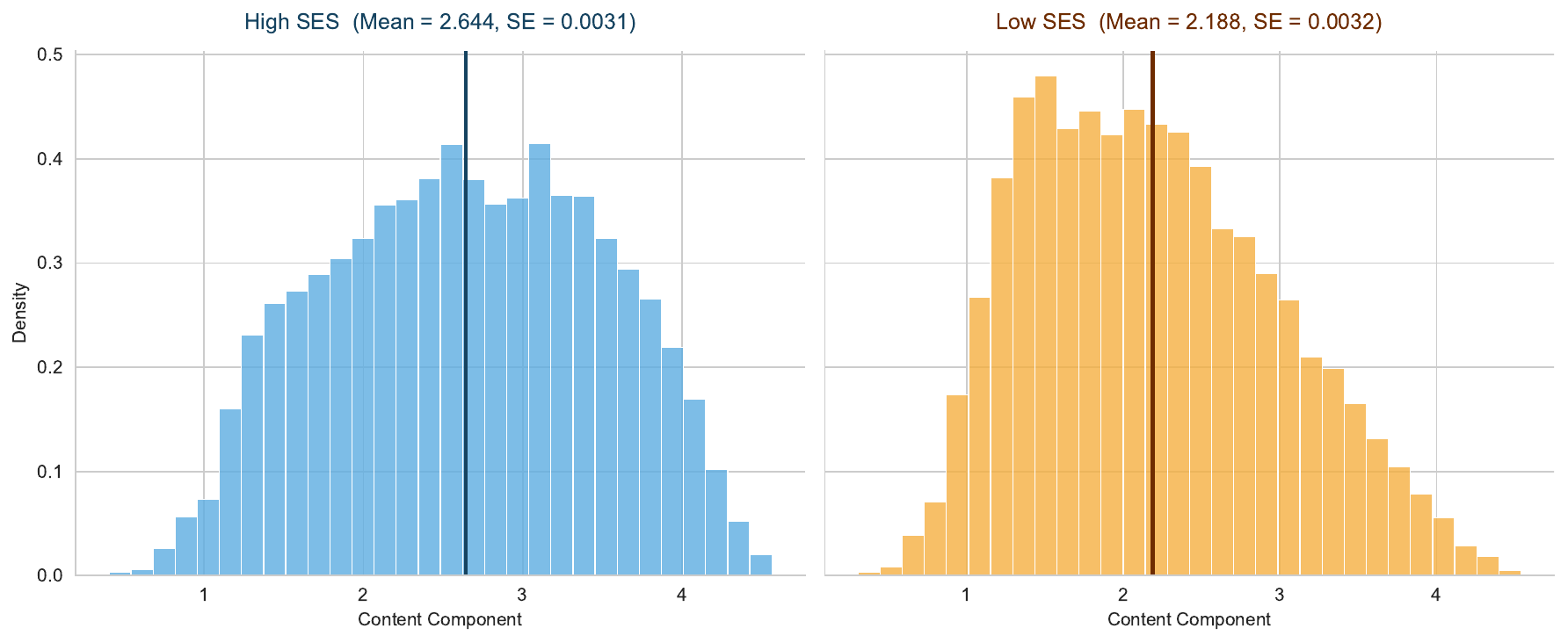}
  \caption*{(a) Content Componenet Distribution by SES}
  \vspace{0.5em}
  \includegraphics[width=\linewidth]{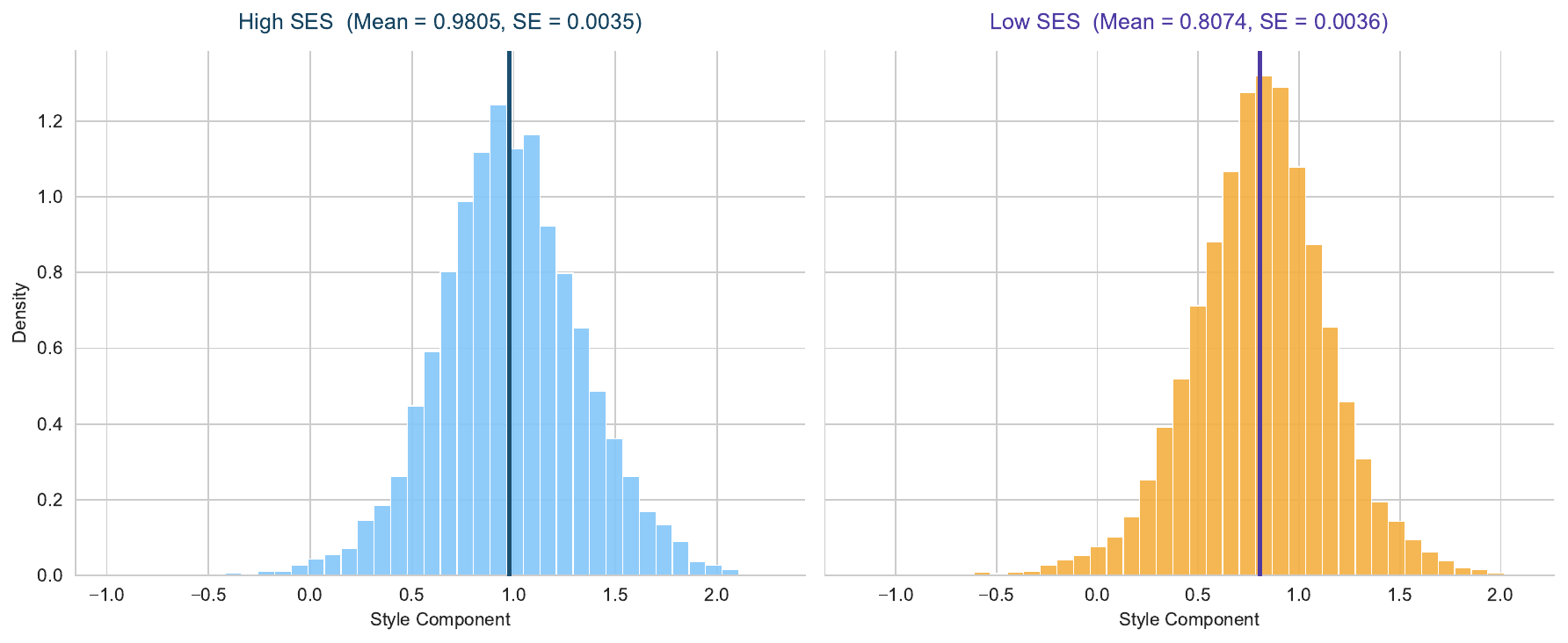}
  \caption*{(b) Style Component distribution by SES }
  \caption{Distribution of "content" and style effects}
  \label{fig:styleAndContentDist}        \begin{minipage}{\linewidth}
    \footnotesize
    \justifying
    \textit{Note:} The figure shows the estimated distributions of the essay-level content component (panel a) and the style component (panel b), plotted separately for high-SES and low-SES students. Each panel reports a histogram for each group, with the vertical line marking the group mean. Because both components are identified only up to an additive constant, the x-axis level is arbitrary up to a common shift; the main objects of interest are the relative shifts and dispersion across SES.
    \end{minipage}
\end{figure}

Figure \ref{fig:styleAndContentDist} displays the estimated distributions of the content and style components, separately for high and low SES students. As both the content component (for each essay) and the style component are identified only up to an additive constant, the x-axis levels in these histograms are arbitrary up to a common shift. We therefore focus on the shapes and relative shifts of the high-vs. low-SES distributions rather than on absolute levels. Panel (a) shows that the content component distributions are clearly separated: the mean content index is 2.644 for high-SES students versus 2.188 for low-SES students. The difference between these two numbers capture the content gap (0.456). The distribution are clearly different, where low SES has high mass of content on lower scores, compare to the high SES stuendes. The support of the content component is wide, and both distributions are realitvly spread out, showing a substantial heterogeneity in content quality within each SES group while maintaining a systematic shift in means. The fact that the variation in the content is large, support the notion that content is the  primary driver of the score gap.  Panel (b) presents the style component distributions, which reveal a smaller but still meaningful separation. High-SES students have a mean style premium of 0.981 relative to the rewrite-induced benchmark, while low-SES students average 0.807 (implying gap of 0.173). Both distributions exhibit tighter dispersion than the content distributions, where deviations from the mean value, are modest in magnitude, with shifts of more than one point being relatively rare. The relatively narrow spread of the style component implies that, the difference in style can have is relatively small, compare to variation in the content.  

\begin{figure}[!h]
    \centering
    \includegraphics[width=\linewidth]{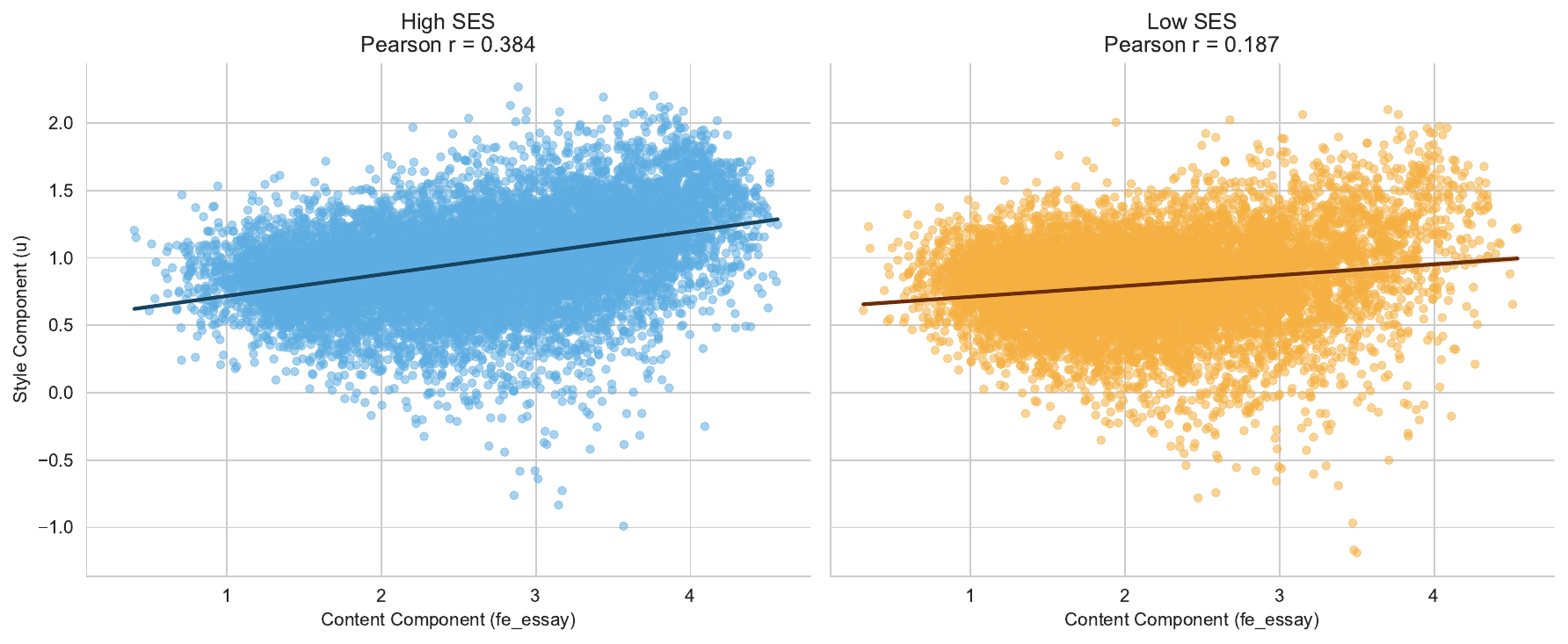}
    \caption{Correlation between Content Component and Style Component}
    \label{fig:styleVsContent}
            \begin{minipage}{\linewidth}
    \footnotesize
    \justifying
    \textit{Note:} The figure plots the relationship between the estimated content component (x-axis; essay fixed effect) and the estimated style component (y-axis; original-style deviation relative to the rewrite benchmark), separately for high-SES (left) and low-SES (right) students. Each point corresponds to an essay, and the fitted line summarizes the within-group linear association.
    \end{minipage}
\end{figure}

Figure \ref{fig:styleVsContent} examines within-student correlation between content and style. Among high-SES students, the Pearson correlation is 0.384, implying that students who score higher on content also tend to receive higher style premia; among low-SES students, the correlation is weaker at 0.187. Overall, the positive correlations indicate that content and style co-move: better ideas tend to be expressed more effectively, making the two dimensions difficult to separate empirically.

The weaker correlation for low-SES students suggests a looser mapping from ideas to expression. One channel is heterogeneity in skill development: some students may construct strong arguments but lack familiarity with academic conventions, while others may master surface mechanics (grammar, templates) without comparable gains in substantive reasoning—either pattern attenuates the content–style link. A second channel is linguistic distance and “code-switching” demands: students whose home language patterns differ from standardized academic English may need to translate and suppress native forms, so sophisticated ideas can coexist with lower measured style, further decoupling the two components.

Next, we turn to how the decomposition compares across subgroups. In Figure \ref{fig:decomMain}, we see how the roles of content, style, and tilt change across gender, race, and grade. The shares are broadly similar by race and gender (the content share ranges from 67\% to 71\%), suggesting that the relative importance of content versus style is not driven by a single demographic subgroup. Where we do see systematic movement is across grade levels: the bars indicate that the content share tends to decline and the style share tends to rise in higher grades (from a content share of 68.5\% in grade 6 to 51.2\% in grade 11 and style component share going from 29.2\% to 42.2\%).  This pattern is consistent with greater stylistic differentiation as students mature: students may increasingly develop distinct voices and specialize in particular forms of writing that are rewarded differently by the scoring rubric; some students may become more adept at adopting the styles most strongly rewarded in school; and, independently, evaluators may place greater weight on stylistic conventions at higher grade levels.

% higher-grade writing increasingly differentiates along conventions of style and polish. At the same time, higher-grade students’ essays may converge in substantive content—e.g. drawing on a more common set of themes, or reaching comparable levels of factual and conceptual coverage, so that differences in what is said shrink even as differences in how it is expressed remain salient. In that environment, a growing share of the observed gap is naturally attributed to style: as content becomes more similar, remaining disparities are increasingly driven by variation in expression, organization, and other rubric-salient conventions.

Finally, Figure \ref{fig:decomPrompts} in the Appendix examines how the decomposition varies across prompts. The prompt-level results suggest that the content–style split is not driven by any single prompt; instead, it fluctuates within a fairly narrow range across tasks. This matters because the raw SES score gap itself is relatively stable across prompts. In most prompts, the gap is primarily explained by content differences, though several prompts exhibit noticeably larger style shares. For example, Distance learning (45\%), Summer projects (41.4\%), Community service (33.2\%), Driverless cars (35\%), and Seeking multiple opinions (31.4\%) all attribute more than 30\% of the gap to style differences.

With the exception of Driverless cars,\footnote{In Driverless cars, students are given an article but are also asked to express their own opinion.} these prompts largely involve “independent” writing: students are asked to produce an essay in response to a question without an accompanying text that anchors what content to include. In such settings, style may play a larger role because reviewers have less of a shared benchmark for what a “complete” answer should contain, making the presentation of ideas more salient.

It's important to emphasize how we read these results. The decomposition is descriptive: it partitions the observed gap under the current joint distribution of essays and scoring practices, and the components naturally map to “levers” (content, expression, scoring rules), but causal interpretations require additional assumptions, as discussed in remark \ref{remark:causal}.  With that caveat, the pattern in Figure \ref{fig:decomMain} points to a clear conclusion: SES disparities in writing scores appear to be driven primarily by differences in substantive content, with a meaningful additional contribution from style, and a smaller role for systematic differences in scoring rules. 

\subsection{Robustness}
% In this section, we assess the robustness of our results. Figure \ref{fig:robustness} compares our main SAT decomposition with four alternative specifications. First, we restrict the sample to essays with SAT scores of 2–5 and consider only rewrites whose scores differ from the original by at most one point. For example, an essay initially scored 3 has three admissible rewrites: SAT-2, SAT-3, and SAT-4. If upward and downward rewrites are symmetric, then $\sum_k \lambda_k \approx 0$, which allows us to difference out any systematic rewrite effects, as discussed in Section \ref{sec:identification}. In this subsample, the total gap is smaller, 0.54 points. The relative importance of content and style is largely unchanged compare to our baseline SAT results, at 69.3\% and 24.6\%, respectively.

\begin{figure}[!h]
    \centering
    \includegraphics[width=\linewidth]{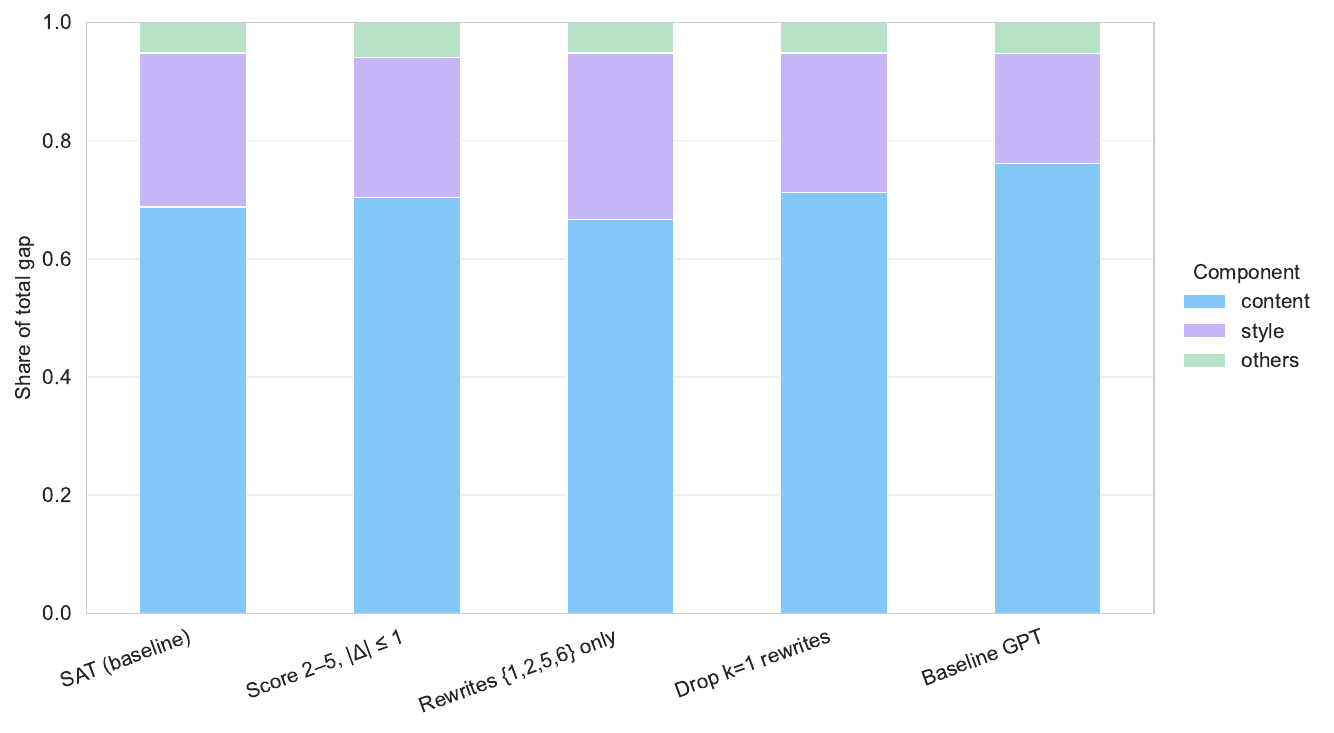}
    \caption{Robustness Analysis}
    \label{fig:robustness}
        \begin{minipage}{\linewidth}
    \footnotesize
    \justifying
    \textit{Note:} The figure presents a stacked bar chart that re-computes the paper’s score-gap decomposition under the baseline SAT-rewrite design and four alternative (“robustness”) implementations. The x-axis lists the five specifications (SAT baseline; restricting to SAT scores 2–5 and using only rewrites that move at most one SAT point; using only rewrite pairs {1,6} and {2,5}; dropping SAT-1 rewrites; and replacing targeted SAT rewrites with a “standard GPT” rewrite), and the y-axis is the share of the total score gap. Each bar is partitioned into three color-coded segments labeled content, style, and others, where the height of each segment indicates that component’s fraction of the total gap in that specification. For exact values and bootstrap standard errors, seee Table \ref{tab:robustness} in the Appendix. 
    \end{minipage}
\end{figure}

In this section, we assess the robustness of our results. Figure \ref{fig:robustness} compares our main SAT decomposition with four alternative specifications. First, we restrict attention to essays whose original human SAT score lies in $\{2,3,4,5\}$. For each such essay with original score $s_i$, we retain only rewrite versions with scores $k \in \{s_i-1, s_i, s_i+1\}$ (e.g., $s_i=3$ implies keeping SAT-2, SAT-3, and SAT-4). We then re-estimate the same fixed-effects regression and SAT decomposition as in Section \ref{sec:identification} on this restricted rewrite panel. If upward and downward rewrites are symmetric, then $\sum_k \lambda_k \approx 0$, allowing us to net out potential rewrite effects. In this subsample, the total gap is smaller, 0.54 points. The relative importance of content and style is largely unchanged compare to our baseline SAT results, at 69.3\% and 24.6\%, respectively.

Next, we repeat the decomposition while restricting attention to rewrites involving SAT scores 1 and 6, and 2 and 5. Appendix Figure \ref{fig:rewritePremious} shows that the average difference between rewrites and baseline essays for scores 1 vs.\ 6 and 2 vs.\ 5 is roughly symmetric, again implying that the rewrite effects net out \(\sum_k \lambda_k \approx 0\). This restriction has little effect on the decomposition: content and style account for 66.5\% and 28.1\%, respectively.

In Figure \ref{fig:diffInDiff} (Section \ref{sec:rewrites}), we saw in the difference-in-differences design that SAT-1 rewrites may be affected by content, as they exhibit slight difference in the effects on  high- and low-SES essays. In the “Drop SAT 1” specification, removing the SAT-1 rewrites has little impact on the decomposition: the content and style shares are 71\% and 23\%, respectively.

Finally, the last bar considers a decomposition that replaces our targeted SAT rewrites with a “standard” GPT rewrite, as described in section \ref{sec:data}, where we simply ask the model to rewrite the essay without specifying a target style level. In this case, the content share is slightly higher (76\%), while the style share is lower (18.5\%). In Appendix \ref{app:baselineDecomp}, we expand our discussion on this alternative rewrite.

Taken together, Figure \ref{fig:robustness} shows that the qualitative conclusion is stable: across specifications, content remains the dominant contributor to the SES gap, with a meaningful but secondary role for style.

\section{Conclusions}

The ability to communicate - ideas, intentions, and concepts - shapes life chances. In schools, workplaces, and professional settings, people are judged not only on what they know or their ideas, but on how well they can express it in the forms institutions recognize and reward. This paper examines a central question in the study of inequality and evaluation: to what extent do observed disparities reflect differences in underlying ideas and reasoning versus differences in individuals’ ability to articulate those ideas in forms institutions reward? In many settings—education, labor markets, and professional evaluation—success depends not only on what people \emph{want} but also on how effectively they express it. When articulation is unevenly distributed, differences in expressive skill may amplify or distort differences in substantive ability. Distinguishing content from style is therefore essential for understanding how inequality arises, and for identifying where policy interventions can be most effective or warranted.

We find that the writing-score gap between high- and low-SES students is driven primarily by differences in content, which account for roughly 69\% of the gap. Differences in style explain about 26\%, with the remaining share attributable to differences in scoring functions. This decomposition is stable across demographic subgroups and most prompts.These results underscore that the ability to develop and communicate ideas clearly is both consequential and unevenly distributed, and that systematic differences in this skill can generate large disparities across many domains. From a policy perspective, the findings suggest that interventions focused solely on surface-level mechanics or standardized expression—whether pedagogical or technological—may substantially narrow, but not eliminate, SES disparities in writing assessment. The dominant role of content points to deeper inequalities in the development of argumentation, reasoning, and organization. At the same time, the sizable contribution of style indicates that academic conventions remain consequential and are themselves shaped by unequal access and exposure.

Beyond writing assessment, the paper contributes a general methodological framework for separating tightly confounded components in observational data. Content and style naturally co-move: stronger ideas are often expressed more effectively, and expressive choices influence how substance is perceived. Traditional approaches—statistical controls or human coding—struggle with this entanglement, relying on strong assumptions, limited feature sets, or subjective judgments.

We show how large language models can serve as research instruments to overcome this limitation. By generating multiple stylistic rewrites of each essay that preserve underlying arguments and evidence, we construct a "generated panel" that induces within-essay variation in presentation while holding content fixed. Exposing each essay to a common distribution of styles allows us to separately identify how scorers respond to substance versus form under mild structural assumptions.

This approach has broad applicability. Many social science questions require disentangling idea quality from presentation—whether in job market papers, persuasive communication, or other creative and professional outputs. LLM-generated variation offers a scalable alternative to costly experiments or restrictive modeling choices.

In sum, SES disparities in writing scores reflect primarily differences in argumentative substance rather than stylistic polish, though both matter. More broadly, the paper illustrates how LLMs can be deployed not just as end-user tools but as instruments for scientific measurement, enabling identification strategies that were previously infeasible.

\bibliographystyle{apalike} 
\bibliography{sample}      

\begin{thebibliography}{}

\bibitem[Arnold et~al., 2022]{arnold2022measuring}
Arnold, D., Dobbie, W., and Hull, P. (2022).
\newblock Measuring racial discrimination in bail decisions.
\newblock {\em American Economic Review}, 112(9):2992--3038.

\bibitem[Blinder, 1973]{Blinder1973}
Blinder, A.~S. (1973).
\newblock Wage discrimination: Reduced form and structural estimates.
\newblock {\em Journal of Human Resources}, 8(4):436--455.

\bibitem[Bohren et~al., 2022]{bohren2022systemic}
Bohren, J.~A., Hull, P., and Imas, A. (2022).
\newblock Systemic discrimination: Theory and measurement.
\newblock Technical Report w29820, National Bureau of Economic Research.

\bibitem[Clark et~al., 2019]{clark-etal-2019-bert}
Clark, K., Khandelwal, U., Levy, O., and Manning, C.~D. (2019).
\newblock What does {BERT} look at? an analysis of {BERT}{'}s attention.
\newblock In {\em Proceedings of the 2019 ACL Workshop BlackboxNLP: Analyzing and Interpreting Neural Networks for NLP}, pages 276--286, Florence, Italy. Association for Computational Linguistics.

\bibitem[Crossley et~al., 2019]{crossley2019taaco2}
Crossley, S.~A., Kyle, K., and Dascalu, M. (2019).
\newblock The tool for the automatic analysis of cohesion 2.0: Integrating semantic similarity and text overlap.
\newblock {\em Behavior Research Methods}, 51(1):14--27.

\bibitem[Crossley et~al., 2016]{crossley2016taaco}
Crossley, S.~A., Kyle, K., and McNamara, D.~S. (2016).
\newblock The tool for the automatic analysis of text cohesion (taaco): Automatic assessment of local, global, and text cohesion.
\newblock {\em Behavior Research Methods}, 48(4):1227--1237.

\bibitem[Crossley et~al., 2024]{crossley2024large}
Crossley, S.~A., Tian, Y., Baffour, P., Franklin, A., Benner, M., and Boser, U. (2024).
\newblock A large-scale corpus for assessing written argumentation: Persuade 2.0.
\newblock {\em Assessing Writing}, 61:100865.

\bibitem[Devlin et~al., 2019a]{devlin2019bert}
Devlin, J., Chang, M.-W., Lee, K., and Toutanova, K. (2019a).
\newblock Bert: Pre-training of deep bidirectional transformers for language understanding.
\newblock In {\em Proceedings of the 2019 conference of the North American chapter of the association for computational linguistics: human language technologies, volume 1 (long and short papers)}, pages 4171--4186.

\bibitem[Devlin et~al., 2019b]{devlin-etal-2019-bert}
Devlin, J., Chang, M.-W., Lee, K., and Toutanova, K. (2019b).
\newblock {BERT}: Pre-training of deep bidirectional transformers for language understanding.
\newblock In {\em Proceedings of the 2019 Conference of the North {A}merican Chapter of the Association for Computational Linguistics: Human Language Technologies, Volume 1 (Long and Short Papers)}, pages 4171--4186, Minneapolis, Minnesota. Association for Computational Linguistics.

\bibitem[Dwork et~al., 2012]{dwork2012fairness}
Dwork, C., Hardt, M., Pitassi, T., Reingold, O., and Zemel, R. (2012).
\newblock Fairness through awareness.
\newblock In {\em Proceedings of the 3rd Innovations in Theoretical Computer Science Conference}, pages 214--226. ACM.

\bibitem[Ethayarajh, 2019]{ethayarajh-2019-contextual}
Ethayarajh, K. (2019).
\newblock How contextual are contextualized word representations? {C}omparing the geometry of {BERT}, {ELM}o, and {GPT}-2 embeddings.
\newblock In {\em Proceedings of the 2019 Conference on Empirical Methods in Natural Language Processing and the 9th International Joint Conference on Natural Language Processing (EMNLP-IJCNLP)}, pages 55--65, Hong Kong, China. Association for Computational Linguistics.

\bibitem[Feder et~al., 2021]{feder2021causalm}
Feder, A., Oved, N., Shalit, U., and Reichart, R. (2021).
\newblock Causalm: Causal model explanation through counterfactual language models.
\newblock {\em Computational Linguistics}, 47(2):333--386.

\bibitem[Fortin et~al., 2011]{fortin2011decomposition}
Fortin, N., Lemieux, T., and Firpo, S. (2011).
\newblock Decomposition methods in economics.
\newblock In {\em Handbook of Labor Economics}, volume~4, pages 1--102. Elsevier.

\bibitem[Hanushek et~al., 2019]{hanushek2019achievement}
Hanushek, E.~A., Peterson, P.~E., Talpey, L.~M., and Woessmann, L. (2019).
\newblock The achievement gap fails to close.
\newblock {\em Education Next}, 19(3):8--17.

\bibitem[Jawahar et~al., 2019a]{jawahar-etal-2019-bert}
Jawahar, G., Sagot, B., and Seddah, D. (2019a).
\newblock What does {BERT} learn about the structure of language?
\newblock In {\em Proceedings of the 57th Annual Meeting of the Association for Computational Linguistics}, pages 3651--3657, Florence, Italy. Association for Computational Linguistics.

\bibitem[Jawahar et~al., 2019b]{jawahar2019does}
Jawahar, G., Sagot, B., and Seddah, D. (2019b).
\newblock What does bert learn about the structure of language?
\newblock In {\em ACL 2019-57th Annual Meeting of the Association for Computational Linguistics}.

\bibitem[Karpukhin et~al., 2020]{karpukhin-etal-2020-dense}
Karpukhin, V., Oguz, B., Min, S., Lewis, P., Wu, L., Edunov, S., Chen, D., and Yih, W.-t. (2020).
\newblock Dense passage retrieval for open-domain question answering.
\newblock In {\em Proceedings of the 2020 Conference on Empirical Methods in Natural Language Processing (EMNLP)}, pages 6769--6781, Online. Association for Computational Linguistics.

\bibitem[Kitagawa, 1955]{Kitagawa1955}
Kitagawa, E.~M. (1955).
\newblock Components of a difference between two rates.
\newblock {\em Journal of the American Statistical Association}, 50(272):1168--1194.

\bibitem[Kusner et~al., 2017]{kusner2017counterfactual}
Kusner, M.~J., Loftus, J., Russell, C., and Silva, R. (2017).
\newblock Counterfactual fairness.
\newblock In {\em Advances in neural information processing systems}, volume~30.

\bibitem[Kyle, 2016]{kyle2016measuring}
Kyle, K. (2016).
\newblock {\em Measuring syntactic development in L2 writing: Fine-grained indices of syntactic complexity and usage-based indices of syntactic sophistication}.
\newblock Doctoral dissertation, Georgia State University.
\newblock Retrieved from ScholarWorks at Georgia State University.

\bibitem[Kyle et~al., 2021]{kyle2021assessing}
Kyle, K., Crossley, S.~A., and Jarvis, S. (2021).
\newblock Assessing the validity of lexical diversity using direct judgements.
\newblock {\em Language Assessment Quarterly}, 18(2):154--170.

\bibitem[Litman et~al., 2021]{litman2021fairness}
Litman, D., Zhang, H., Correnti, R., Matsumura, L.~C., and Wang, E. (2021).
\newblock A fairness evaluation of automated methods for scoring text evidence usage in writing.
\newblock In {\em International Conference on Artificial Intelligence in Education}, pages 255--267, Cham. Springer International Publishing.

\bibitem[Ludwig and Mullainathan, 2023]{ludwig2023machine}
Ludwig, J. and Mullainathan, S. (2023).
\newblock Machine learning as a tool for hypothesis generation.
\newblock Technical Report w31017, National Bureau of Economic Research.

\bibitem[Morgan and Hu, 2024]{morgan2024explaining}
Morgan, P.~L. and Hu, E.~H. (2024).
\newblock Explaining achievement gaps: The role of socioeconomic factors.
\newblock Research report, Thomas B. Fordham Institute.

\bibitem[{National Center for Education Statistics}, 2012]{nationsreport2012}
{National Center for Education Statistics} (2012).
\newblock The nation’s report card: Writing 2011.
\newblock Technical Report NCES 2012-470, Institute of Education Sciences, U.S. Department of Education.

\bibitem[Oaxaca, 1973]{Oaxaca1973}
Oaxaca, R. (1973).
\newblock Male--female wage differentials in urban labor markets.
\newblock {\em International Economic Review}, 14(3):693--709.

\bibitem[Potter et~al., 2025]{potter2025assessing}
Potter, A., Shortt, M., Goldshtein, M., and Roscoe, R.~D. (2025).
\newblock Assessing academic language in tenth grade essays using natural language processing.
\newblock {\em Assessing Writing}, 64:100921.

\bibitem[Reardon, 2011]{reardon2011widening}
Reardon, S.~F. (2011).
\newblock The widening academic achievement gap between the rich and the poor: New evidence and possible explanations.
\newblock In Duncan, G.~J. and Murnane, R.~J., editors, {\em Whither Opportunity? Rising Inequality, Schools, and Children's Life Chances}, chapter~5, pages 91--116. Russell Sage Foundation, New York.

\bibitem[Reimers and Gurevych, 2019]{reimers-gurevych-2019-sentence}
Reimers, N. and Gurevych, I. (2019).
\newblock Sentence-{BERT}: Sentence embeddings using {S}iamese {BERT}-networks.
\newblock In {\em Proceedings of the 2019 Conference on Empirical Methods in Natural Language Processing and the 9th International Joint Conference on Natural Language Processing (EMNLP-IJCNLP)}, pages 3982--3992, Hong Kong, China. Association for Computational Linguistics.

\bibitem[Rogers et~al., 2020]{rogers-etal-2020-primer}
Rogers, A., Kovaleva, O., and Rumshisky, A. (2020).
\newblock A primer in {BERT}ology: What we know about how {BERT} works.
\newblock {\em Transactions of the Association for Computational Linguistics}, 8:842--866.

\bibitem[Tenney et~al., 2019]{tenney-etal-2019-bert}
Tenney, I., Das, D., and Pavlick, E. (2019).
\newblock {BERT} rediscovers the classical {NLP} pipeline.
\newblock In {\em Proceedings of the 57th Annual Meeting of the Association for Computational Linguistics}, pages 4593--4601, Florence, Italy. Association for Computational Linguistics.

\bibitem[van~der Maaten and Hinton, 2008]{vanDerMaaten2008tsne}
van~der Maaten, L. and Hinton, G. (2008).
\newblock Visualizing data using t-sne.
\newblock In {\em Advances in Neural Information Processing Systems}, volume~21.

\bibitem[Vig et~al., 2020]{vig2020investigating}
Vig, J., Gehrmann, S., Belinkov, Y., Qian, S., Nevo, D., Singer, Y., and Shieber, S. (2020).
\newblock Investigating gender bias in language models using causal mediation analysis.
\newblock In {\em Advances in Neural Information Processing Systems 33 (NeurIPS 2020)}.

\bibitem[Wang and Chiu, 2023]{wang2023multi}
Wang, Z. and Chiu, M.~M. (2023).
\newblock Multi-discourse modes in student writing: Effects of combining narrative and argument discourse modes on argumentative essay scores.
\newblock {\em Applied Linguistics}, page amac073.

\bibitem[Zenker and Kyle, 2021]{zenker2021investigating}
Zenker, F. and Kyle, K. (2021).
\newblock Investigating minimum text lengths for lexical diversity indices.
\newblock {\em Assessing Writing}, 47:100505.

\end{thebibliography}
\newpage
\appendix
\section{Additional Figures}
\begin{figure}[H]
    \centering
    \includegraphics[width=\linewidth]{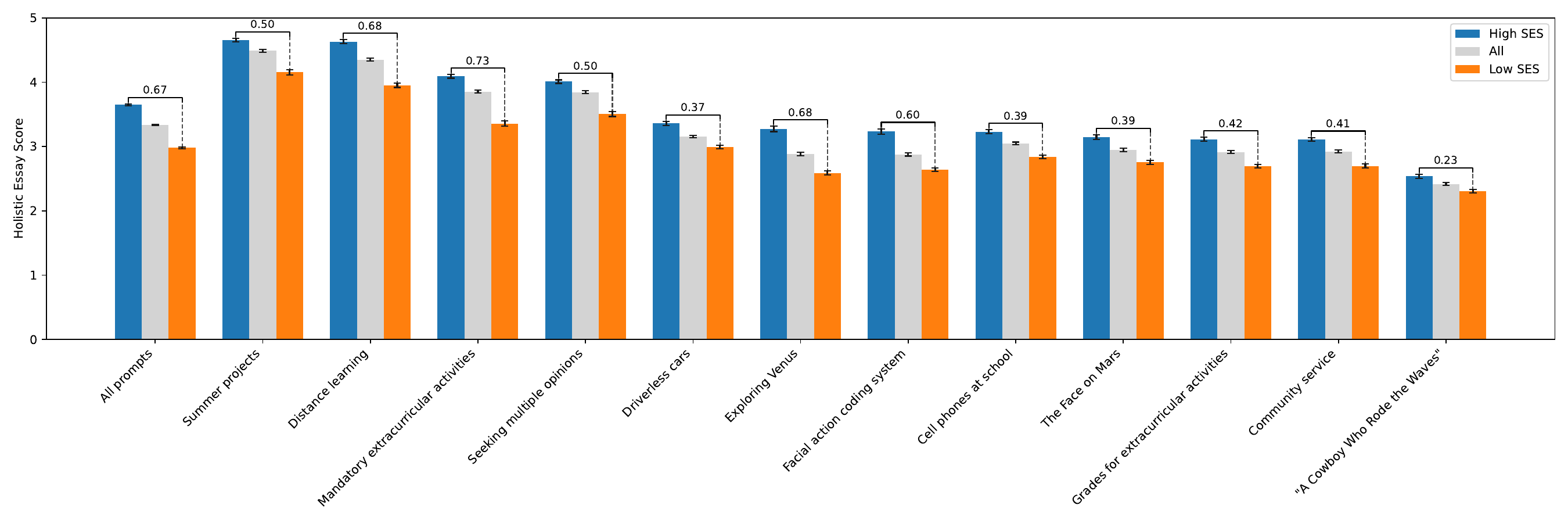}
    \caption{Score Gap by Prompt }
    \label{fig:scoreGap_by_prompt}
    \begin{minipage}{\linewidth}
    \footnotesize
    \justifying
    \textit{Note:}
    The figure shows mean holistic essay scores (y-axis) by writing prompt (x-axis), with grouped bars for high-SES, low-SES, and the full sample (“All”), including error bars. Numeric annotations above prompt groups indicate the high–low difference displayed for each prompt.
    \end{minipage}
\end{figure}

\begin{figure}[H]
    \centering
    \includegraphics[width=\linewidth]{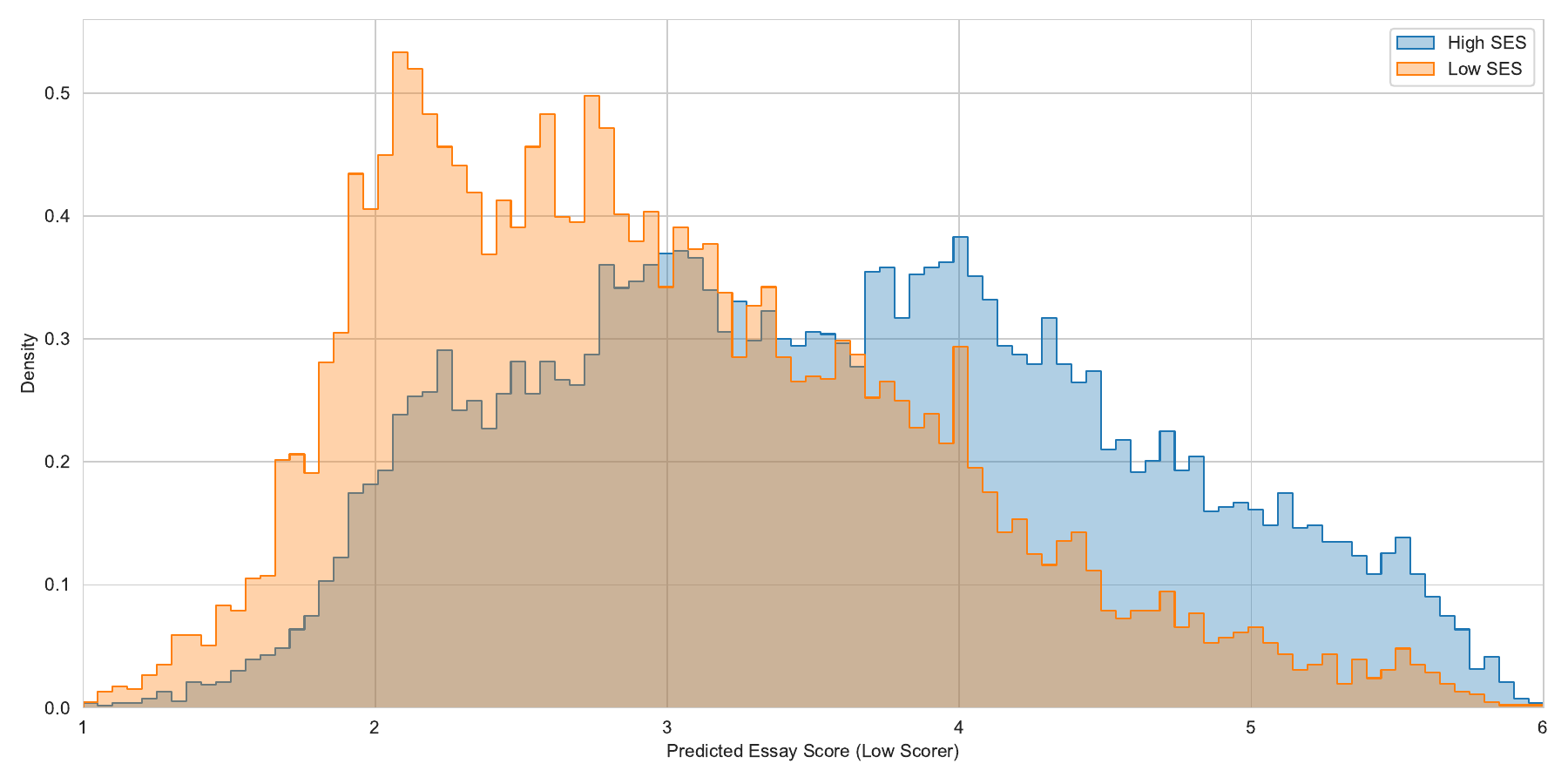}
    \caption{Predicted Score for Students Essays with the low Scorer}
    \label{fig:predictedLowScorerDist}
    \begin{minipage}{\linewidth}
    \footnotesize
    \justifying
    \textit{Note:} The figure shows overlaid density/histogram distributions of predicted essay scores from the “Low Scorer,” separately for high-SES and low-SES students (x-axis: predicted essay score; y-axis: density).
\end{minipage}
\end{figure}

\newpage
\begin{landscape}
\begin{figure}[H]
    \centering
    \includegraphics[width=\linewidth]{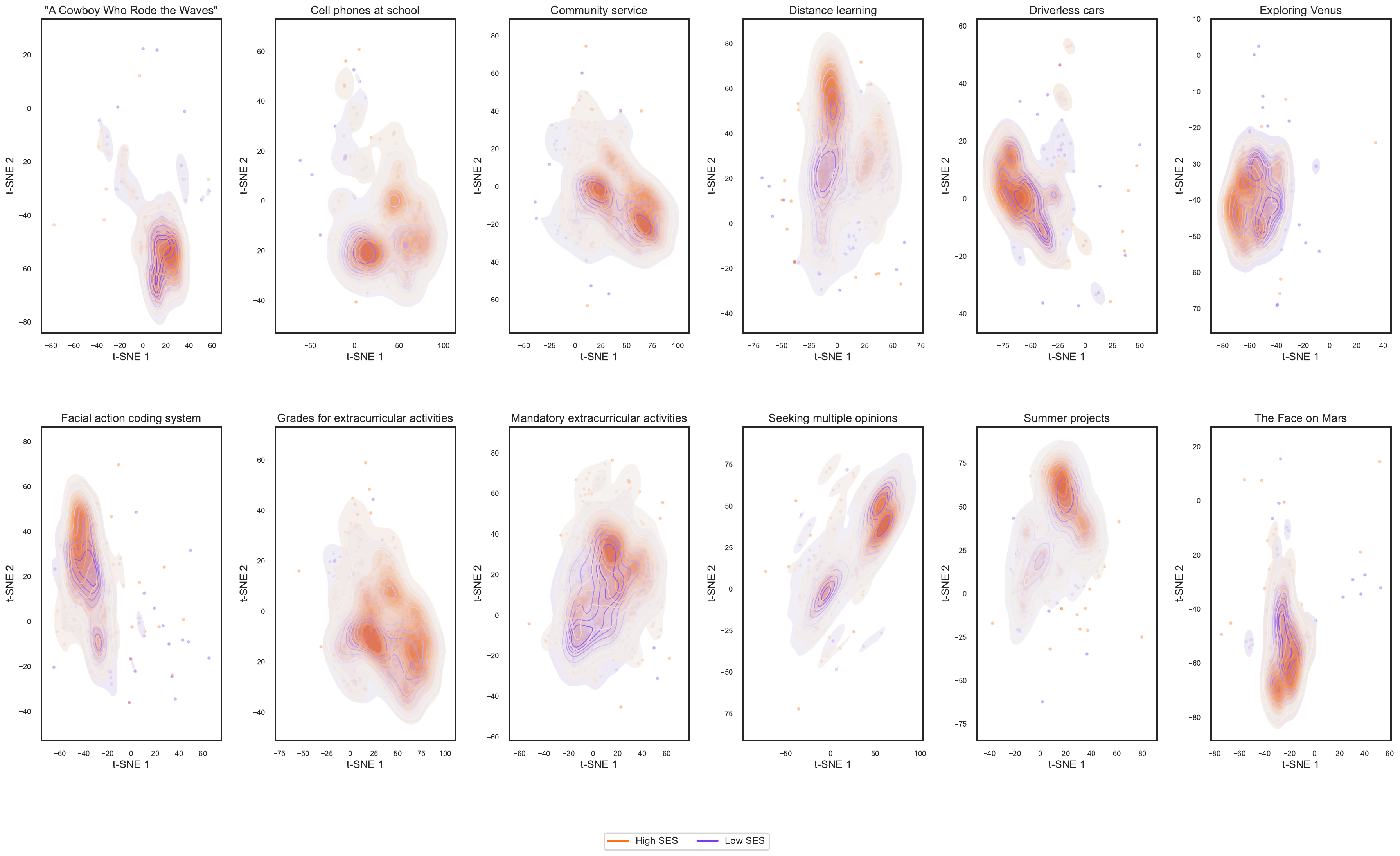}
    \caption{Distribution of Two-Dimensional Textual Embeddings for High- and Low-SES Students Across Writing Prompts}
    \label{fig:tsneGridPrompt}
    \begin{minipage}{\linewidth}
    \footnotesize
    \justifying
    \textit{Note:} The figure shows a grid of t-SNE density plots, one panel per writing prompt, with axes t-SNE 1 and t-SNE 2. Each panel overlays high-SES and low-SES density shading/contours to visualize where essays from each group fall in the embedded space for that specific prompt.        
    \end{minipage}

\end{figure}

\end{landscape}
\newpage
\begin{figure}[H]
    \centering
    \includegraphics[width=\textwidth]{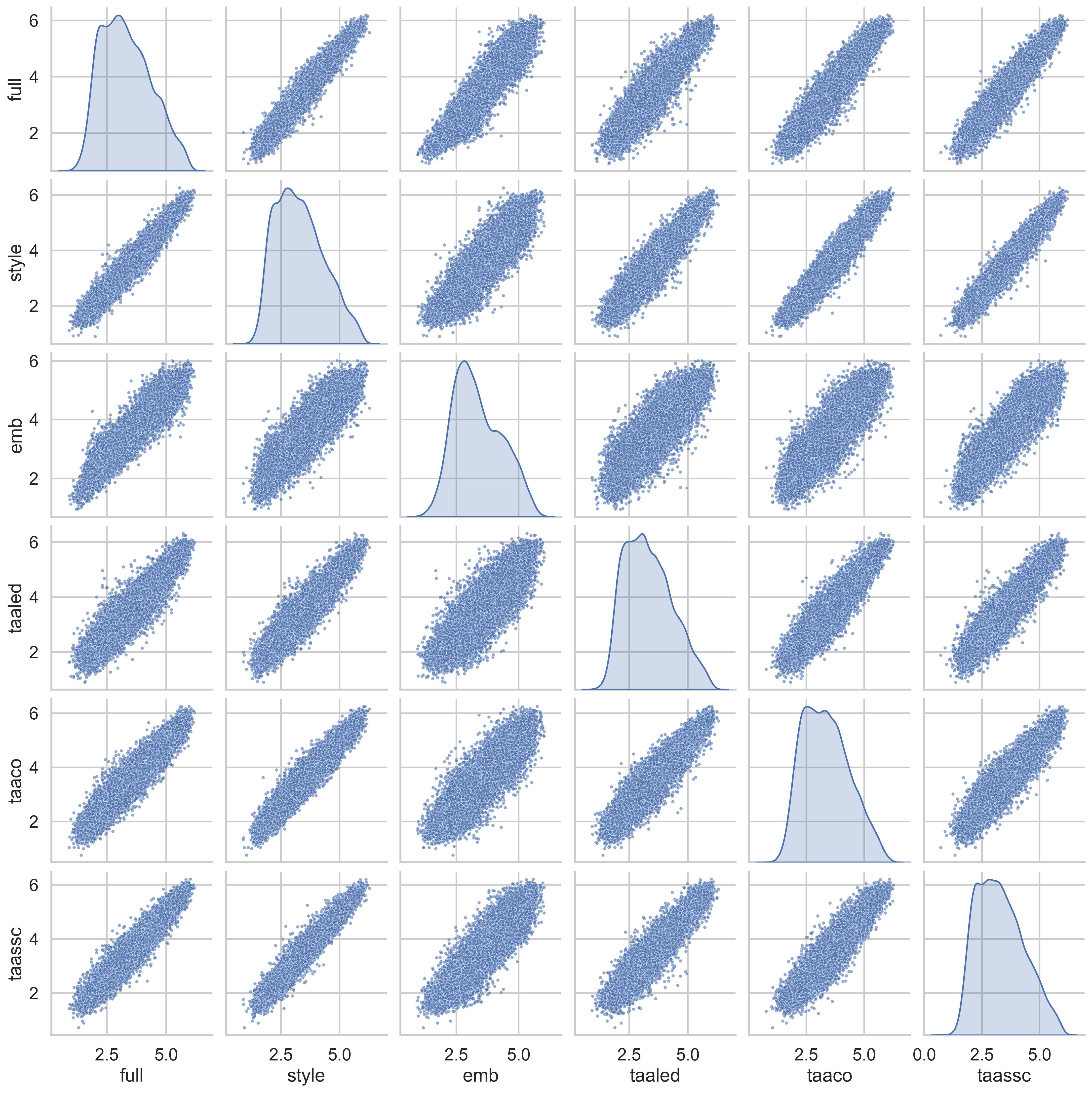}
    \caption{Correlation between predicted values, given a subset of the explaining variables - High Scorer }
    \label{fig:highScorer_PredictionCorrelation}
\begin{minipage}{\linewidth}
\footnotesize
\justifying
\textit{Note:} The figure shows a scatterplot matrix (pair plot) comparing predicted \emph{High Scorer} scores across models/specifications. Each off-diagonal panel plots the predictions from two specifications against each other (one point per essay), while the diagonal panels show the marginal distribution of predictions for each specification. \texttt{Full} uses all explanatory variables; \texttt{Style} uses only our style variables; and \texttt{emb} uses only embedding-based features. The specifications \texttt{taaled}, \texttt{taaco}, and \texttt{taassc} use subsets of the explanatory variables, as described in Section~\ref{sec:data}.
\end{minipage}

\end{figure}

\begin{figure}[H]
    \centering
    \includegraphics[width=\textwidth]{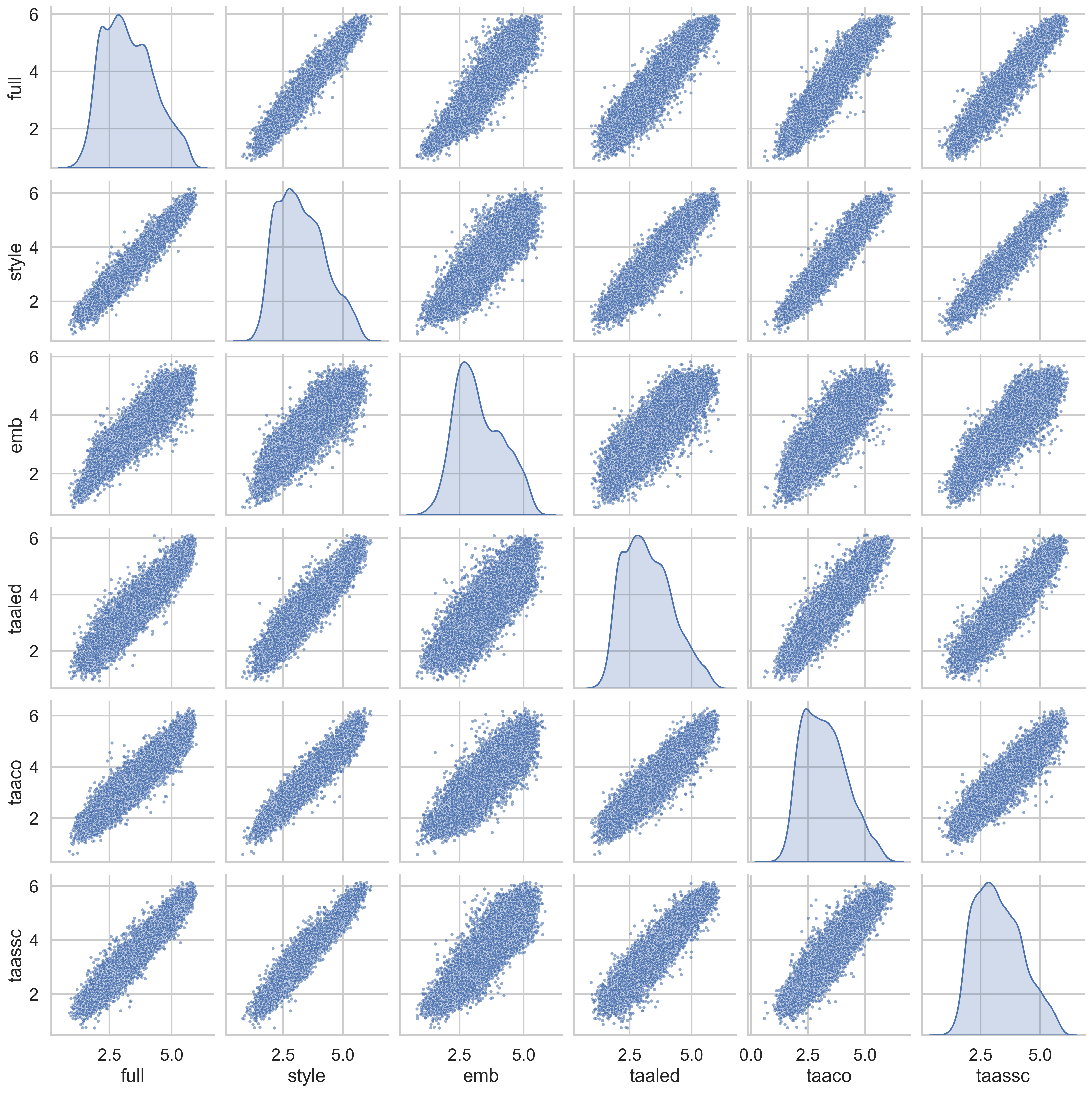}
    \caption{Correlation between predicted values, given a subset of the explaining variables - Low Scorer}
    \label{fig:lowScorer_PredictionCorrelation}
    \begin{minipage}{\linewidth}
\footnotesize
\justifying
\textit{Note:} The figure shows a scatterplot matrix (pair plot) comparing predicted \emph{Low Scorer} scores across models/specifications. Each off-diagonal panel plots the predictions from two specifications against each other (one point per essay), while the diagonal panels show the marginal distribution of predictions for each specification. \texttt{Full} uses all explanatory variables; \texttt{Style} uses only our style variables; and \texttt{emb} uses only embedding-based features. The specifications \texttt{taaled}, \texttt{taaco}, and \texttt{taassc} use subsets of the explanatory variables, as described in Section~\ref{sec:data}.
\end{minipage}

\end{figure}

\begin{figure}[H]
    \centering
    \begin{subfigure}[t]{0.48\textwidth}
        \centering
        \includegraphics[width=\linewidth]{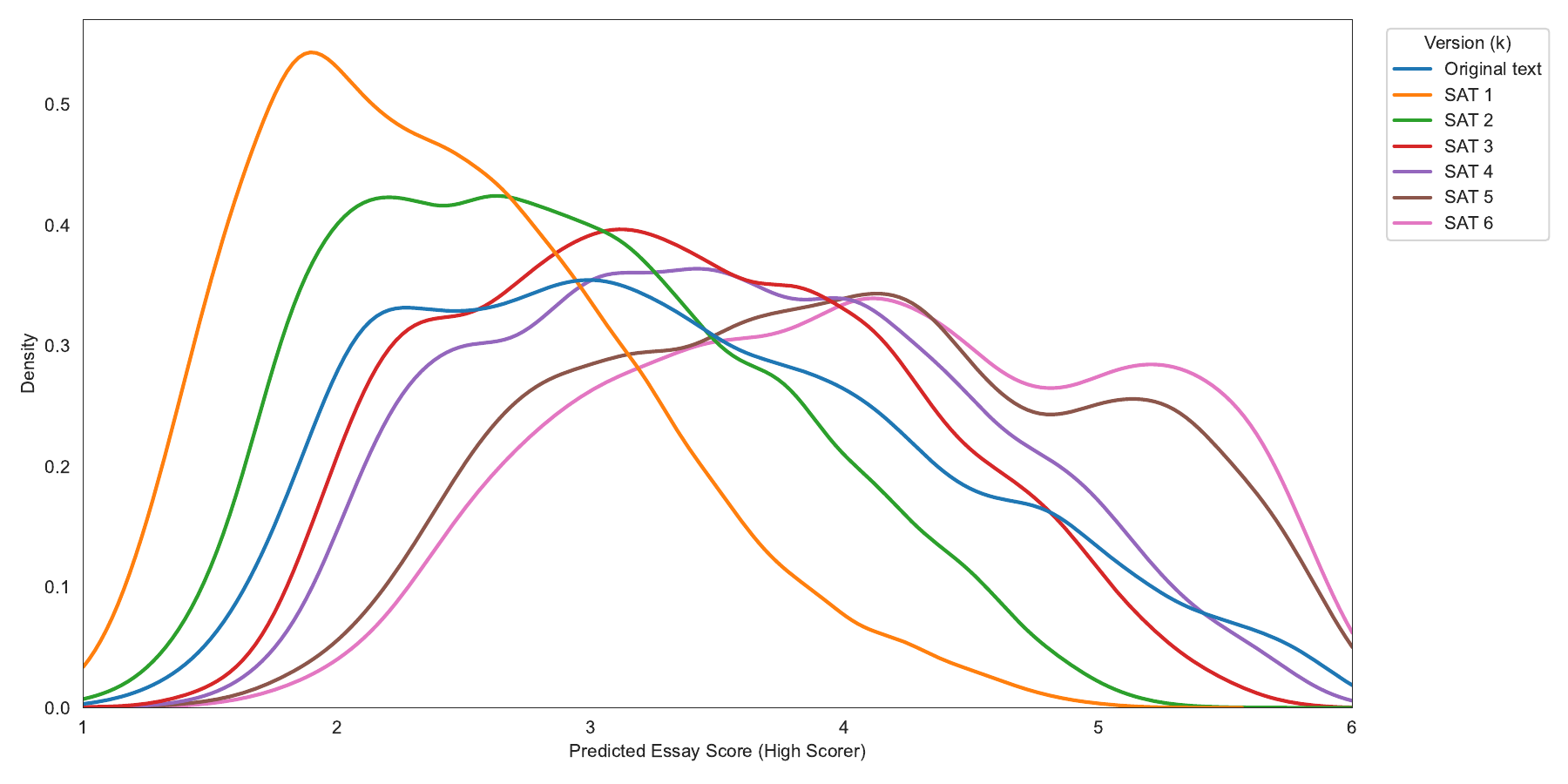}
        \caption{Predicted Scores By Rewrite for Using the High Scorer}
        \label{fig:kdensity_by_k_high}
    \end{subfigure}
    \hfill
    \begin{subfigure}[t]{0.48\textwidth}
        \centering
        \includegraphics[width=\linewidth]{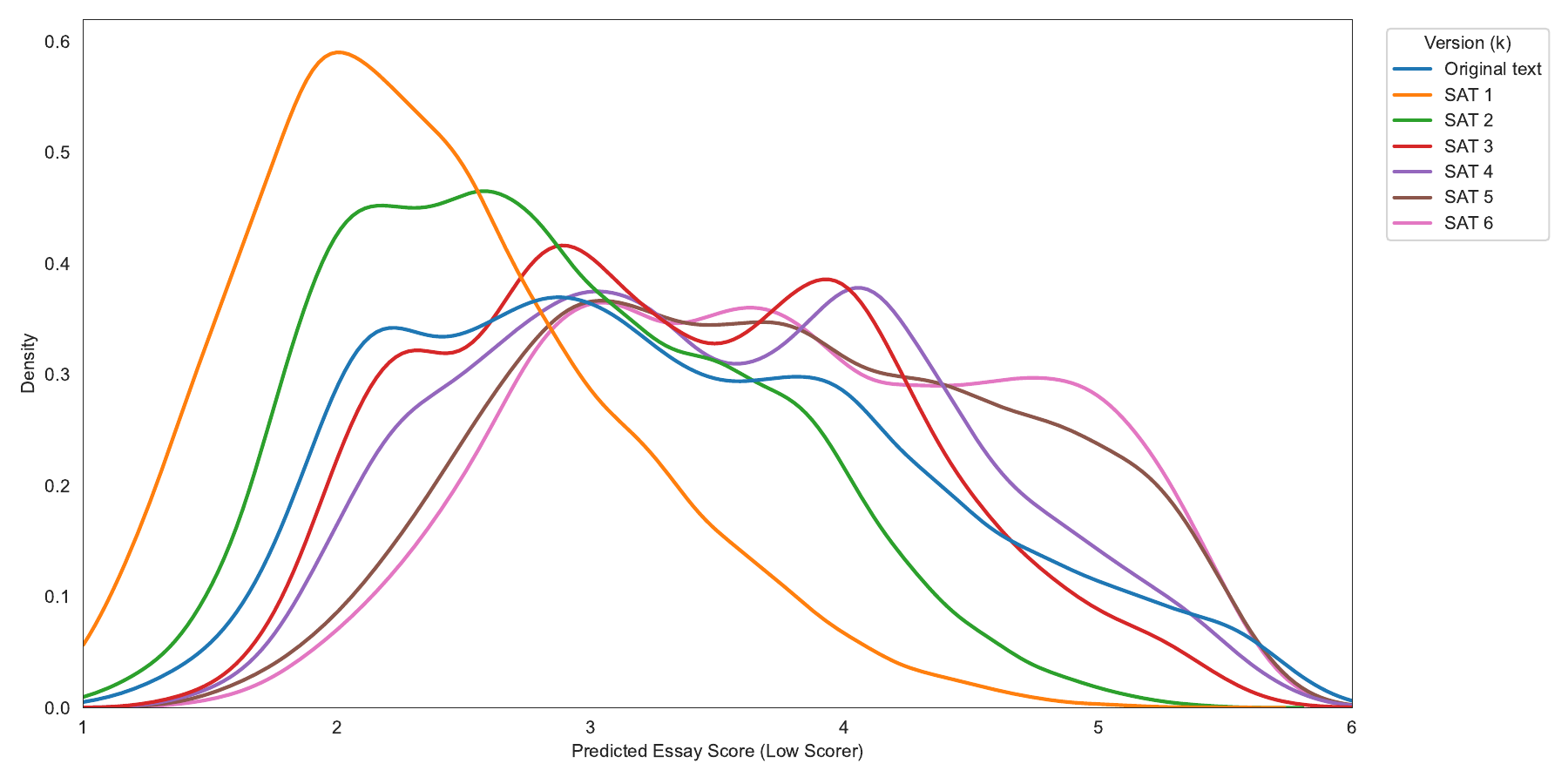}
        \caption{Predicted Scores By Rewrite for Using the Low Scorer }
        \label{fig:kdensity_by_k_low}
    \end{subfigure}
    \caption{Predicted Scores for Essay Rewrites by the Two Scorers}
    \label{fig:kdensity_by_k}
    \begin{minipage}{\linewidth}
\footnotesize
\justifying
\textit{Note:}The figure shows kernel density curves of predicted essay scores for essay rewrites, separately for two scorers. Panel (a) plots densities over predicted scores for the High Scorer; panel (b) plots the analogous densities for the Low Scorer. Each panel overlays curves for the original text and rewrite versions (SAT 1 through SAT 6), with a legend mapping colors to versions.
\end{minipage}
\end{figure}

\begin{figure}[H]
    \centering
    \includegraphics[width=\linewidth]{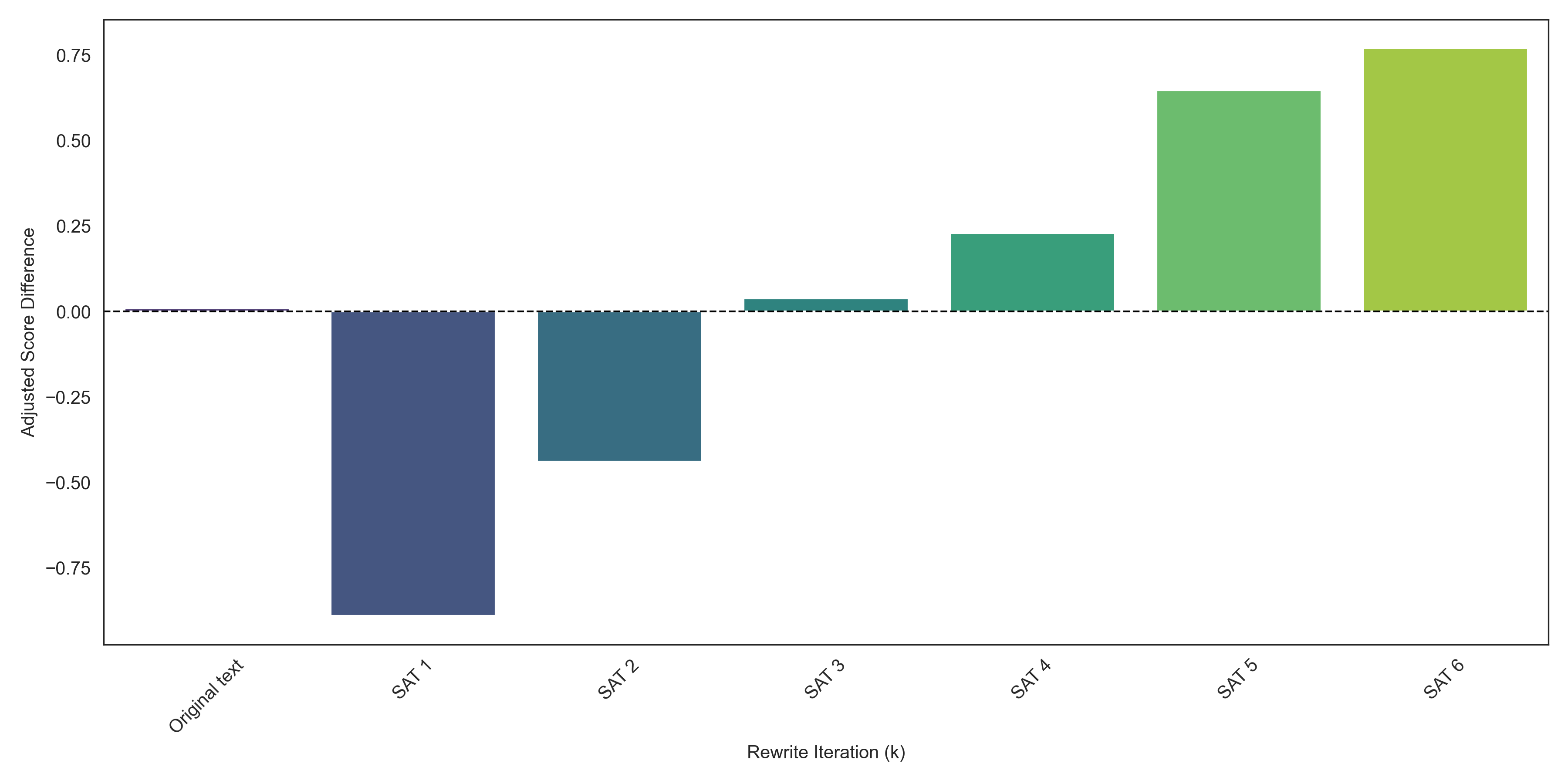}
    \caption{Difference in average scores between the SAT rewrites and the original essay score}
    \label{fig:rewritePremious}
    \begin{minipage}{\linewidth}
\footnotesize
\justifying
\textit{Note:} The figure shows a bar chart of the difference in average score between each SAT rewrite iteration and the original essay score. The x-axis lists rewrite iterations (Original text, SAT 1–SAT 6), the y-axis reports the (adjusted) score difference, and a horizontal dashed line marks zero.
    \end{minipage}
\end{figure}

\begin{figure}[H]
    \centering
    \includegraphics[width=\linewidth]{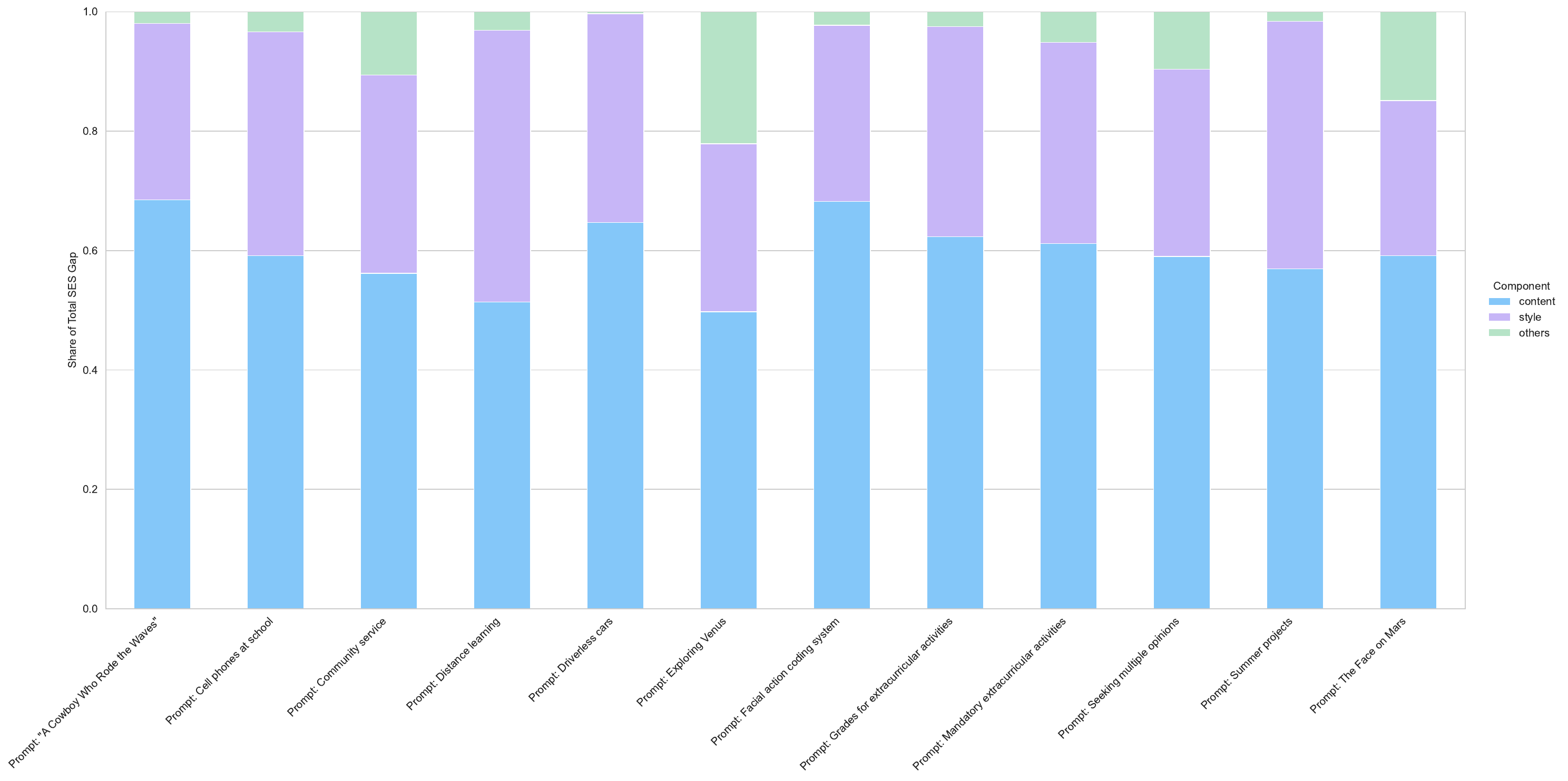}
    \caption{Content and Style Decomposition, By Prompt}
    \label{fig:decomPrompts}
    \begin{minipage}{\linewidth}
\footnotesize
\justifying
\textit{Note:} The figure shows a set of stacked bars that break the overall high–low SES difference in writing scores into three components—content, style, and others—and reports these components as shares of the total SES gap within each subgroup. Each bar corresponds to a writing prompt, and the y-axis runs from 0 to 1 so that the full height of a bar represents 100\% of that subgroup’s SES gap. Within each bar, the colored segments indicate the fraction attributed to the content component, the fraction attributed to the style component, and the remaining fraction grouped as “others” (the residual portion of the decomposition). The plot is displaying how the gap is allocated across components rather than the gap’s size in score points. For exact values and standard errors see Table  \ref{tab:decompositionTableMain}, in the Appendix.   
\end{minipage}
\end{figure}

\section{Additional Tables}
\begin{table}[H]
\centering
\resizebox{\textwidth}{!}{%
  \begin{tabular}{llllllll}
\toprule
 & Total Gap & Content & Style & Other & Share Content & Share Style & Share Other \\
\midrule
All & \shortstack{0.663\\(0.015)} & \shortstack{0.456\\(0.012)} & \shortstack{0.173\\(0.005)} & \shortstack{0.034\\(0.002)} & \shortstack{0.688\\(0.007)} & \shortstack{0.261\\(0.006)} & \shortstack{0.052\\(0.004)} \\
Male & \shortstack{0.657\\(0.020)} & \shortstack{0.462\\(0.017)} & \shortstack{0.168\\(0.007)} & \shortstack{0.027\\(0.003)} & \shortstack{0.703\\(0.009)} & \shortstack{0.256\\(0.009)} & \shortstack{0.041\\(0.005)} \\
Female & \shortstack{0.682\\(0.021)} & \shortstack{0.460\\(0.018)} & \shortstack{0.181\\(0.007)} & \shortstack{0.041\\(0.003)} & \shortstack{0.674\\(0.010)} & \shortstack{0.266\\(0.009)} & \shortstack{0.060\\(0.005)} \\
White & \shortstack{0.569\\(0.022)} & \shortstack{0.385\\(0.019)} & \shortstack{0.143\\(0.008)} & \shortstack{0.041\\(0.005)} & \shortstack{0.677\\(0.013)} & \shortstack{0.252\\(0.013)} & \shortstack{0.071\\(0.009)} \\
Non-White & \shortstack{0.746\\(0.021)} & \shortstack{0.533\\(0.017)} & \shortstack{0.181\\(0.007)} & \shortstack{0.032\\(0.003)} & \shortstack{0.714\\(0.007)} & \shortstack{0.243\\(0.007)} & \shortstack{0.043\\(0.004)} \\
Grade 6 & \shortstack{0.238\\(0.031)} & \shortstack{0.163\\(0.029)} & \shortstack{0.070\\(0.014)} & \shortstack{0.005\\(0.009)} & \shortstack{0.685\\(0.063)} & \shortstack{0.296\\(0.062)} & \shortstack{0.019\\(0.038)} \\
Grade 8 & \shortstack{0.577\\(0.017)} & \shortstack{0.386\\(0.015)} & \shortstack{0.157\\(0.006)} & \shortstack{0.034\\(0.003)} & \shortstack{0.669\\(0.011)} & \shortstack{0.272\\(0.011)} & \shortstack{0.059\\(0.006)} \\
Grade 10 & \shortstack{0.518\\(0.023)} & \shortstack{0.316\\(0.020)} & \shortstack{0.161\\(0.009)} & \shortstack{0.041\\(0.004)} & \shortstack{0.610\\(0.017)} & \shortstack{0.311\\(0.016)} & \shortstack{0.079\\(0.008)} \\
Grade 11 & \shortstack{0.484\\(0.032)} & \shortstack{0.248\\(0.025)} & \shortstack{0.204\\(0.016)} & \shortstack{0.032\\(0.008)} & \shortstack{0.512\\(0.026)} & \shortstack{0.422\\(0.025)} & \shortstack{0.066\\(0.017)} \\
Prompt="A Cowboy Who Rode the Waves" & \shortstack{0.238\\(0.031)} & \shortstack{0.163\\(0.030)} & \shortstack{0.070\\(0.015)} & \shortstack{0.005\\(0.009)} & \shortstack{0.685\\(0.065)} & \shortstack{0.296\\(0.067)} & \shortstack{0.019\\(0.037)} \\
Prompt=Cell phones at school & \shortstack{0.393\\(0.033)} & \shortstack{0.233\\(0.028)} & \shortstack{0.147\\(0.016)} & \shortstack{0.013\\(0.008)} & \shortstack{0.592\\(0.036)} & \shortstack{0.375\\(0.037)} & \shortstack{0.033\\(0.020)} \\
Prompt=Community service & \shortstack{0.383\\(0.034)} & \shortstack{0.215\\(0.031)} & \shortstack{0.127\\(0.014)} & \shortstack{0.041\\(0.008)} & \shortstack{0.562\\(0.043)} & \shortstack{0.332\\(0.037)} & \shortstack{0.106\\(0.023)} \\
Prompt=Distance learning & \shortstack{0.568\\(0.045)} & \shortstack{0.292\\(0.033)} & \shortstack{0.259\\(0.023)} & \shortstack{0.018\\(0.010)} & \shortstack{0.514\\(0.029)} & \shortstack{0.455\\(0.029)} & \shortstack{0.031\\(0.018)} \\
Prompt=Driverless cars & \shortstack{0.340\\(0.035)} & \shortstack{0.220\\(0.031)} & \shortstack{0.119\\(0.015)} & \shortstack{0.001\\(0.007)} & \shortstack{0.647\\(0.044)} & \shortstack{0.350\\(0.044)} & \shortstack{0.003\\(0.021)} \\
Prompt=Exploring Venus & \shortstack{0.640\\(0.043)} & \shortstack{0.319\\(0.035)} & \shortstack{0.180\\(0.018)} & \shortstack{0.142\\(0.008)} & \shortstack{0.498\\(0.027)} & \shortstack{0.281\\(0.022)} & \shortstack{0.221\\(0.019)} \\
Prompt=Facial action coding system & \shortstack{0.542\\(0.042)} & \shortstack{0.370\\(0.036)} & \shortstack{0.160\\(0.016)} & \shortstack{0.012\\(0.007)} & \shortstack{0.682\\(0.027)} & \shortstack{0.295\\(0.026)} & \shortstack{0.023\\(0.012)} \\
Prompt=Grades for extracurricular activities & \shortstack{0.394\\(0.035)} & \shortstack{0.246\\(0.031)} & \shortstack{0.138\\(0.015)} & \shortstack{0.010\\(0.007)} & \shortstack{0.623\\(0.039)} & \shortstack{0.352\\(0.038)} & \shortstack{0.025\\(0.019)} \\
Prompt=Mandatory extracurricular activities & \shortstack{0.696\\(0.044)} & \shortstack{0.426\\(0.035)} & \shortstack{0.235\\(0.019)} & \shortstack{0.036\\(0.009)} & \shortstack{0.612\\(0.023)} & \shortstack{0.337\\(0.023)} & \shortstack{0.051\\(0.014)} \\
Prompt=Seeking multiple opinions & \shortstack{0.522\\(0.038)} & \shortstack{0.308\\(0.033)} & \shortstack{0.164\\(0.017)} & \shortstack{0.050\\(0.009)} & \shortstack{0.590\\(0.032)} & \shortstack{0.314\\(0.030)} & \shortstack{0.096\\(0.019)} \\
Prompt=Summer projects & \shortstack{0.450\\(0.039)} & \shortstack{0.256\\(0.032)} & \shortstack{0.186\\(0.018)} & \shortstack{0.007\\(0.010)} & \shortstack{0.570\\(0.033)} & \shortstack{0.414\\(0.032)} & \shortstack{0.016\\(0.022)} \\
Prompt=The Face on Mars & \shortstack{0.408\\(0.039)} & \shortstack{0.242\\(0.035)} & \shortstack{0.106\\(0.017)} & \shortstack{0.061\\(0.008)} & \shortstack{0.592\\(0.042)} & \shortstack{0.260\\(0.037)} & \shortstack{0.149\\(0.024)} \\
\bottomrule
\end{tabular}

}
\caption{The Score Gap Decomposition}
\label{tab:decompositionTableMain}
    \begin{minipage}{\linewidth}
    \footnotesize
    \justifying
    \textit{Note:} This table decomposes the gap in holistic writing scores between high- and low-SES students. “Total gap,” “Content,” “Style,” and “Other” report the decomposition in levels, while “Share content,” “Share style,” and “Share other” report the corresponding shares of the gap attributable to each component. The first row reports results for the full sample; the remaining rows report results by subpopulation and prompt. Standard errors (in parentheses) are obtained via a bootstrap with 500 replications.
    \end{minipage}
\end{table}

\begin{table}[H]
\centering
\resizebox{\textwidth}{!}{%
  \begin{tabular}{llllllll}
\toprule
 & Total Gap & Content & Style & Other & Share Content & Share Style & Share Other \\
\midrule
SAT (baseline) & \shortstack{0.663\\(0.015)} & \shortstack{0.456\\(0.012)} & \shortstack{0.173\\(0.005)} & \shortstack{0.034\\(0.002)} & \shortstack{0.688\\(0.007)} & \shortstack{0.261\\(0.006)} & \shortstack{0.052\\(0.004)} \\
Score $2-5$, $|\Delta|\leq 1$ & \shortstack{0.547\\(0.013)} & \shortstack{0.385\\(0.012)} & \shortstack{0.129\\(0.005)} & \shortstack{0.032\\(0.002)} & \shortstack{0.704\\(0.008)} & \shortstack{0.237\\(0.008)} & \shortstack{0.059\\(0.005)} \\
Rewrites {1,2,5,6} only & \shortstack{0.663\\(0.014)} & \shortstack{0.442\\(0.011)} & \shortstack{0.187\\(0.006)} & \shortstack{0.034\\(0.002)} & \shortstack{0.666\\(0.007)} & \shortstack{0.282\\(0.007)} & \shortstack{0.052\\(0.004)} \\
Drop k=1 rewrites & \shortstack{0.663\\(0.014)} & \shortstack{0.472\\(0.013)} & \shortstack{0.157\\(0.005)} & \shortstack{0.034\\(0.002)} & \shortstack{0.712\\(0.007)} & \shortstack{0.237\\(0.007)} & \shortstack{0.052\\(0.004)} \\
Baseline GPT & \shortstack{0.657\\(0.013)} & \shortstack{0.501\\(0.013)} & \shortstack{0.122\\(0.005)} & \shortstack{0.035\\(0.002)} & \shortstack{0.762\\(0.008)} & \shortstack{0.186\\(0.007)} & \shortstack{0.053\\(0.004)} \\
\bottomrule
\end{tabular}

}
\caption{The Score Gap Decomposition - Robustness}
\label{tab:robustness}
    \begin{minipage}{\linewidth}
    \footnotesize
    \justifying
\textit{Note:} Each row corresponds to a different way of constructing the rewrite-based variation used in the decomposition. SAT (baseline) is the paper’s main specification: for each original essay, the LLM is prompted to produce a sequence of targeted rewrites intended to match SAT essay-score writing levels  (i.e., “SAT-1” through “SAT-6” rewrites), and the decomposition is computed using this full set of targeted SAT rewrites. Score 2–5, \(|\Delta|\le 1\) restricts attention to essays whose SAT score lies between 2 and 5 and uses only comparisons between rewrite levels that differ by at most one SAT point (adjacent rewrite levels), so that the rewrite-pair difference satisfies \(|k-k'|\le 1\). Rewrites \({1,2,5,6}\) only re-computes the decomposition using only the more extreme targeted rewrite levels—keeping SAT-1, SAT-2, SAT-5, and SAT-6 rewrites (equivalently emphasizing comparisons such as 1 vs 6 and 2 vs 5). Drop (k=1) rewrites excludes the SAT-1 rewrite from the targeted rewrite set and recomputes the decomposition using only SAT-2 through SAT-6. Baseline GPT replaces the targeted SAT rewrites with a single “standard GPT” rewrite that asks the model to rewrite the essay without specifying a target SAT/style level, and then applies the same decomposition procedure using this alternative rewrite variation. Columns report both level and share versions of the decomposition. Total Gap is the estimated high-SES minus low-SES difference in mean holistic writing scores in that specification. Content, Style, and Other are the additive components of the total gap in score units: “Content” is the portion attributed to the content component, “Style” is the portion attributed to the style component, and “Other” is the residual component from the decomposition (the part not captured by the content and style components). Share Content, Share Style, and Share Other divide each component by Total Gap, so the three shares sum to 1 within each row. Parentheses report bootstrap standard errors based on 500 replications.

    \end{minipage}
\end{table}

\begin{table}[H]
\centering
\resizebox{\textwidth}{!}{%
  \begin{tabular}{lllllll}
\toprule
 & Total Gap & Content & Style & Share Content & Share Style & Share Other \\
\midrule
All & \shortstack{0.671\\(0.013)} & \shortstack{0.520\\(0.013)} & \shortstack{0.114\\(0.005)} & \shortstack{0.775\\(0.008)} & \shortstack{0.170\\(0.007)} & \shortstack{0.055\\(0.004)} \\
Male & \shortstack{0.660\\(0.019)} & \shortstack{0.524\\(0.018)} & \shortstack{0.106\\(0.007)} & \shortstack{0.794\\(0.011)} & \shortstack{0.161\\(0.011)} & \shortstack{0.045\\(0.005)} \\
Female & \shortstack{0.693\\(0.018)} & \shortstack{0.526\\(0.018)} & \shortstack{0.124\\(0.006)} & \shortstack{0.759\\(0.011)} & \shortstack{0.179\\(0.010)} & \shortstack{0.062\\(0.005)} \\
White & \shortstack{0.576\\(0.023)} & \shortstack{0.443\\(0.023)} & \shortstack{0.089\\(0.008)} & \shortstack{0.769\\(0.016)} & \shortstack{0.155\\(0.015)} & \shortstack{0.076\\(0.008)} \\
Non-White & \shortstack{0.758\\(0.022)} & \shortstack{0.617\\(0.021)} & \shortstack{0.106\\(0.007)} & \shortstack{0.815\\(0.009)} & \shortstack{0.140\\(0.009)} & \shortstack{0.045\\(0.004)} \\
Grade 6 & \shortstack{0.225\\(0.029)} & \shortstack{0.164\\(0.033)} & \shortstack{0.053\\(0.017)} & \shortstack{0.728\\(0.086)} & \shortstack{0.234\\(0.083)} & \shortstack{0.038\\(0.040)} \\
Grade 8 & \shortstack{0.583\\(0.017)} & \shortstack{0.443\\(0.018)} & \shortstack{0.105\\(0.007)} & \shortstack{0.761\\(0.012)} & \shortstack{0.179\\(0.012)} & \shortstack{0.060\\(0.006)} \\
Grade 10 & \shortstack{0.538\\(0.023)} & \shortstack{0.383\\(0.023)} & \shortstack{0.110\\(0.008)} & \shortstack{0.712\\(0.019)} & \shortstack{0.205\\(0.016)} & \shortstack{0.083\\(0.009)} \\
Grade 11 & \shortstack{0.492\\(0.032)} & \shortstack{0.282\\(0.029)} & \shortstack{0.176\\(0.015)} & \shortstack{0.572\\(0.032)} & \shortstack{0.358\\(0.029)} & \shortstack{0.070\\(0.015)} \\
Prompt="A Cowboy Who Rode the Waves" & \shortstack{0.225\\(0.030)} & \shortstack{0.164\\(0.033)} & \shortstack{0.053\\(0.018)} & \shortstack{0.728\\(0.088)} & \shortstack{0.234\\(0.083)} & \shortstack{0.038\\(0.041)} \\
Prompt=Cell phones at school & \shortstack{0.400\\(0.031)} & \shortstack{0.266\\(0.032)} & \shortstack{0.121\\(0.016)} & \shortstack{0.664\\(0.041)} & \shortstack{0.302\\(0.041)} & \shortstack{0.034\\(0.019)} \\
Prompt=Community service & \shortstack{0.388\\(0.035)} & \shortstack{0.229\\(0.036)} & \shortstack{0.117\\(0.016)} & \shortstack{0.591\\(0.050)} & \shortstack{0.302\\(0.045)} & \shortstack{0.107\\(0.022)} \\
Prompt=Distance learning & \shortstack{0.571\\(0.043)} & \shortstack{0.327\\(0.040)} & \shortstack{0.223\\(0.021)} & \shortstack{0.572\\(0.039)} & \shortstack{0.390\\(0.039)} & \shortstack{0.037\\(0.017)} \\
Prompt=Driverless cars & \shortstack{0.339\\(0.034)} & \shortstack{0.278\\(0.036)} & \shortstack{0.061\\(0.015)} & \shortstack{0.818\\(0.048)} & \shortstack{0.180\\(0.046)} & \shortstack{0.002\\(0.020)} \\
Prompt=Exploring Venus & \shortstack{0.649\\(0.045)} & \shortstack{0.369\\(0.042)} & \shortstack{0.135\\(0.017)} & \shortstack{0.569\\(0.033)} & \shortstack{0.207\\(0.026)} & \shortstack{0.224\\(0.019)} \\
Prompt=Facial action coding system & \shortstack{0.591\\(0.041)} & \shortstack{0.453\\(0.040)} & \shortstack{0.120\\(0.015)} & \shortstack{0.766\\(0.028)} & \shortstack{0.203\\(0.027)} & \shortstack{0.031\\(0.011)} \\
Prompt=Grades for extracurricular activities & \shortstack{0.399\\(0.032)} & \shortstack{0.281\\(0.034)} & \shortstack{0.110\\(0.016)} & \shortstack{0.704\\(0.045)} & \shortstack{0.274\\(0.045)} & \shortstack{0.021\\(0.018)} \\
Prompt=Mandatory extracurricular activities & \shortstack{0.701\\(0.043)} & \shortstack{0.517\\(0.040)} & \shortstack{0.143\\(0.020)} & \shortstack{0.738\\(0.028)} & \shortstack{0.205\\(0.027)} & \shortstack{0.057\\(0.014)} \\
Prompt=Seeking multiple opinions & \shortstack{0.518\\(0.041)} & \shortstack{0.341\\(0.041)} & \shortstack{0.128\\(0.018)} & \shortstack{0.658\\(0.041)} & \shortstack{0.248\\(0.036)} & \shortstack{0.095\\(0.020)} \\
Prompt=Summer projects & \shortstack{0.459\\(0.040)} & \shortstack{0.294\\(0.038)} & \shortstack{0.157\\(0.016)} & \shortstack{0.641\\(0.040)} & \shortstack{0.341\\(0.034)} & \shortstack{0.017\\(0.022)} \\
Prompt=The Face on Mars & \shortstack{0.406\\(0.039)} & \shortstack{0.271\\(0.041)} & \shortstack{0.073\\(0.018)} & \shortstack{0.668\\(0.051)} & \shortstack{0.179\\(0.045)} & \shortstack{0.153\\(0.023)} \\
\bottomrule
\end{tabular}

}
\caption{The Score Gap Baseline Decomposition}
\label{tab:decompositionAppendixTableMain}
    \begin{minipage}{\linewidth}
    \footnotesize
    \justifying
    \textit{Note:} This table decomposes the gap in holistic writing scores between high- and low-SES students, using the Baseline decomposition from section \ref{app:baselineDecomp}, and compare the neutral GPT style. “Total gap,” “Content,” “Style,” and “Other” report the decomposition in levels, while “Share content,” “Share style,” and “Share other” report the corresponding shares of the gap attributable to each component. The first row reports results for the full sample; the remaining rows report results by subpopulation and prompt. Standard errors (in parentheses) are obtained via a bootstrap with 500 replications.
    \end{minipage}
\end{table}

\section{Baseline Decomposition}\label{app:baselineDecomp}

Our main identification in the main text relies on three assumptions: separability (Assumption \ref{ass:sep}), rewrite content fidelity (Assumption \ref{ass:id-fidelity}), and additivity of rewrite effects (Assumption \ref{ass:id-style}). In this section, we offer an alternative, baseline-based interpretation that relaxes Assumptions \ref{ass:sep} and \ref{ass:id-style}. 

Under this interpretation, the decomposition should be read as measuring content and style differences \emph{relative to a chosen reference style}. Concretely, if a researcher or policymaker is willing to adopt a “neutral” baseline writing style (e.g., a standardized LLM rewrite), then the content component is the score gap that remains after rewriting both groups into that baseline, and the style component is the differential premium (or penalty) that each group’s original style receives relative to the baseline. 

This framing maps directly to a policy counterfactual. For example, consider the case where we allow (or require) students to submit a GPT-polished version of their essay. In that world, both groups’ texts are pushed toward the same reference style, so style-related advantages or penalties are muted. The decomposition then tells you how much of the observed gap would shrink under this policy (the style component) and how much would remain even after stylistic standardization (the content component).

To see this interpretation, we begin by defining new notation. For group $G\in\{H,L\}$, let
\[
\mu_G^{S,\mathrm{orig}} := \mathbb{E}\big[S(\text{original text})\mid G\big],
\qquad
\mu_G^{S,\mathrm{neu}} := \mathbb{E}\big[S(T(\text{original text}))\mid G\big],
\]
where $T$ rewrites to GPT-style (``neutral''). Scoring functions: $S^{(H)}$ and $S^{(L)}$. The observed score gap is given by the following expression:
\[
\Delta_{\mathrm{obs}} \;=\; \mu_H^{S^{(H)},\mathrm{orig}} \;-\; \mu_L^{S^{(L)},\mathrm{orig}} .
\]

We now derive the decomposition term. Add and subtract $\mu_L^{S^{(H)},\mathrm{orig}}$:
\begin{align*}
\Delta_{\mathrm{obs}}
&= \big(\mu_H^{S^{(H)},\mathrm{orig}} - \mu_L^{S^{(H)},\mathrm{orig}}\big)
\;+\;
\underbrace{\textcolor{TiltColor}{\big(\mu_L^{S^{(H)},\mathrm{orig}} - \mu_L^{S^{(L)},\mathrm{orig}}\big)}}_{\textcolor{TiltColor}{\text{Scoring-function tilt (vs.\ $S^{(H)}$)}}}.
\end{align*}

Now add and subtract the neutralized means \emph{under the same scorer $S^{(H)}$}:
\begin{align*}
\mu_H^{S^{(H)},\mathrm{orig}} - \mu_L^{S^{(H)},\mathrm{orig}}
&=
\underbrace{\textcolor{ContentColor}{\big(\mu_H^{S^{(H)},\mathrm{neu}} - \mu_L^{S^{(H)},\mathrm{neu}}\big)}}_{\textcolor{ContentColor}{\text{Content}^{(H)}}}
\\ &\qquad
+\;
\underbrace{\textcolor{StyleColor}{\big(\mu_H^{S^{(H)},\mathrm{orig}} - \mu_H^{S^{(H)},\mathrm{neu}}\big)}}_{\textcolor{StyleColor}{\text{H style premium under }S^{(H)}}}
\;-\;
\underbrace{\textcolor{StyleColor}{\big(\mu_L^{S^{(H)},\mathrm{orig}} - \mu_L^{S^{(H)},\mathrm{neu}}\big)}}_{\textcolor{StyleColor}{\text{L style premium under }S^{(H)}}}.
\end{align*}
Taking the difference of the two blue terms gives the net
\[
\textcolor{StyleColor}{\text{Style}^{(H)}} \;=\; 
\textcolor{StyleColor}{\big(\mu_H^{S^{(H)},\mathrm{orig}} - \mu_H^{S^{(H)},\mathrm{neu}}\big)}
\;-\;
\textcolor{StyleColor}{\big(\mu_L^{S^{(H)},\mathrm{orig}} - \mu_L^{S^{(H)},\mathrm{neu}}\big)}.
\]

Here the style can be thought of as the premium of the high and low SES writing style compare to the neutral level. 

Finally combining the above gives us:
\[
\Delta_{\mathrm{obs}}
=
\underbrace{\textcolor{ContentColor}{\big(\mu_H^{S^{(H)},\mathrm{neu}} - \mu_L^{S^{(H)},\mathrm{neu}}\big)}}_{\textcolor{ContentColor}{\text{Content}^{(H)}}}
\;+\;
\underbrace{\textcolor{StyleColor}{\Big[\big(\mu_H^{S^{(H)},\mathrm{orig}} - \mu_H^{S^{(H)},\mathrm{neu}}\big) - \big(\mu_L^{S^{(H)},\mathrm{orig}} - \mu_L^{S^{(H)},\mathrm{neu}}\big)\Big]}}_{\textcolor{StyleColor}{\text{Style}^{(H)}}}
\;+\;
\underbrace{\textcolor{TiltColor}{\big(\mu_L^{S^{(H)},\mathrm{orig}} - \mu_L^{S^{(L)},\mathrm{orig}}\big)}}_{\textcolor{TiltColor}{\text{Scoring-function tilt (vs.\ $S^{(H)}$)}}}
\]

Therefore, the main difference is that the style is measured as premium compare to the benchmark style, and content is measured as the difference in means given the same style distribution. Rewriting the content in integral form:
\begin{align}
\mu_H^{S^{(H)},\mathrm{neu}} - \mu_L^{S^{(H)},\mathrm{neu}}
&= \int_{C,S} S(C,S)\, P_{\mathrm{Baseline}}(S\mid C)\, P_H(C)\, dC\, dS \nonumber\\
&\quad - \int_{C,S} S(C,S)\, P_{\mathrm{Baseline}}(S\mid C)\, P_L(C)\, dC\, dS \nonumber\\
&= \int_{C,S} S(C,S)\, P_{\mathrm{Baseline}}(S\mid C)\, \bigl(P_H(C)-P_L(C)\bigr)\, dC\, dS ,
\end{align}
where \(P_G(C)\) denotes the (group-specific) distribution of content \(C\), and \(P_{\text{Baseline}}(S\mid C)\) is the conditional distribution of style induced by the neutral rewrite \(T\) (i.e., the “baseline style rule,” possibly allowing multiple acceptable realizations of style for the same content). This representation makes clear that the content term holds style fixed at the baseline and varies only the content distribution across groups. The content component captures the part of the score gap that would remain if both groups were expressed in the same neutral writing style.

By contrast, the style term isolates the relative advantage of each group’s original style compared to the neutral baseline, holding fixed the scorer $S^{(H)}$. In the policy counterfactual where all students submit LLM-polished essays, these “style premia” shrink mechanically. The post-policy observed gap can then be written as
\[
\mu_H^{S^{(H)},\mathrm{neu}}
-
\mu_L^{S^{(L)},\mathrm{neu}}
\;=\;
\textcolor{ContentColor}{\text{Content}^{(H)}}
\;+\;
\bigl(
\mu_L^{S^{(H)},\mathrm{neu}}
-
\mu_L^{S^{(L)},\mathrm{neu}}
\bigr).
\]
The first term,
\textcolor{ContentColor}{\(\text{Content}^{(H)}\)},
is the residual gap, driven by differences in content, that persists after standardizing writing style.
The second term captures the remaining difference attributable to the scorer functions, evaluated on neutralized text (a tilt component under neutralization).
Under this definition, universal neutralization removes the style component mechanically, while the overall change in the observed gap may additionally reflect how scorer differences manifest on GPT-polished essays.

% In the policy counterfactual where students submit GPT-polished essays, these “style premia” shrink mechanically because \(\mu_G^{S^{(H)},\mathrm{orig}}\) is replaced by \(\mu_G^{S^{(H)},\mathrm{neu}}\). The predicted change in the observed gap under universal neutralization, under the high score ranker, is therefore exactly \(-,\textcolor{StyleColor}{\text{Style}^{(H)}}\), while \(\textcolor{ContentColor}{\text{Content}^{(H)}}\) is the residual gap that persists even after standardizing style. It's important to note that the total gap would also be affected by how the different rankers are going to responsd to the LLM rewrites. 

\subsection{Decomposition Results}

\begin{figure}[H]
    \centering
    \includegraphics[width=\linewidth]{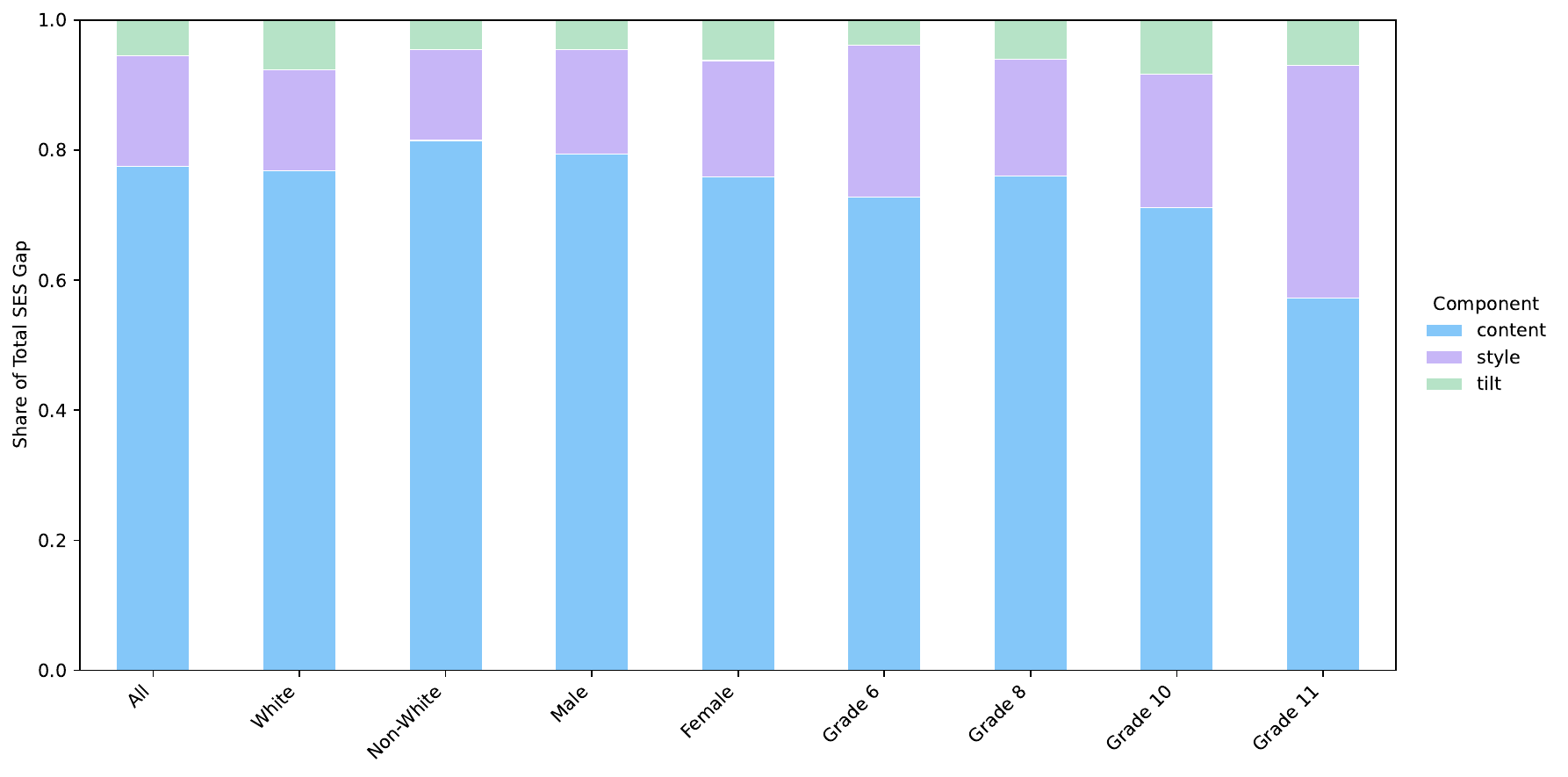}
    \caption{Baseline Decomposition}
    \label{fig:baselineDecomp}
\end{figure}

We now turn to the decomposition results using the standard GPT rewrites, as described in Section \ref{sec:data}. Figure \ref{fig:baselineDecomp} presents the decomposition. The pattern closely mirrors Figure \ref{fig:decomMain} in the main text, highlighting that differences in content are the primary contributor to the gap. Even after equating writing style, substantial differences remain, largely driven by content.

\begin{figure}[H]
    \centering
    \includegraphics[width=\linewidth]{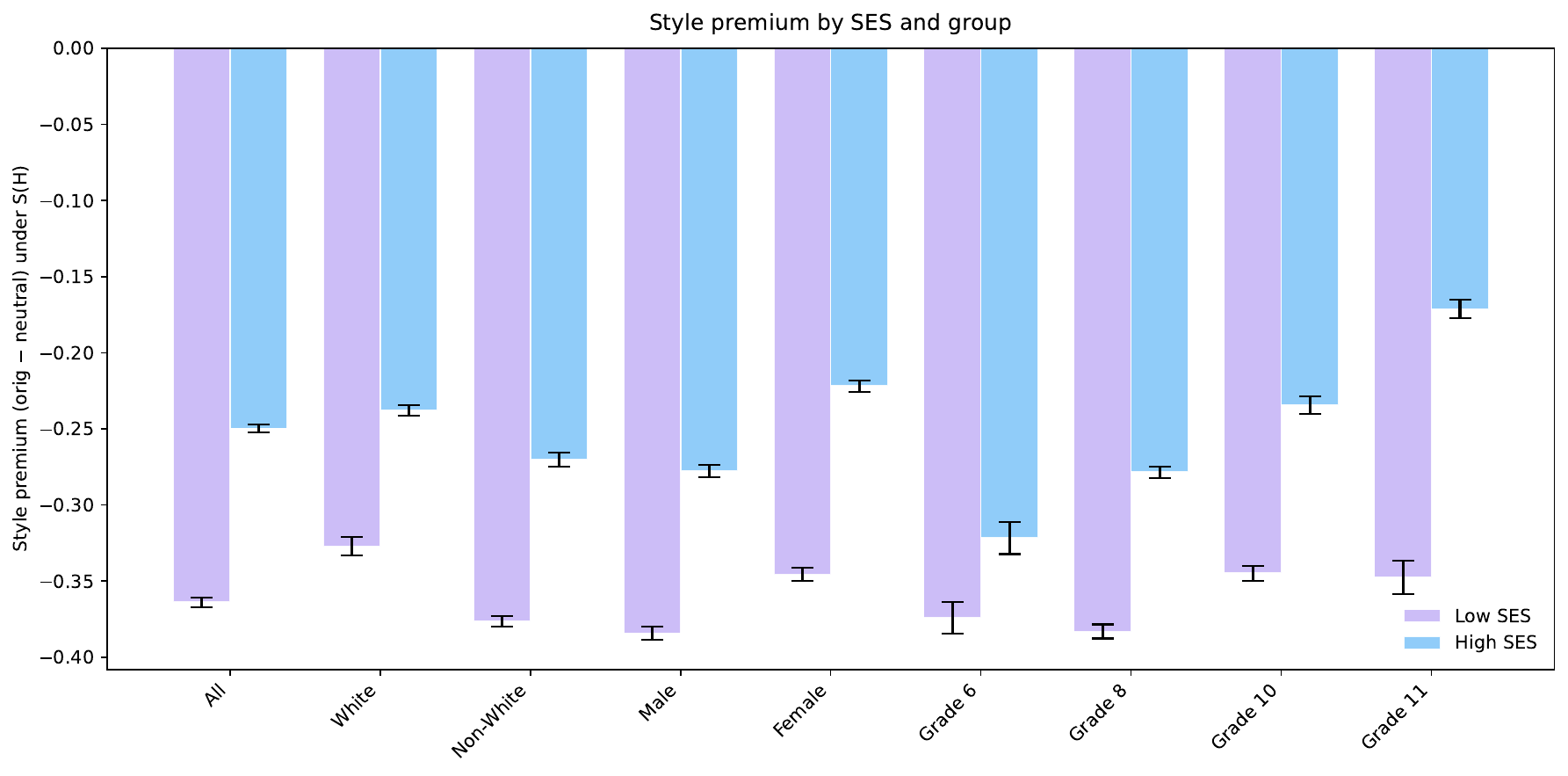}
    \caption{Style Premium Across Groups}
    \label{fig:StylePremium}
\end{figure}

Figure \ref{fig:StylePremium} shows the style premium across groups for low- and high-SES students. Across all students, writing quality is lower than in the neutralized GPT version, implying that GPT improves style on average. The improvement is larger for low-SES students across all groups, consistent with our main result that low-SES writing style is weaker and contributes to the gap. The figure also suggests that if all essays were processed through an LLM-style “style equalizer,” low-SES students would benefit more than high-SES students.

\section{Variable descriptions}\label{app:styleVariables}

\begin{enumerate}

\item \textbf{Word Count (Volume)}: Total number of words in the essay. Higher values indicate longer texts.

\item \textbf{Avg Word Length (Lexical Sophistication)}: Average length of words (typically characters per word). Higher values often reflect longer, potentially more advanced vocabulary (though it can also be influenced by proper nouns and morphology).

\item \textbf{Mean Length of Clause (Syntactic Complexity)}: Average number of words per clause. Higher values indicate more elaborated clause structure and greater within-clause syntactic complexity.

\item \textbf{Mean Length of T-Unit (Sentence Complexity)}: Average length (in words) of a T-unit: one main clause plus any subordinate clauses attached to it. Higher values indicate more sentence-level elaboration through embedding/subordination.

\item \textbf{Mean Verbal Dependencies}: Average number of syntactic dependents attached to verbs (e.g., objects, auxiliaries, modifiers). Higher values suggest richer predicate structure and greater syntactic elaboration around verbs.

\item \textbf{Proportion of Infinitives}: Share of verb constructions using infinitives (e.g., ``to go'', ``to see''). Higher values indicate more infinitival usage, often associated with goal/intent framing and non-finite syntax.

\item \textbf{Proportion of Non-Finite Clauses}: Share of clauses that are non-finite (infinitival, gerund, participial). Higher values indicate more embedding without full tense marking (often more syntactic compression).

\item \textbf{Lexical density (tokens)}: Proportion of tokens that are content words (nouns/verbs/adjectives/adverbs) rather than function words. Higher values indicate more information-dense, typically less conversational writing.

\item \textbf{Lexical diversity (MTLD; all words)}: Vocabulary diversity measure designed to be relatively robust to text length. Higher values indicate more varied word use (less repetition).

\item \textbf{Lexical diversity (MATTR; lemmas)}: Moving-average type--token ratio computed on lemmas (so inflectional variants are collapsed, e.g., \emph{run/runs/running}). Higher values indicate more lexical variety net of inflection.

\item \textbf{Content-word overlap (adjacent sentences)}: Degree to which adjacent sentences reuse the same content words. Higher values indicate stronger local cohesion through repetition (very high values can also reflect redundancy).

\item \textbf{Semantic cohesion (Word2Vec; adjacent sentences)}: Semantic similarity between adjacent sentences computed using Word2Vec-style embeddings. Higher values indicate stronger topical/semantic continuity across neighboring sentences.

\item \textbf{Nominalizations}: Rate of nominalized forms (nouns derived from verbs/adjectives, often with suffixes like ``-tion'', ``-ment''). Higher values are commonly associated with more abstract/academic register, sometimes at the expense of directness.

\item \textbf{Pronoun-to-noun ratio}: Pronouns relative to nouns. Higher values indicate more pronoun-based reference (often more narrative/conversational but sometimes less explicit); lower values indicate more noun-based reference (often more explicit/formal).

\end{enumerate}

\section{Writing Prompts}\label{app:writingPrompts}

The essays in our dataset are written in response to a fixed set of prompts provided by the
\textit{Persuade 2.0} corpus. These prompts define the topic and argumentative context of each essay and are stored under the variable \texttt{prompt\_name}. Table~\ref{tab:writing-prompts} lists all prompts used in the dataset, along with the number and proportion of essays associated with each prompt.

\begin{table}[h]
\centering
\begin{tabular}{lrr}
\hline
\textbf{Prompt} & \textbf{Count} & \textbf{Share (\%)} \\
\hline
Facial action coding system              & 2,167 & 10.4 \\
Distance learning                        & 2,157 & 10.3 \\
Driverless cars                          & 1,886 & 9.0 \\
Exploring Venus                          & 1,862 & 8.9 \\
Summer projects                          & 1,750 & 8.4 \\
Mandatory extracurricular activities    & 1,670 & 8.0 \\
Cell phones at school                    & 1,656 & 7.9 \\
Grades for extracurricular activities   & 1,626 & 7.8 \\
The Face on Mars                         & 1,583 & 7.6 \\
Seeking multiple opinions                & 1,552 & 7.4 \\
Community service                        & 1,542 & 7.4 \\
A Cowboy Who Rode the Waves          & 1,372 & 6.6 \\
\hline
\end{tabular}
\caption{Writing prompts in the Persuade 2.0 dataset and their frequencies. Percentages are computed
relative to the total number of essays.}
\label{tab:writing-prompts}
\end{table}

In addition to the prompt labels summarized in Table~\ref{tab:writing-prompts}, the dataset includes the full assignment descriptions associated with each prompt under the variable \texttt{assignment}. These assignments specify the task format (e.g., argumentative or explanatory), the intended audience (e.g., principal, senator, general reader), and the main requirements, such as using details from provided articles or addressing counterarguments. For example, essays written under the \textit{Facial action coding system} prompt require students to argue whether emotion-recognition technology should be used in classrooms based on an accompanying article. Other prompts, such as \textit{Driverless cars}, ask students to construct evidence-based arguments grounded in informational texts, while prompts like \textit{Seeking multiple opinions} or \textit{Summer projects} encourage explanatory or persuasive writing based on personal reasoning and examples. These assignment-level instructions provide a consistent structure within each prompt while varying topic and audience, allowing us to control for stylistic variation rather than differences in writing objectives.

For reference, we list below the full assignment instructions associated with each prompt.

\subsection*{Full Assignment Instructions}

\begin{itemize}
  \item \textbf{Facial action coding system.}  
  In the article ``Making Mona Lisa Smile,'' the author describes how a new technology called the Facial Action Coding System enables computers to identify human emotions. Using details from the article, write an essay arguing whether the use of this technology to read the emotional expressions of students in a classroom is valuable.

  \item \textbf{Distance learning.}  
  Some schools offer distance learning as an option for students to attend classes from home by way of online or video conferencing. Do you think students would benefit from being able to attend classes from home? Take a position on this issue. Support your response with reasons and examples.

  \item \textbf{Driverless cars.}  
  In the article ``Driverless Cars are Coming,'' the author presents both positive and negative aspects of driverless cars. Using details from the article, create an argument for or against the development of these cars.

  \item \textbf{Exploring Venus.}  
  In ``The Challenge of Exploring Venus,'' the author suggests studying Venus is a worthy pursuit despite the dangers it presents. Using details from the article, write an essay evaluating how well the author supports this idea.

  \item \textbf{Summer projects.}  
  Some schools require students to complete summer projects to ensure continued learning during the break. Should these projects be teacher-designed or student-designed? Take a position and support it with reasons and examples.

  \item \textbf{Mandatory extracurricular activities.}  
  Your principal has decided that all students must participate in at least one extracurricular activity. Do you agree or disagree with this decision? Use specific details and examples to support your position.

  \item \textbf{Cell phones at school.}  
  Your principal is reconsidering the school's cell phone policy. Write a letter convincing her which policy you believe is better and support your position with specific reasons.

  \item \textbf{Grades for extracurricular activities.}  
  Your principal is considering requiring at least a grade B average for participation in sports or other activities. Write a letter arguing for or against this policy change.

  \item \textbf{The Face on Mars.}  
  After reading ``Unmasking the Face on Mars,'' write an argumentative essay convincing someone that the Face is a natural landform, using evidence from the article.

  \item \textbf{Seeking multiple opinions.}  
  Explain why seeking multiple opinions can help someone make a better choice. Use specific details and examples.

  \item \textbf{Community service.}  
  Write a letter to your principal taking a position on whether students should be required to perform community service and support your argument with examples.

  \item \textbf{A Cowboy Who Rode the Waves.}  
  After reading ``A Cowboy Who Rode the Waves,'' write an argument from Luke’s point of view convincing others to participate in the Seagoing Cowboys program.

\end{itemize}

\section{Rewrite Prompts}\label{app:RewritePrompt}

To generate rewrites, we use a structured set of instruction-based prompts that explicitly constrain the model to modify only stylistic features: such as vocabulary, sentence structure, and mechanical correctness, while preserving the original essay’s ideas, arguments, and examples. The prompts are parameterized to target SAT writing score levels (1–6), alongside a neutral rewrite prompt and a corrective prompt used when our evaluation function/prompt verification flags content drift. 

\paragraph{Expected Output}

For all rewrite prompts, the expected output is a single rewritten essay string, with no explanations or metadata.

\subsection{SAT-Conditioned Rewrite Prompts}

All SAT-conditioned rewrites share a fixed system prompt that defines global constraints on content preservation, while the user prompt varies to specify the target SAT style level. This separation mirrors common practice in LLM prompting, where system-level instructions enforce invariant constraints and user-level instructions encode variation.

\paragraph{System Prompt (shared across SAT rewrites)}

\begin{Verbatim}[breaklines,breakanywhere,fontsize=\small]
You are an expert SAT essay writing tutor and text editor.
Your task is to rewrite student essays by changing only their style, not their content.

Content must be preserved:
- Keep the same thesis, stance, arguments, and examples.
- Do not add or remove ideas, arguments, or examples.
- Do not change the underlying facts, claims, or conclusions.

You are allowed to change only the style and the written text according to the instructions.

When you respond:
- Follow the style instructions given in the user message.
- Output ONLY the rewritten essay text, with no explanations, no headings, and no commentary.
\end{Verbatim}

\paragraph{User Prompt Template (SAT Levels 1–6)}

\paragraph{SAT Score 1 (Lowest)}

\begin{Verbatim}[breaklines,breakanywhere,fontsize=\small]
Rewrite the following essay by changing only its style (not its content) so that it matches a
lowest SAT style level (score 1) in vocabulary, sentence structure, and mechanics.

Specifically, rewrite the text, while maintaining all the content intact, such that it:
- displays fundamental problems in vocabulary, with extremely limited or frequently inappropriate
  word choice that sometimes results in unclear or confusing statements;
- demonstrates severe flaws in sentence structure, with pervasive fragments, run-ons, and sentences
  that are often very hard to parse;
- contains pervasive errors in grammar, usage, and mechanics that persistently interfere with
  meaning and regularly make the text difficult or impossible to understand.

Remember: do not change the ideas, arguments, or examples.
Only change wording, sentence structure, and correctness.

ESSAY:
[ESSAY_TEXT]
\end{Verbatim}

\paragraph{SAT Score 2 (Very Weak)}

\begin{Verbatim}[breaklines,breakanywhere,fontsize=\small]
Rewrite the following essay by changing only its style (not its content) so that it matches a
very weak SAT style level (score 2) in vocabulary, sentence structure, and mechanics.

Specifically, rewrite the text, while maintaining all the content intact, such that it:
- uses very limited and often vague vocabulary, with word choices that are regularly imprecise or
  inappropriate;
- shows serious problems in sentence structure, including frequent fragments, run-on sentences,
  or very awkward constructions;
- contains many serious errors in grammar, usage, and mechanics that often interfere with meaning
  and make the text difficult to follow.

Remember: do not change the ideas, arguments, or examples.
Only change wording, sentence structure, and correctness.

ESSAY:
[ESSAY_TEXT]
\end{Verbatim}

\paragraph{SAT Score 3 (Developing)}

\begin{Verbatim}[breaklines,breakanywhere,fontsize=\small]
Rewrite the following essay by changing only its style (not its content) so that it matches a
developing SAT style level (score 3) in vocabulary, sentence structure, and mechanics.

Specifically, rewrite the text, while maintaining all the content intact, such that it:
- uses basic and often repetitive vocabulary, sometimes imprecise or slightly inappropriate;
- relies heavily on simple and repetitive sentence structures, with attempts at more complex
  sentences that are often awkward;
- contains frequent errors in grammar, usage, and mechanics that distract the reader and sometimes
  blur or weaken the meaning, though the main ideas remain mostly understandable.

Remember: do not change the ideas, arguments, or examples.
Only change wording, sentence structure, and correctness.

ESSAY:
[ESSAY_TEXT]
\end{Verbatim}

\paragraph{SAT Score 4 (Moderate)}

\begin{Verbatim}[breaklines,breakanywhere,fontsize=\small]
Rewrite the following essay by changing only its style (not its content) so that it matches a
moderate SAT style level (score 4) in vocabulary, sentence structure, and mechanics.

Specifically, rewrite the text, while maintaining all the content intact, such that it:
- uses mostly basic but generally appropriate vocabulary, with some repetition or vagueness;
- relies mainly on simple or moderately varied sentence structures, with some awkward or clumsy
  sentences;
- contains noticeable errors in grammar, usage, and mechanics that may distract at times but
  rarely obscure the overall meaning.

Remember: do not change the ideas, arguments, or examples.
Only change wording, sentence structure, and correctness.

ESSAY:
[ESSAY_TEXT]
\end{Verbatim}

\paragraph{SAT Score 5 (Strong)}

\begin{Verbatim}[breaklines,breakanywhere,fontsize=\small]
Rewrite the following essay by changing only its style (not its content) so that it matches a
strong SAT style level (score 5) in vocabulary, sentence structure, and mechanics.

Specifically, rewrite the text, while maintaining all the content intact, such that it:
- uses generally effective vocabulary, with some variety and mostly accurate word choice;
- shows some variety in sentence structure, with mostly clear and controlled sentences;
- contains only occasional errors in grammar, usage, and mechanics, none of which seriously
  interfere with meaning.

Remember: do not change the ideas, arguments, or examples.
Only change wording, sentence structure, and correctness.

ESSAY:
[ESSAY_TEXT]
\end{Verbatim}

\paragraph{SAT Score 6 (High)}

\begin{Verbatim}[breaklines,breakanywhere,fontsize=\small]
Rewrite the following essay by changing only its style (not its content) so that it matches a
high SAT style level (score 6) in vocabulary, sentence structure, and mechanics.

Specifically, rewrite the text, while maintaining all the content intact, such that it:
- exhibits skillful use of language, with varied, precise, and appropriate vocabulary;
- demonstrates meaningful variety in sentence structure, with fluent, well-controlled sentences;
- is almost entirely free of errors in grammar, usage, and mechanics, so that any minor errors do
  not distract or interfere with meaning.

Remember: do not change the ideas, arguments, or examples.
Only change wording, sentence structure, and correctness.

ESSAY:
[ESSAY_TEXT]
\end{Verbatim}

\subsection{Neutral Rewrite Prompt}

To provide a baseline rewrite reflecting a generic LLM paraphrasing style without explicit
proficiency targeting, we use a minimal neutral rewrite prompt. This prompt preserves meaning
and intent while allowing unconstrained stylistic reformulation.

\begin{verbatim}
Your task is to rewrite the following text while preserving its original meaning,
intent, and content.
Do not change, add, or remove any information.

Text:
[TEXT]

Output: Only the rewritten text.
\end{verbatim}

\subsection{Corrective Rewrite Prompt}

In cases where an automated verification step flags a rewrite as failing to preserve content,
we apply a corrective rewrite prompt. This prompt re-injects the original instruction together
with the incorrect output and explicitly emphasizes strict semantic fidelity. As discussed in
Section 3.2, this fallback mechanism was used few times due to the low frequency of content
violations.

\begin{verbatim}
The previous rewrite attempt did not preserve the meaning of the original text.
Your task is to try again.

Here is the original rewriting instruction:
[PROMPT + Essay]

Here is the last (incorrect) output:
[OUTPUT]

Rewrite the text again, this time strictly preserving the meaning of the original
while following the style instructions from the prompt.
Do not add, remove, or change information.
Only adjust style as requested.

OUTPUT: Only the corrected rewritten text.
\end{verbatim}

\section{Verification Prompt}\label{app:verificationPrompt}

To automatically assess whether rewritten texts preserve the original content, we use a dedicated verification prompt designed for binary semantic equivalence evaluation. The evaluator is instructed to ignore stylistic differences entirely and focus solely on whether the rewritten texts maintain the same meaning, arguments, and factual content as the original. The prompt supports batch evaluation, allowing multiple rewrites of the same source essay to be assessed simultaneously. The expected output is a list of binary indicators denoting content preservation.

\subsubsection{Evaluation Prompt Template}

\begin{verbatim}
You are a text evaluation specialist with expertise in socioeconomic style transfer.

-- INPUT DESCRIPTION --
1) Original text:
"""
[ORIGINAL_TEXT]
"""

2) Rewritten text(s):
"""
[REWRITTEN_TEXTS]
"""

-- YOUR TASK --
Compare the rewritten text(s) against the original and decide whether
content was preserved in the rewrites.

Ignore any stylistic changes. Focus only on whether the meaning
and content are preserved.

-- OUTPUT FORMAT --
Output a list of only YES or NO (uppercase, no additional commentary)
for each text.

Example: ['YES', 'NO', 'NO']
\end{verbatim}

\paragraph{Expected Output}  
A list of strings of the form:
\begin{verbatim}
['YES', 'NO', ...]
\end{verbatim}

\end{document}